\documentclass[fleqn,usenatbib]{mnras}

\usepackage{newtxtext,newtxmath}
\usepackage[T1]{fontenc}

\DeclareRobustCommand{\VAN}[3]{#2}
\let\VANthebibliography\thebibliography
\def\thebibliography{\DeclareRobustCommand{\VAN}[3]{##3}\VANthebibliography}

\usepackage{graphicx}	% Including figure files
\usepackage{amsmath}	% Advanced maths commands
\usepackage{subcaption}
\usepackage{enumerate}
\usepackage{csquotes}
\usepackage[dvipsnames]{xcolor}
\usepackage{gensymb}
\usepackage{mwe}
\usepackage{float}
\usepackage{soul}
\usepackage{comment}
\title[Rotation of multiple populations in GCs]{The role of rotation on the formation of second generation stars in globular clusters}%{On the role of globular cluster rotation on the second generation formation within globular clusters}

\author[E. Lacchin et al.]{
E. Lacchin$^{1,2,3}$\thanks{E-mail: elena.lacchin@inaf.it},
F. Calura $^{1}$, %, C. Nipoti $^{2}$\\ 
E. Vesperini $^{4}$,
A. Mastrobuono-Battisti $^{3}$\\
$^{1}$ INAF - OAS, Osservatorio di Astrofisica e Scienza dello Spazio di Bologna, via Gobetti 93/3, I-40129 Bologna, Italy\\
$^{2}$ Department of Physics and Astronomy, University of Bologna, via Gobetti 93/3, 40129 Bologna, Italy\\
$^{3}$ GEPI, Observatoire de Paris, PSL Research University, CNRS, Place Jules Janssen, 9R19P Meudon,  France\\
$^{4}$ Department of Astronomy, Indiana University, Swain West, 727 E. 3rd Street, Bloomington, IN 47405, USA
}
\date{Accepted XXX. Received YYY; in original form ZZZ}
\pubyear{2022}

\begin{document}
\label{firstpage}
\pagerange{\pageref{firstpage}--\pageref{lastpage}}
\maketitle

\defcitealias{calura2019}{C19}
 
\begin{abstract}

By means of 3D hydrodynamic simulations, we explore the effects of rotation in the formation of second-generation (SG) stars in globular clusters (GC).
Our simulations follow the SG formation in a first-generation (FG) internally rotating GC; SG stars form out of FG asymptotic giant branch (AGB) ejecta and external pristine gas accreted by the system.
We have explored two different initial rotational velocity profiles for the FG cluster and two different inclinations of the rotational axis with respect to the direction of motion of the external infalling gas, whose density has also been varied. For a low ($10^{-24}{\rm  g\ cm^{-3}}$) external gas density,
a disk of SG helium-enhanced stars is formed. The SG is characterized by distinct chemo-dynamical phase space patterns: it shows a more rapid rotation than the FG with the helium-enhanced SG subsystem rotating more rapidly than the moderate helium-enhanced one.
 In models with high external gas density ($10^{-23}{\rm  g\ cm^{-3}}$), the inner SG disc is disrupted by the early arrival of external gas and only a small fraction of highly enhanced helium stars preserves the rotation acquired at birth. Variations in the inclination angle between the rotation axis and the direction of the infalling gas and the velocity profile can slightly alter the extent of the stellar disc and the rotational amplitude.
%No significant variation has been found in the timespan of our simulations when changing the inclination angle between the rotation axis and the direction of the infalling gas, while different velocity profiles can slightly alter the extent of the stellar disc and the rotational amplitude.
The results of our simulations illustrate the complex link between dynamical and chemical properties of multiple populations and provide new elements for the interpretation of observational studies and future investigations of the dynamics of multiple-population GCs.

%By means of 3D hydrodynamic simulations, we explore the effects of rotation in the formation of second generation (SG) stars in a massive  proto-globular cluster (GC) and their dynamical and chemical evolution. We assume that SG stars were formed out of the rotating first generation asymptotic giant branch (AGB) stars plus external, non-rotating gas accreted by the system. We have tested two different rotational velocity profiles for FG stars and two different inclinations of their rotational axis with respect to the direction of the accreted gas, whose density has also been varied. For a low external gas density, a disk of helium enhanced stars is formed, at variance with the high density one where {\bf the disk} is disrupted after the arrival of external gas. When the disk survives, SG stars rotational amplitude is significantly larger than FG stars one and moreover, extremely helium enhanced SG stars are rotating faster than poorly enriched ones. No significant variation has been found in the timespan of our simulation changing the inclination angle, while different velocity profiles can alter the extent of the disk and the rotational intensity. 

\end{abstract}

\begin{keywords}
hydrodynamics - methods: numerical  - globular cluster: general - galaxies: star formation - stars: kinematics and dynamics
\end{keywords}

%%%%%%%%%%%%%%%%%%%%%%%%%%%%%%%%%%%%%%%%%%%%%%%%%%

%%%%%%%%%%%%%%%%% BODY OF PAPER %%%%%%%%%%%%%%%%%%

\section{Introduction}
\label{sec:Intro}

%It is already well known that globular clusters (GCs) host multiple stellar populations. From both spectroscopic and photometric studies, it has been observed star-to-star variations in the chemical composition of GC stars when light elements are concerned, all of them linked by (anti)correlations.
Spectroscopic and photometric studies have now clearly revealed that globular clusters host multiple stellar populations with different chemical compositions. A significant fraction of GC stars show, in fact, \enquote{anomalous} chemical composition rarely found in the field, such as an enhancement in He, Al and Na and a depletion in O, Mg \citep{piotto2005,carretta2009a,milone2017,gratton2019,masseron2019,marino2019}.
Stars are generally divided, depending on their chemical composition, into two or more groups: the first population  (or first generation; hereafter FG) share the same chemical pattern with field stars, while the second population (or second generation; hereafter SG) display anomalous abundances. Multiple populations within the same GC have been detected not only in the Milky Way GCs, but also in external galaxies such as the Magellanic Clouds \citep{mucciarelli2009}, Fornax galaxy \citep{larsen2014} and M31 \citep{nardiello2019}.% even though differences are seen when focusing on the fraction of SG stars. In the Galactic GCs, its value reaches $\sim 0.7$ while in the Magellanic Clouds clusters the FG is found to be the dominant component \citep{dalessandro2016,milone2020,dondoglio2021}.

So far, a complete picture explaining the origin of anomalous stars, and therefore the formation of multiple stellar populations within GCs, is still lacking. Several scenarios have been proposed to address this topic, nevertheless none of them is able to reproduce all observational constraints \citep{renzini2015,bastian2018,gratton2019}. The main difference between the various scenarios resides in the identification of the stars providing the processed gas out of which SG stars formed. Possible sources of processed gas, suggested in the literature, include asymptotic giant branch (AGB) stars \citep{dercole2008,dercole2016}, fast-rotating massive stars \citep{decressin2007}, massive stars \citep{elmegreen2017}, supermassive stars \citep{denissenkov&hartwick2014,gieles2018}, massive interacting binaries \citep{demink2009,bastian2013}, black holes accretion disks \citep{breen2018} and stellar mergers \citep{wang2020}.

Among the aforementioned scenarios, the most thoroughly studied to date is the AGB one, which is also the one we are adopting in the present work. In this scenario, after the formation of FG stars, FG massive stars clear the system from both the gas leftover by the FG formation and the one ejected by massive stars themselves \citep{calura2015}. SG are then formed out of the ejecta of FG AGB stars plus pristine gas with same chemical composition of the gas from which FG were formed \citep{dercole2010,dercole2012,dantona2016}.

\citet{dercole2008} performed, for the first time, hydrodynamic simulations of a star-forming GC in the AGB framework finding that the gas ejected by AGB stars collects in a cooling flow towards the cluster centre. SG stars form in the central regions of the FG system out of this gas. The predicted central concentration of SG stars relative to the FG population has been observed in several clusters where some memory of the initial differences between generations has been preserved. Various studies have also shown that AGB ejecta alone do not reproduce the observed anticorrelations, such as the Na-O one \citep{carretta2009a}. In order to produce the observed chemical patterns, AGB ejecta should be diluted with pristine gas (namely gas with the same chemical composition as the one out of which FG were formed) while SG stars are formed \citep{dercole2012,dercole2016,dantona2016}. Dilution of processed gas with pristine gas is required by many of the aforementioned scenarios \citep{gratton2019}. 
\citet{calura2019} performed the first series of 3D hydrodynamic simulations of a massive proto-GC in the AGB framework, following up the work of \citet{dercole2008} which is instead based on 1D simulations. They take into account the radiative cooling and modelled, together with the AGB feedback, the accretion of pristine gas. They found, in addition to \citet{dercole2008} results, that the most helium enriched SG stars, which are also the first to be formed, are characterized by a more compact spatial distribution than the less helium enriched ones. This trend is consistent with that found in  observational studies of the SG populations spatial distribution \citep{johnson2012,simioni2016}.

In most studies on GC formation, clusters are modelled as non-rotating systems; however, in the last few years an increasing number of GCs have been found to show signatures of internal rotation (see e.g. \citealt{pancino2007,bellazzini2012,fabricius2014,lardo2015,boberg2017,bianchini2018,ferraro2018, lanzoni2018a,lanzoni2018b,kamann2018,sollima2019,vasiliev2021}). 
The observed internal rotation is generally found to be quite moderate with typical ratios between the rotational amplitudes to the central velocity dispersion $(V_{\rm rot}/\sigma_0)$ ranging from 0.05 to about 0.7 \citep{bellazzini2012,fabricius2014,kamann2018}. Present-day rotation, however, is most likely the remnant of a stronger early rotation \citep{henaultbrunet2012,mapelli2017} which has been then lessened under the effects of two-body relaxation \citep{bianchini2018,kamann2018,sollima2019} and tidal forces exerted by the host galaxy. During the long term evolution of the cluster, the combined effect of angular momentum transport from the cluster centre outwards and of escaping stars carrying it away, leads to a loss of angular momentum and therefore to a decrease of the rotational amplitude (see e.g. \citealt{einsel1999,tiongco2017,livernois2022} ).

It has been shown that internal rotation can strongly affect GC long term evolution, in  particular both reducing the relaxation timescale (or shortening the time of core collapse), and therefore accelerating its dynamical evolution, and the mass-loss rate \citep{einsel1999,kim2002,kim2004,ernst2007,kim2008,hong2013,mastrobuonobattisti2021}. Moreover, rotation affects also the present-day morphology \citep{hong2013}, as well as the dynamics of multiple stellar populations \citep{mastrobuonobattisti2013,mastrobuonobattisti2016,henaultbrunet2015,tiongco2019,mastrobuonobattisti2021}.

The observational study of the kinematics of multiple stellar populations is still in its early stages, but a few investigations have already revealed differences in the rotation amplitudes between stellar populations \citep{lee2015,lee2017,lee2018,dalessandro2021,cordoni2020,szigeti2021}, with the SG component rotating faster than the FG, with the exception of $\omega$ Cen where \citet{bellini2018} found opposite results. 
Recently, \citet{cordero2017} and \citet{kamann2020} analyzed the rotational patterns of two GCs subpopulations and found that extreme SG stars (Na-enhanced and extremely O-depleted) are characterized by a larger rotational amplitude than intermediate SG ones (moderately Na-enhanced and O-depleted). 
In some cases, instead, no difference as been detected in the rotational patters of distinct populations \citep{milone2020,cordoni2020,cordoni2020b,szigeti2021}. This rotational homogeneity between different populations could hint that these systems are dynamically evolved, where stars are already well kinematically mixed.

Rotation is, therefore, a fundamental physical process which needs to be included in theoretical models, however, very few studies have investigated its effects on the formation of multiple populations in GCs so far.
\citet{bekki2010}, \citet{bekki2011} and more recently \citet{mckenzie2021}, studied the effects of rotation on the SG formation in the AGB framework, through 3D hydrodynamic simulations. They found that SG and FG stars display significant differences both in the rotational velocity ($V_{\rm rot}$) and in the velocity dispersion ($\sigma$) profiles, with SG stars rotating faster but with a smaller dispersion. Moreover, SG stars show a flattened and compact spatial distribution which should be smoothed out during the cluster evolution in order to match the observations. These initial kinematical differences between various stellar populations are found to significantly decrease during the subsequent evolution of the cluster on two-body relaxation time-scales, which points towards a spatial and kinematical mixing of the populations \citep{vesperini2013,mastrobuonobattisti2013,henaultbrunet2015,mastrobuonobattisti2016,tiongco2019,vesperini2021}.

%However, \citet{bekki2010,bekki2011} and \citet{mckenzie2021} do not follow the evolution of any chemical element and no infall is assumed as a source of dilution of the AGB ejecta, a strict requirement for the AGB scenario to work. 

In this work, we extend the study of \citet[hereafter \citetalias{calura2019}]{calura2019} to investigate the effects of FG rotation on the formation of SG stars in a massive proto-GC, through a series of 3D hydrodynamic simulations. In our simulations, we will explore how the interplay between the kinematics of the AGB ejecta released by a rotating FG system and the accreted pristine gas affect the final kinematic properties of SG stars and the dependence of these properties on the chemical composition for different SG subpopulations.

The paper is organized as follows: in Section \ref{sec:simuset} we present the setup of our simulations focusing on the implementation of the rotating FG component. 
Section \ref{sec:results} contains the results of out models and describes the dynamics of both the gas and the stellar component. In Section \ref{sec:discussion} we discuss the outcomes of the simulations and compare them with previous results in the literature. Finally, we summarize our conclusion in Section \ref{sec:conclusions}. 

\section{SIMULATION SETUP}

\label{sec:simuset}
%\textcolor{blue}{Our aim is to study the effects of Type Ia SN explosions on the SF of a young GC with an initial mass of ${\rm M_{FG}=10^7M_{\odot}}$.density of the pristine gas!!!!!!!!!! ${\rm \rho_{pg}=10^{-24} g/cm^3} $ and ${\rm \rho_{pg}=10^{-23} g/cm^3} $ }

The initial setup of this work is similar to the one of \citetalias{calura2019} with the difference that we are here assuming a rotational FG. Therefore, all our simulations start ${t_{\rm AGB}=\rm 39Myr}$ after the FG formation, once all massive stars have already exploded as core-collapse supernovae (CC-SNe) and the system is completely cleared out of both SN-enriched ejecta and pristine gas \citep{calura2015}. At this time, the intermediate mass stars are undergoing their AGB phase, returning mass and energy in the gas-free system. %For computational reasons, however, we assume that the computational box is filled with a negligible amount of gas. 

We perform 3D hydrodynamic simulations using a customized version of the adaptive mesh refinement (AMR) code RAMSES \citep{teyssier2002}, which uses an unsplit second-order Godunov method to solve the Euler equations of hydrodynamics. The gas obeys the adiabatic equation of state for an ideal monoatomic gas $P\propto \rho^{\gamma}$, where $P$ and $\rho$ are the pressure and density of the gas and the adiabatic index $\gamma$ is set to $5/3$. We take into account various astrophysical processes such as the mass and energy return from AGB stars, radiative cooling and star formation. For simplicity, the FG system is modeled as a static \citet{plummer1911} density profile with mass $M_{\rm FG}=10^7 {\rm M_{\rm \odot}}$ and Plummer radius $a=23$pc. On the other hand, SG stars are instead modelled as collisionless particles; their dynamic evolution is derived by means of a Particle-Mesh solver.
This paper is focused on the formation and early dynamics of multiple populations and an investigation of the origin of the various chemical patterns is beyond the scope of this work. Here we only trace the evolution of the helium abundance of both the gas and the stars and use this abundance to identify the various SG subgroups.
%Moreover, we trace the evolution of the helium abundance of both the gas and the stars and use this abundance to identify various subgroups of Sg populations. 
In all our simulations, we adopt a size of the computational box of ${\rm L^3=(292\ pc)^3}$ uniformly divided into $(512)^3$ cells, which corresponds to a spatial resolution of 0.6 pc.

\begin{table}
%\scalebox{0.9}{
\centering
\caption{Simulations parameters.}
\hspace{-0.8cm}
\resizebox{1.05\columnwidth}{!}{
\renewcommand{\arraystretch}{1.40}
\begin{tabular}{llc} 
\hline
\hline
Parameter & Description & Values \\
\hline
$M_{\rm FG}$      & FG stellar mass                           & $10^7 {\rm M_{\odot}}$\\
$a$    & FG Plummer radius     & 23 pc\\
$\rho_{\rm pg} $  & Density of the pristine gas                      & ${\rm 10^{-24,-23} g\ cm ^{-3}} $ \\
$v_{\rm pg}  $    & Pristine gas velocity relative to the cluster    & ${\rm 20\ km\ s^{-1}}$   \\
$Z_{\rm pg}  $    & Metallicity of the pristine gas                  & 0.001  \\
$X_{\rm He} $     & Pristine gas helium mass fraction          & 0.246  \\
$T_{\rm pg} $     & Temperature of the pristine gas                  & $10^4 {\rm  K}$ \\
$T_{\rm floor}$   & Minimum temperature                              & $10^3{\rm K}$ \\
$t_{\star} $      & Star formation time-scale                        & 0.1 Gyr \\
\hline
$v_{\rm pk}$      & FG velocity at $R_{\rm pk}$ for analytic profile (Eq. \ref{eq:anarot}) & $2.5\ {\rm km\ s^{-1}}$ \\
$v_{\rm SB}$      & FG velocity at $a$ for solid body profile & $2.5\ {\rm km\ s^{-1}}$ \\
\hline
\hline
\end{tabular}
}
\label{tab:param}

\end{table}

\begin{table*}
%\centering
\caption{Models description. {\it Columns}: (1) name of the model, (2) axis around which the cluster rotates, (3) type of velocity profile ,(4) density of the pristine gas, (5) starting time of the infall (the initial time of the simulations is fixed at $t_{\rm AGB}=39$Myr) , (6) spatial resolution.} 
\begin{tabular}{cccccc} 
\hline
\hline
 Model & Rotational axis&Velocity profile&$\rho_{\rm pg} {\rm (g\ cm^{-3})} $&${\rm t_{inf} (Myr)} $& Resolution (pc)\\
\hline

%{\rm cool\_SF\_hinf\_ana\_x}$ & ${\rm v_{pk}=2.5km/s}$; along $x$ axis&$10^{-23} $& 1&1.1\\
%${\rm cool\_SF\_hinf\_ana\_z}$& along $z$ axis&$10^{-23} $& 1&1.1\\
%${\rm cool\_SF\_iinf\_ana\_x}$& along $x$ axis&$5\times 10^{-24} $& 9&1.1\\
%${\rm cool\_SF\_iinf\_ana\_z}$& along $z$ axis&$5\times 10^{-24} $& 9&1.1\\
%${\rm cool\_SF\_linf\_ana\_x}$& along $x$ axis&$10^{-24} $& 21&1.1\\
%${\rm cool\_SF\_linf\_ana\_z}$& along $z$ axis&$10^{-24} $& 21&1.1\\
%hline
%${\rm cool\_SF\_hinf\_sb\_x}$& ${\rm v_{r_h}=2.5km/s}$; along $x$ axis&$10^{-23} $& 1&1.1\\
%${\rm cool\_SF\_hinf\_sb\_z}$ & along $z$ axis&$10^{-23} $& 1&1.1\\
%${\rm cool\_SF\_linf\_sb\_x}$& along $x$ axis&$10^{-24} $& 21&1.1\\
%${\rm cool\_SF\_linf\_sb\_z}$& along $z$ axis&$10^{-24} $& 21&1.1\\
%\hline
%${\rm cool\_SF\_hinf}$ & -&$10^{-23} $& 1&1.1\\
%${\rm cool\_SF\_linf}$ & -&$10^{-24} $& 21&1.1\\
%\hline
%\hline
${\rm HDanax}$& along $x$ &analytic &$10^{-23} $& 1&0.57\\
${\rm HDanaz}$& along $z$ &analytic &$10^{-23} $& 1&0.57\\
%${\rm cool\_SF\_hinf\_ana\_x}$& SF, rot ana; LMAX9 BOX9e20& $10^7$ & along $x$ axis&$5\times 10^{-24} $& 9&0.57\\
%${\rm cool\_SF\_hinf\_ana\_z}$& SF, rot ana;  LMAX9 BOX9e20& $10^7$ & along $z$ axis&$5\times 10^{-24} $& 9&0.57\\
%${\rm Iinf\_ana\_x}$& along $x$ &analytic &$5\times 10^{-24} $& 9&0.57\\
%${\rm Iinf\_ana\_z}$ & along $z$ &analytic &$5\times 10^{-24} $& 9&0.57\\
${\rm LDanax}$& along $x$ &analytic &$10^{-24} $& 21&0.57\\
${\rm LDanaz}$ & along $z$ &analytic &$10^{-24} $& 21&0.57\\
\hline
%${\rm cool\_SF\_hinf\_sb\_x}$& SF, solid body; LMAX9 BOX9e20& $10^7$ & ${\rm v_{(r_h}=2.5km/s}$; along $x$ axis&$10^{-23} $& 1&0.57\\
%${\rm cool\_SF\_hinf\_sb\_z}$& SF, solid body;  LMAX8 BOX9e20& $10^7$ & along $z$ axis&$10^{-23} $& 1&0.57\\
%${\rm cool\_SF\_linf\_sb\_x}$& SF, solid body; LMAX8 BOX9e20& $10^7$ & along $x$ axis&$10^{-24} $& 21&0.57\\
${\rm LDsbz}$ &along $z$& solid body &$10^{-24} $& 21&0.57\\
\hline
${\rm HD}$& -&-&$10^{-23} $& 1&0.57\\
%${\rm Iinf}$& -&-&$5\times 10^{-24} $& 9&0.57\\
${\rm LD}$ & -&-&$10^{-24} $& 21&0.57\\
\hline
\hline

\end{tabular}
\label{tab:simu}
\end{table*}

\subsection{Initial setup}
As in \citetalias{calura2019}, we follow the scenario of \citet{dercole2016} and therefore assume that the cluster is located in the disk of a high-$z$ star-forming galaxy \citep{kravtsov&gnedin2005,bekki2012,kruijssen2015}. In addition, we assume that the cluster is orbiting in its host galaxy, which leads to an asymmetric accretion of gas from the side towards which the system is moving. In models, this is accomplished by placing the cluster at the centre of the computational box and, at time $t_{\rm inf}$, gas starts to flow from one of the boundaries. Such accretion leads to a dilution of the AGB ejecta with pristine gas which, as discussed in the Sec. \ref{sec:Intro}, is required in all the models to match the chemical trends derived by observations. The beginning of the infall of pristine gas, however, does not correspond to the beginning of the simulation. As in \citet{dercole2016}, we assume that FG CC-SNe explosions have carved a large cavity around the cluster, an hypothesis confirmed also by 3D hydrodynamic simulations of the expansion of SN driven bubbles located in galactic disks \citep{hopkins2012,creasey2013,walch2015,KIM2017}.

Once the wind ram pressure of the expanding shell equals the pressure of the surrounding ISM, the bubble stalls, losing its original structure. The maximum radius reached by the cavity, which corresponds to the stalling radius, is:

\begin{equation}
 R_{\rm eq,2}=41.43\left( \frac{L_{41}}{n_0 V_{w,8}(\sigma^2_{0,6}+v^2_{\rm pg,6})}\right)^{1/2} 
\end{equation}
 where $L_{41}$ is the mechanical luminosity of CC-SNe of FG stars in units of $10^{41} {\rm erg\ s^{-1}}$ whose value reflects the number of FG CC-SNe and therefore on the initial cluster mass. In our case, with a $M_{\rm FG}=10^7 {\rm M_{ \odot}}$ its value is around unity for a standard initial mass function (IMF) like \citet{kroupa2001}.
 The quantity $n_0 $ represents the ISM number density while $V_{w,8}$ is the velocity of the wind in units of ${\rm 10^{8} cm\ s^{-1}}$ (see \citealt{dercole2016} for detailed calculations of $L_{41}$ and $V_{w}$). The two velocities $\sigma_{0,6} \sim 1$ and $v_{\rm pg,6} \sim 2$, both expressed in units of ${\rm 10^{6} cm\ s^{-1}}$, represent the velocity dispersion of the cluster within the galaxy, namely the isothermal sound speed and the velocity of the recollapsing ISM relative to the system, respectively.

After the bubble has stalled, the ISM gas starts refilling the cavity with a velocity comparable with the local sound speed. The time at which the ISM gas reaches the system depends on the stalling radius $R_{\rm eq}$ through:

\begin{equation}
t_{\rm inf}=t_{\rm SNe}+\frac{R_{\rm eq}}{\sigma_{0}+v_{\rm pg}}
\label{eq:infall}
\end{equation}

where $t_{\rm SN}=30$Myr is the lifetime of the smallest star which explodes as a CC-SN, so after $t_{\rm SNe}$ no more FG massive stars ($m>8{\rm M_{\odot}}$) are going off. 
We have run simulations with the same values of  the pristine gas density adopted in \citet{calura2019}: a low-density case with ${\rm \rho_{pg}=10^{-24} g\ cm^{-3}} $ and a high-density one ${\rm \rho_{pg}=10^{-23} g\ cm^{-3}} $, representing, respectively, the typical gas density in a dwarf galaxy and in a high-redshift disc.
%We have here performed simulations adopting two different values for the pristine gas density: a low density case with ${\rm \rho_{pg}=10^{-24} g\ cm^{-3}} $ and a high density one assuming ${\rm \rho_{pg}=10^{-23} g\ cm^{-3}}$. {\bf These values correspond to the average ISM density at the present time and a value 10 times greater, motivated by the fact that in the early Universe the ISM density is likely to have been higher.\textcolor{red}{citare qualcosa? d'ercole16 aveva preso $10^{-22}$? }}.
Since ${\rm n_0 \sim \rho_{pg}/m_p}$ we derive that, for a pristine gas density of ${\rm \rho_{pg}=10^{-24} g\ cm^{-3}} $ the infall begins around 60 Myr after the FG was formed, while for a higher ISM density, it starts after 40Myr. 

At the beginning of the simulation, the boundaries of all six faces of our computational box are set to be outflow. At $t_{\rm inf}$, the pristine gas flows into the box though the $yz$ plane at negative $x$. 

Table \ref{tab:param} contains a summary of the main parameters assumed for our simulations, whereas Table \ref{tab:simu} reports a description of the models we have performed with a particular emphasis on the rotational prescriptions we have adopted.

We point out that our model is aimed at exploring the formation of SG stars in a massive cluster with a present-day mass $\approx 10^6 {\rm M_{\odot}}$. In a future investigation, we will explore the implications of initial rotation for a broader range of initial conditions including clusters with different masses and structural parameters (see e.g. \citealt{yaghoobi2022} for a study of the formation of multiple populations in non-rotating clusters with different initial masses).

\subsection{Physical ingredients}
\label{sec:phy_ing}

We summarize here the main physical ingredients which are included in our simulations, with a particular focus on the implementation of rotation in the FG system (for more details on the basic setup, see \citetalias{calura2019} and \citealt{lacchin2021}). 

The star formation is here modelled sub-grid as described by \citet{rasera&teyssier2006}. Stars are formed only in cells where the gas temperature is lower than $2\times 10^4$ K, which means only if the gas is in its neutral form. In addition, the velocity field should be convergent and, therefore, $\nabla \cdot v <0 $. For computational reasons, not all the gas in a cell is eligible for star formation; in all our simulations, 90 per cent of the gas is allowed to be converted into star particles according to the \citet{schmidt1959} law:

\begin{equation}
\dot{\rho}_{\rm \star, SG} =\frac{\rho}{t_{\star}}
\end{equation}

where $t_{\star}$ corresponds to the time scale of star formation and is proportional to the free-fall time. In all our runs, we have assumed $t_{\star}=0.1$ Gyr.
To every newborn particle, we associate a mass $M_{\rm p}=N\ m_0$ where $m_0=0.1  \mathrm{M_{\odot}}$ is the minimum mass and $N $ is derived from the Poisson distribution assuming a mean of $\lambda_{\rm p}= \left( \frac{\rho \Delta x ^3}{m_0}\right) \frac{\Delta t}{t_{\star}} $. In every cell, only one star particle is allowed to form at each timestep, which is then located in the cell centre. Both the chemical composition and the velocity of each newborn particle are equal to those of the gas in the parent cell. To conserve mass, momentum and energy, all these quantities are properly removed from the parental cell once a star particle is formed. 

The FG is instead modelled as a static component spatially distributed following the analytic Plummer mass density profile with $M_{\rm FG}=10^7{\rm M_{\odot}}$ and $a=23 $ pc. The system is evolved for 65 Myr to compare our results with the one of \citetalias{calura2019}.

To model the mass return of FG AGB stars, a source term is added to the mass conservation equation. The mass injected by AGB stars per unit time and volume is:

\begin{equation}
\dot{\rho}_{\rm \star,AGB}=\alpha \rho_{\star}
\end{equation}
 where $\alpha$ is the specific mass return rate as a function of time of the FG component, while $\rho_{\star}$ is the local density of FG stars. 
 
The energetic feedback of AGB stars is also modelled as a source term and takes the form of: 

\begin{equation}
S=0.5\alpha \rho_{\star} (3 \sigma^2+ v^2+v_{\rm wind}^2)
\end{equation}

where $\sigma$ represents the FG velocity dispersion, $v$ is the velocity of the gas while $v_{\rm wind}$ is the wind velocity of the AGB stars, which we assume to be $\sim 2\times 10^6 {\rm cm\ s^{-1}}$\citep{dercole2008}. 

Here, we trace the helium composition both of the gas and of SG stars. We assume, for AGB stars the yields of \citet{ventura&dantona2011}, therefore the helium mass fraction of the AGB ejecta spans between 0.36 for the gas released by the most massive AGBs to 0.32 for AGB ejecta at the end of the simulations. No iron is assumed to be produced inside AGBs, therefore the iron composition of their ejecta is the same as the one of the pristine gas 

The source terms $\dot{\rho}_{\rm \star,AGB}$ and S are added at each timestep to the density and energy of the fluid, respectively, as well as the cooling term which takes into account the loss of energy due to radiation (see \citealt{few2014}). The pristine gas inflowing the system has a fixed temperature of $T=10^4$K, a typical value for the warm photoionised ISM in a star-forming galaxy \citep{haffner2009}.% We also apply a temperature floor of $T=10^3$K as in \citetalias{calura2019}.

\subsection{Rotation}

Signatures of internal rotation are found both in simulations of star cluster formation through the collapse of giant molecular clouds \citep{mapelli2017,ballone2020,chen2021} and from observations of young clusters (see e.g. \citealt{fischer1992,fischer1993,henaultbrunet2012}).

In addition to the setup of \citetalias{calura2019}, here we assume that the FG system is characterized by the presence of internal rotation. This is implemented by imparting a rotation velocity to the AGB ejecta following the rotational curve radial profile suggested by \citet{lyndenbell1967} and found to provide a good description of the observed rotation profile in several star clusters (\citealt{mackey2013,Kacharov2014,bianchini2018,kamann2020,dalessandro2021,leanza2022}) of the form:

\begin{equation}
    v_{\rm rot}=\frac{2 v_{\rm pk} R}{R_{\rm pk}} \left( 1+\left( \frac{R}{R_{\rm pk}}\right)^2 \right)^{-1}
    \label{eq:anarot}
\end{equation}
 $R_{\rm pk}$ represents the location of the peak of the profile, while $v_{\rm pk}$ is the value of the rotational amplitude at the peak. Here, we have set $R_{\rm pk}=a$ and $v_{\rm pk}=2.5\ {\rm km \ s^{-1}}$; the present-day peak rotational velocity of old clusters is, in most cases, smaller than the value adopted here, a consequence of the effects of long-term dynamical evolution leading to a gradual decrease in the strength of internal rotation (see e.g. \citealt{tiongco2017} and references therein).

 In Appendix \ref{sec:appendix} we also report the results of models assuming a solid body rotation. Such profile has been adopted by some works in literature such as \citet{bekki2010,bekki2011} and \citet{mckenzie2021}. The angular velocity $\omega$ has been derived assuming a velocity at the Plummer radius $ v_{rot}(r=a)= 2.5\ {\rm km \ s^{-1}}$.

%%%%%%%%%%%%%%%%%%%%%%%%%%%%%%%%%%%%%%%%%%%%%%%%%%%%%%%%%%%%%%%%%%%%%%%%%%%%%%%%%%%%%%%%%%%%%%%%
%%%%%%%%%%%%%%%%%%%%%%%%%%%%%%%%%%%%%%%%%%%%%%%%%%%%%%%%%%%%%%%%%%%%%%%%%%%%%%%%%%%%%%%%%%%%%%%%
%%%%%%%%%%%%%%%%%%%%%%%%%%%%%%%%%%%%%%%%%%%%%%%%%%%%%%%%%%%%%%%%%%%%%%%%%%%%%%%%%%%%%%%%%%%%%%%%
\section{RESULTS}
\label{sec:results}
In this section we present the results of the models assuming that the FG internal rotation follows the radial profile in Eq. \ref{eq:anarot}. 
 We have explored two different orientations for the rotational axis (parallel and perpendicular to the infall, see \citealt{tiongco2018} and \citealt{tiongco2022} for $N$-body studies of the long-term evolution of clusters with different orientations of the rotation axis) studying their effects for various pristine gas densities ($10^{-24}-10^{-23}{\rm g\cdot cm^{-3}}$). All model parameters are reported in Table \ref{tab:simu}.
 
 The results obtained by adopting a solid-body rotation are very similar to those found assuming the profile in Eq. \ref{eq:anarot}; for this reason, here we present only the results obtained for the models with the analytical profile in Eq. \ref{eq:anarot}. We discuss the comparison with the models adopting a solid-body rotation in Sec. \ref{sec:appendix}.

%%%%%%%%%%%%%%%%%%%%%%%%%%%%%%%%%%%%%%%%%%%%%%%%%%%%%%%%%%%%%%%%%%%%%%%%%%%%%%%%%%%%%%%%%%%%%%%%

\begin{figure*}
        \centering

        \includegraphics[width=0.953\textwidth]{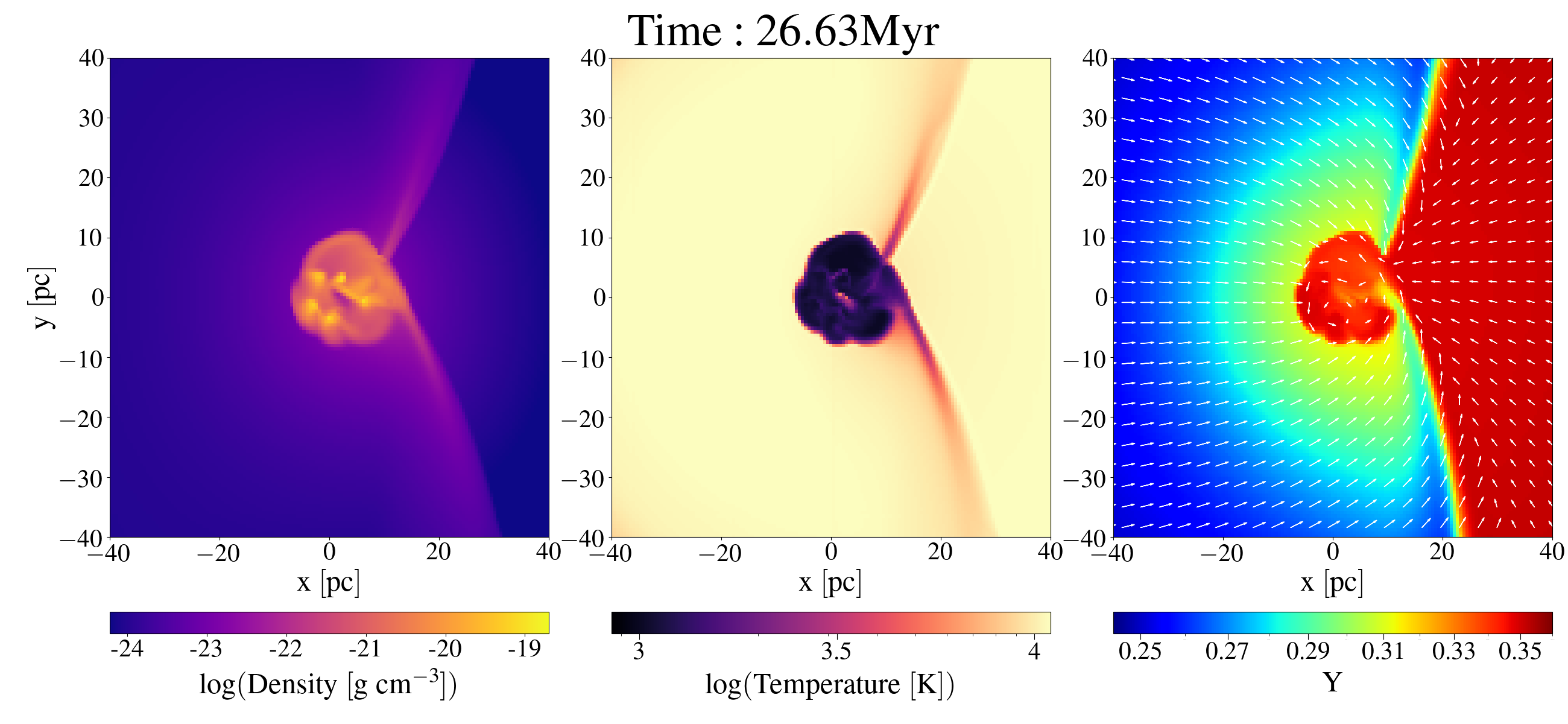}
        \\
        \includegraphics[width=0.953\textwidth]{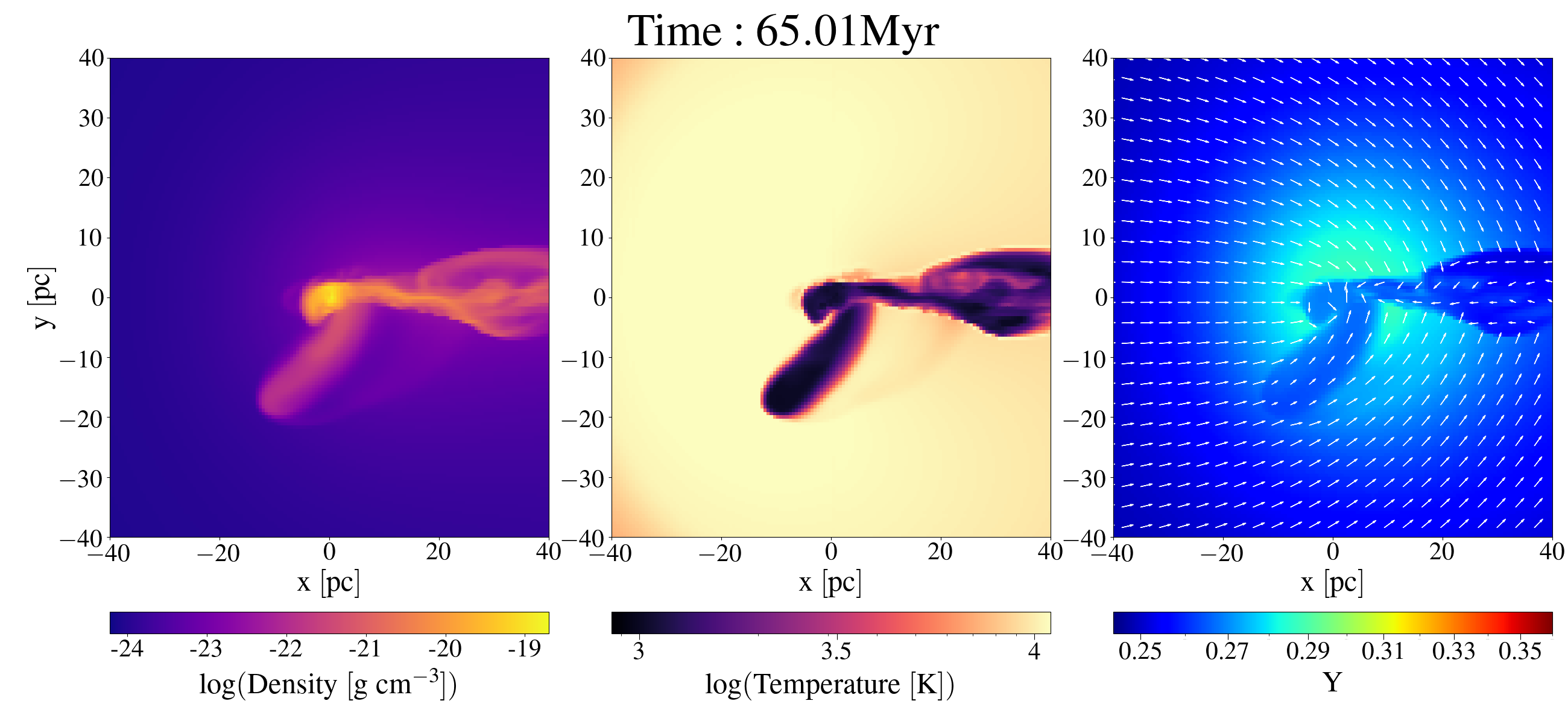}

\caption{Two-dimensional maps of the gas density (left-hand panels) and of the temperature (central panels) and the helium mass fraction (right-hand panels) on the x-y plane for the {\tt LDanaz} simulation. The corresponding evolutionary time of each set of panels is reported at the top. The white arrows in the helium mass fraction maps represent the gas velocity field.}
  \label{fig:maps_LDanaz}
\end{figure*}
%%%%%%%%%%%%%%%%%%%%%%%%%%%%%%%%%%%%%%%%%%%%%%%%%%%%%
\begin{figure*}
        \centering

       \includegraphics[width=0.324\textwidth]{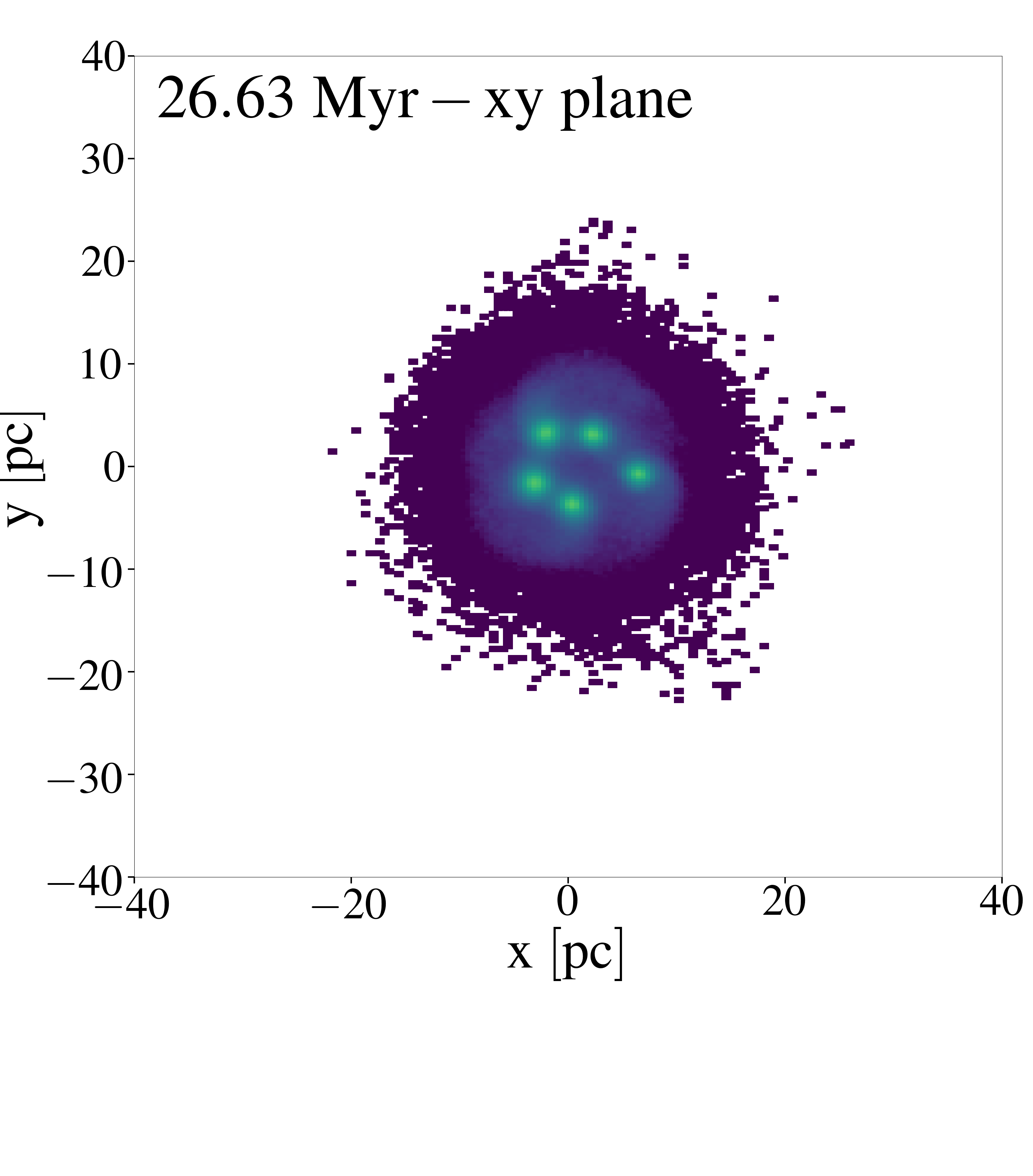}        
        \includegraphics[width=0.324\textwidth]{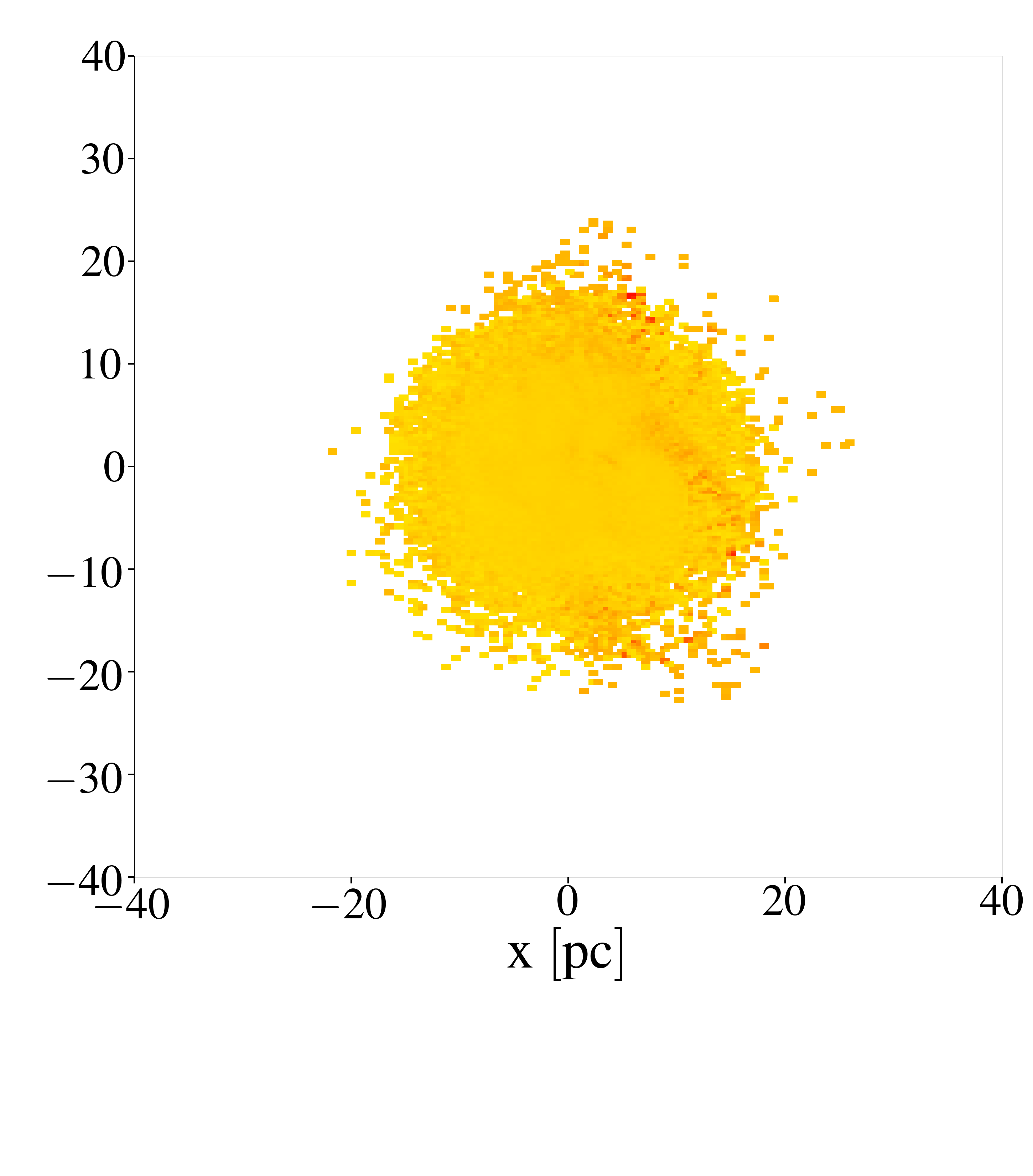}
         \includegraphics[width=0.324\textwidth]{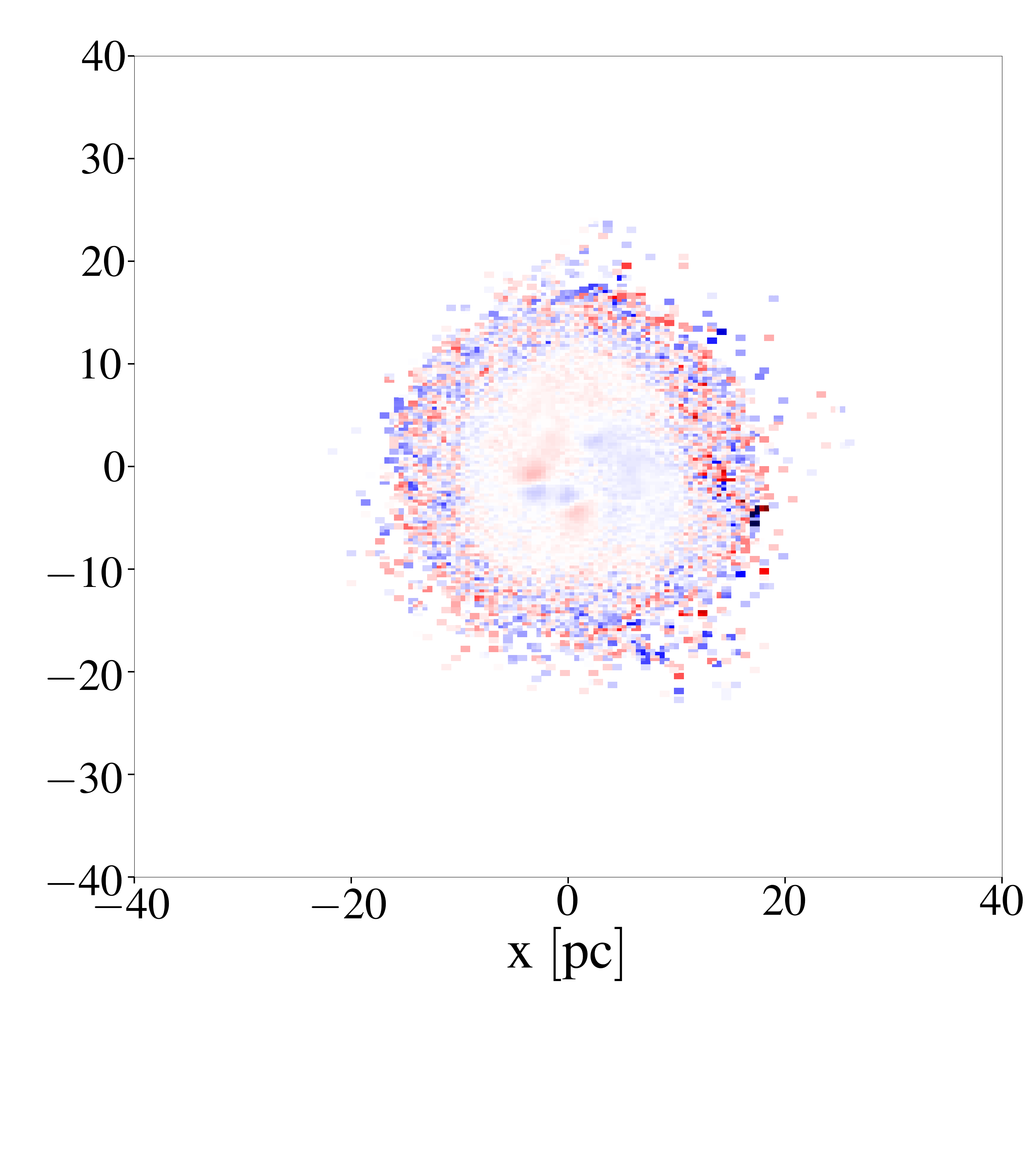}
      \\
       \vspace{-1.1cm}
      
        \includegraphics[width=0.324\textwidth]{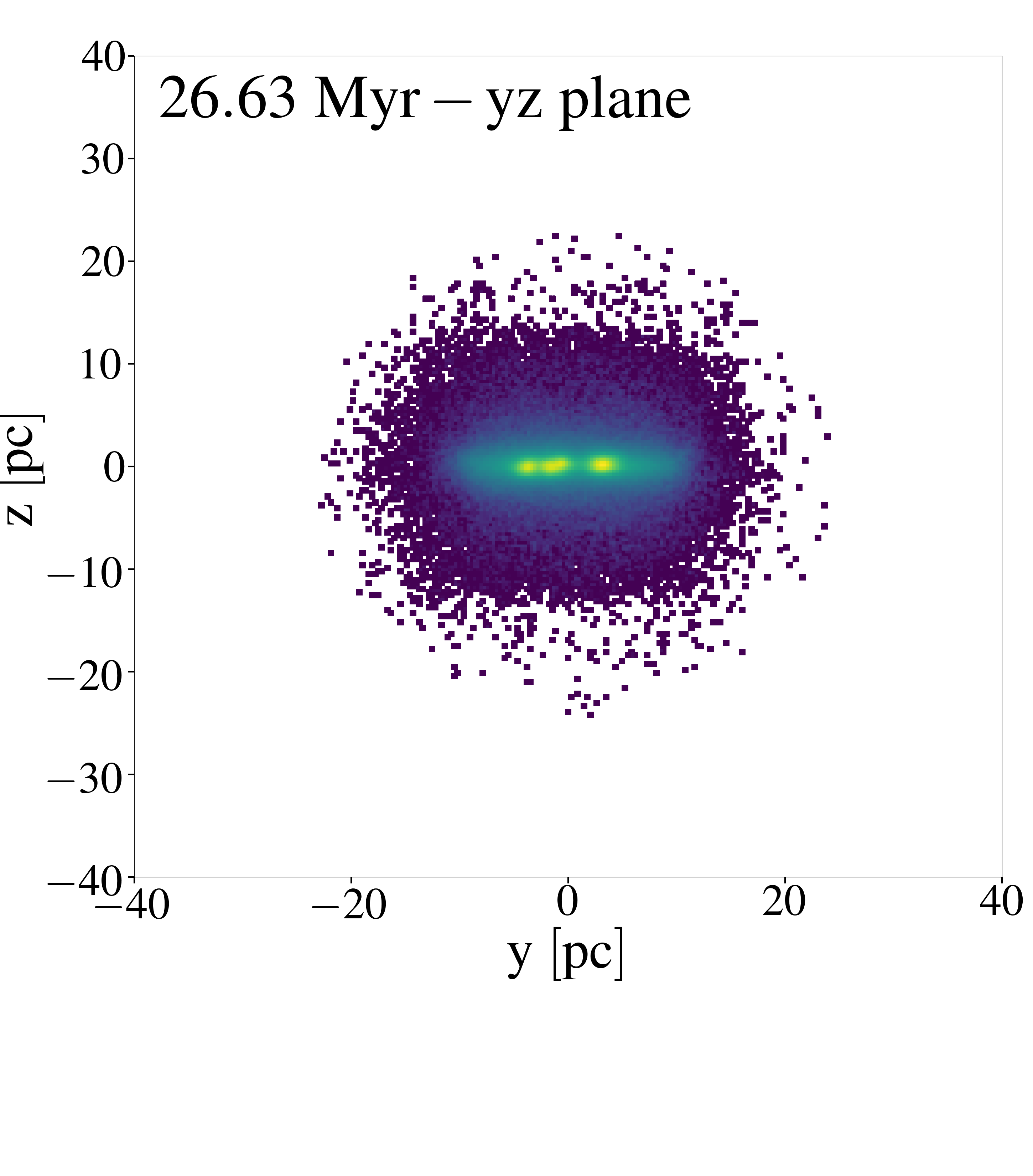}
        \includegraphics[width=0.324\textwidth]{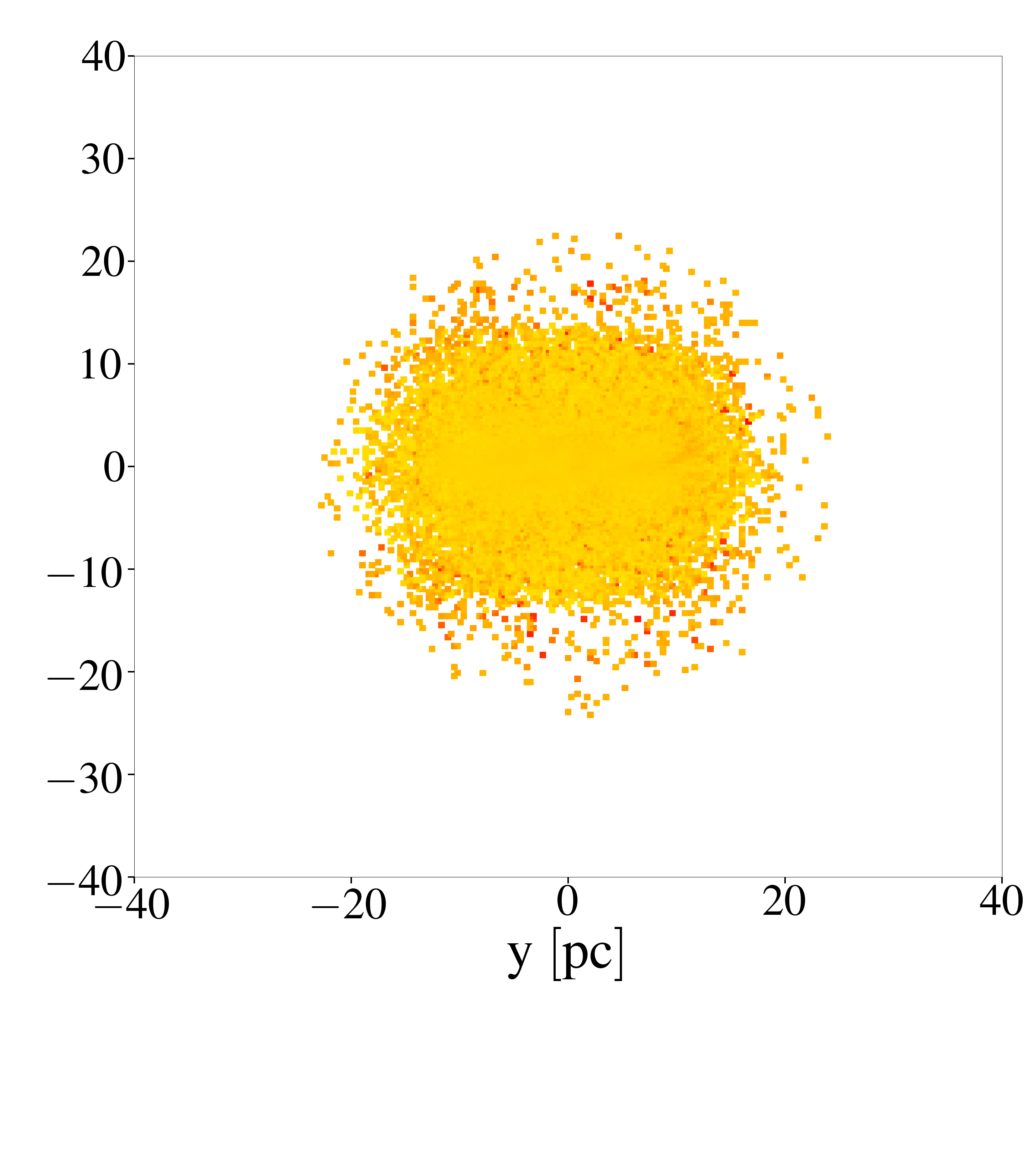}
           \includegraphics[width=0.324\textwidth]{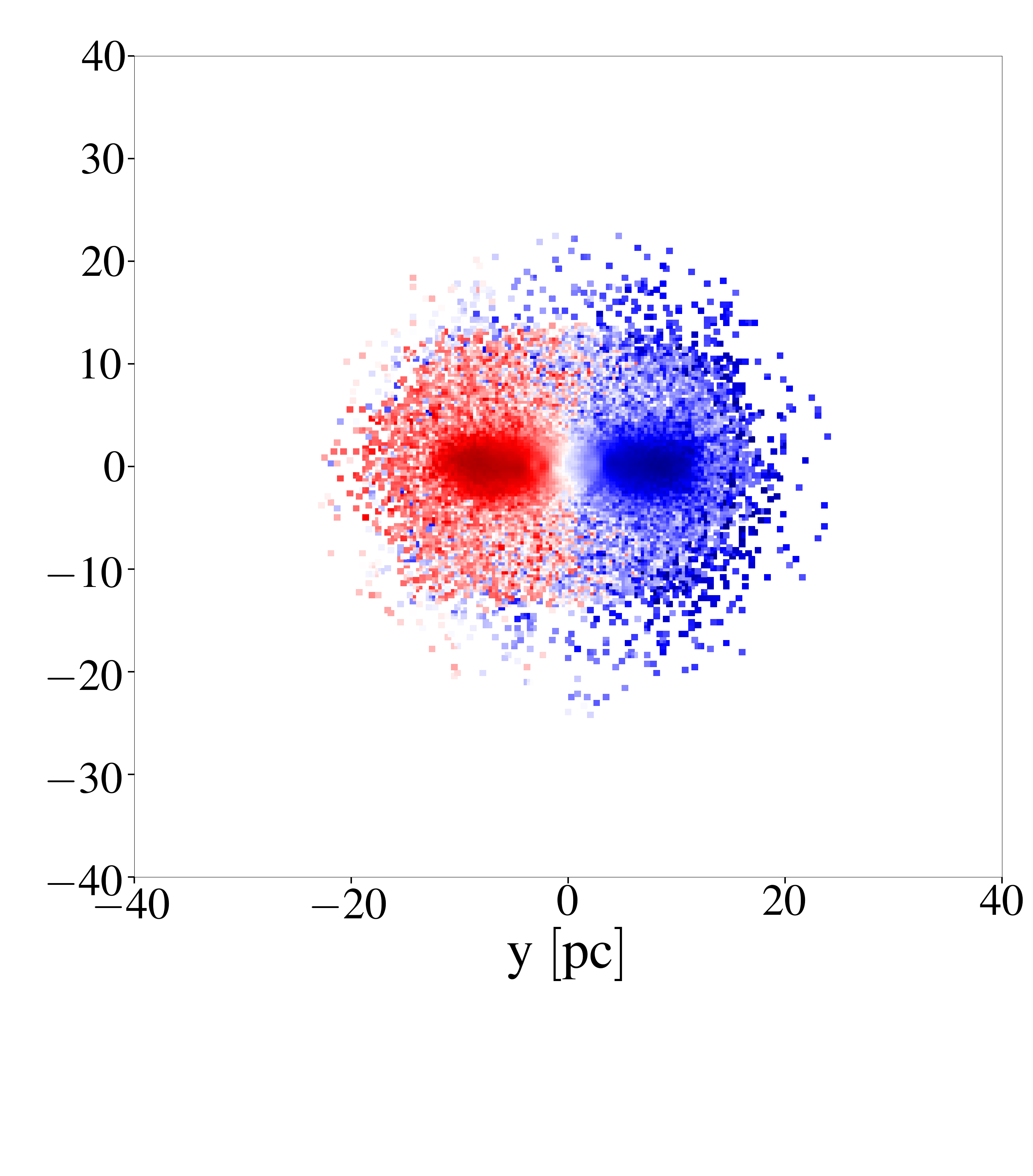}
        \\
        \vspace{-1.1cm}
          \includegraphics[width=0.324\textwidth]{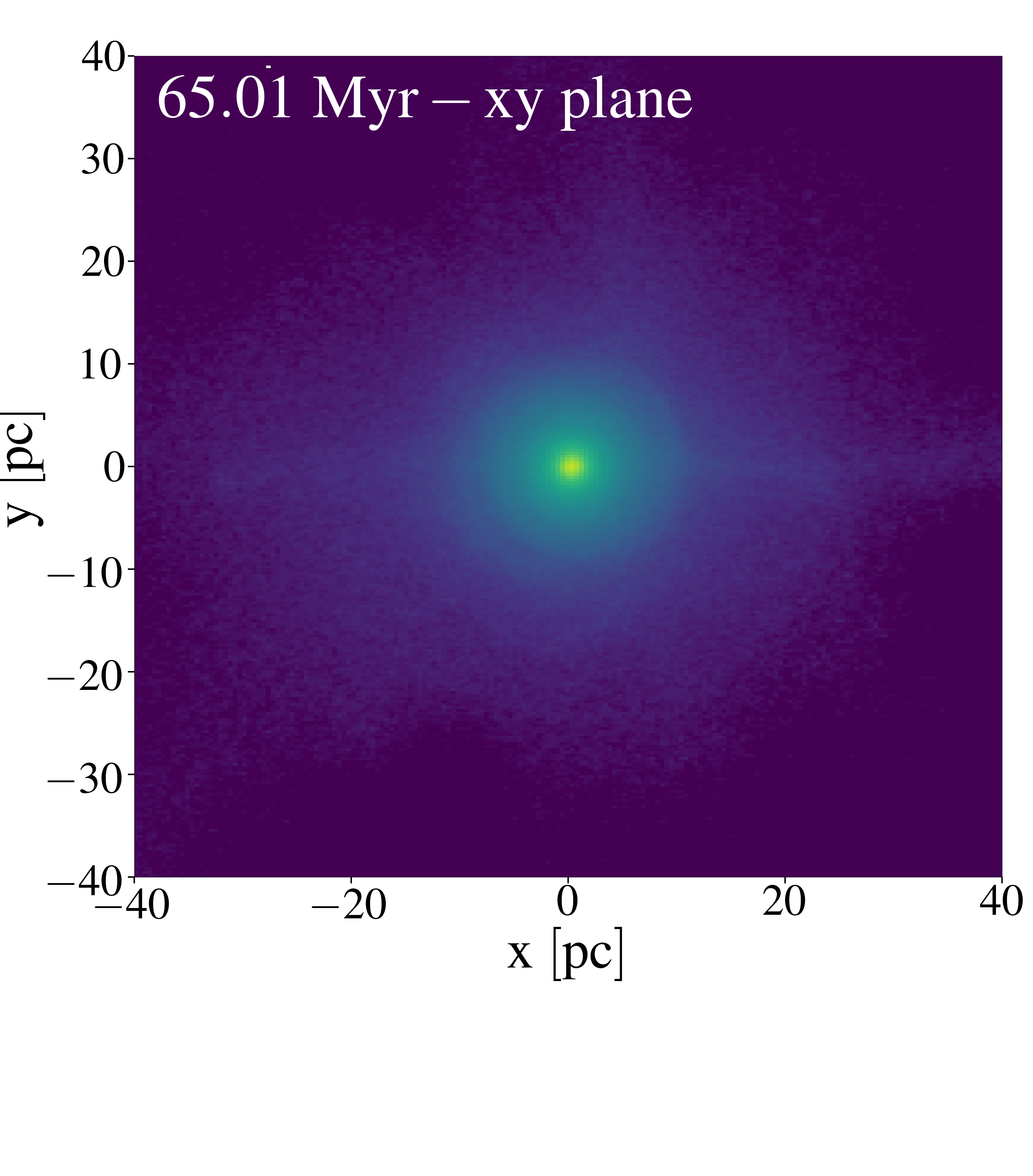}
        \includegraphics[width=0.324\textwidth]{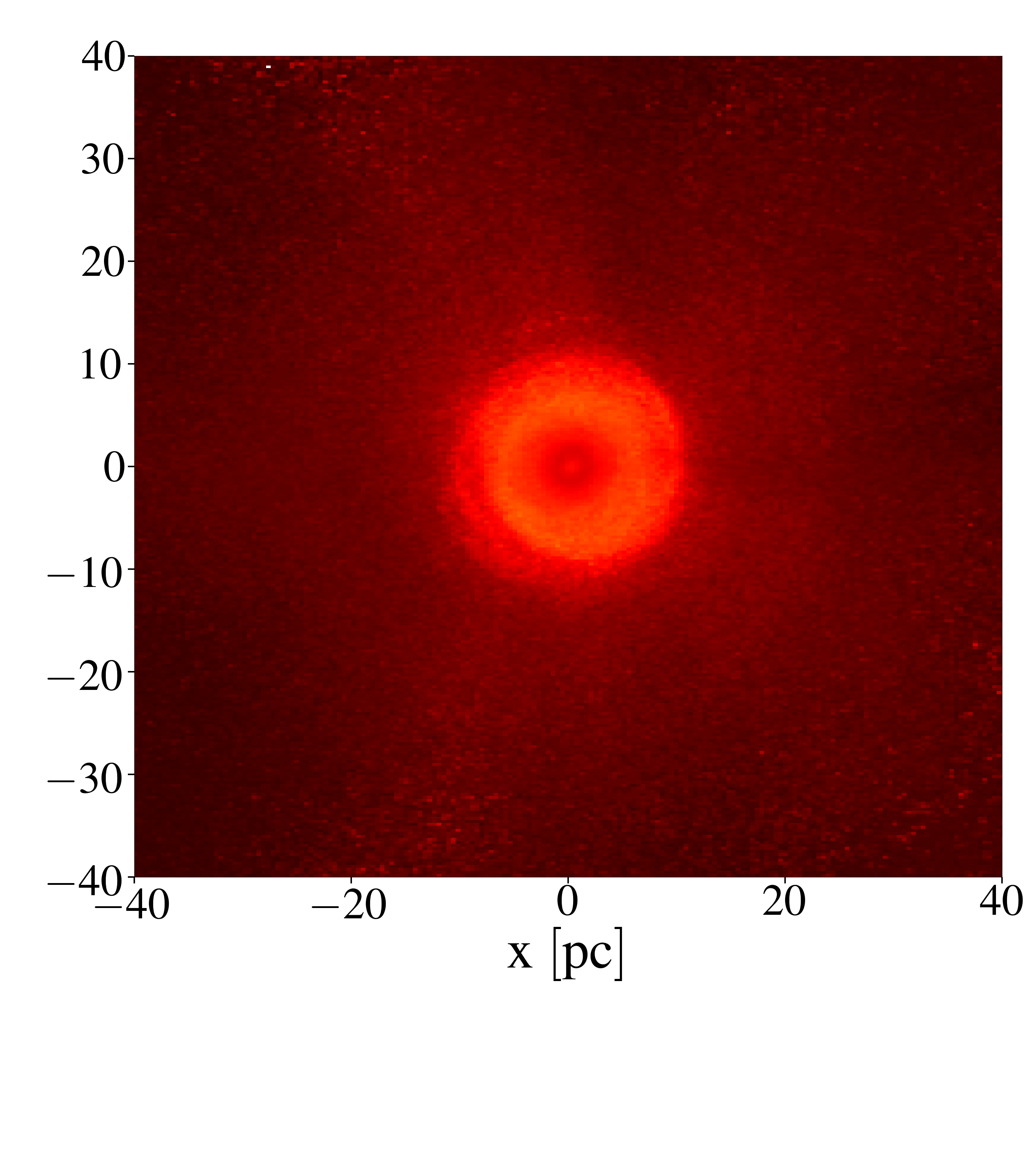}
        \includegraphics[width=0.324\textwidth]{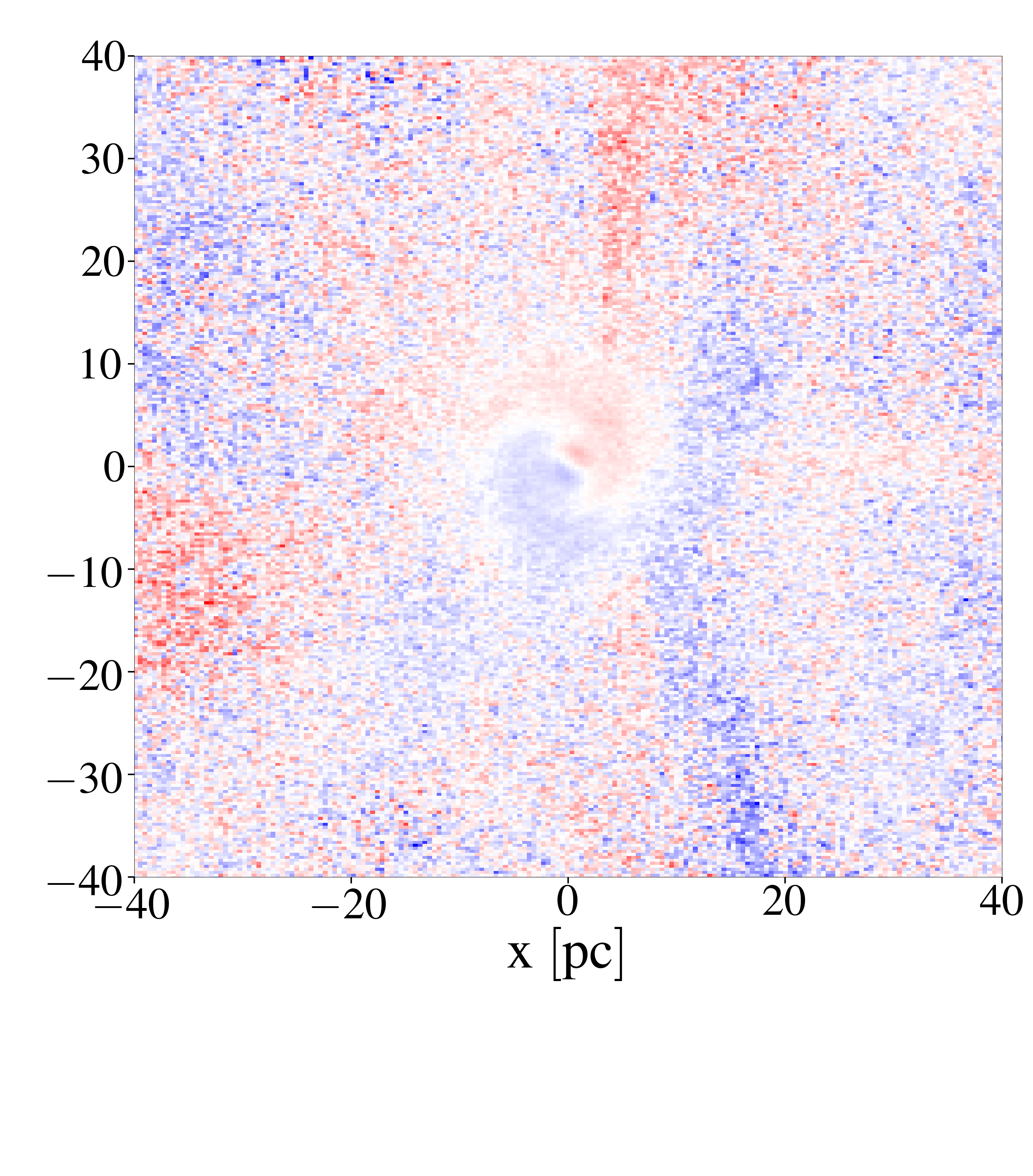}
        \\
        \vspace{-1.1cm}
        \includegraphics[width=0.324\textwidth]{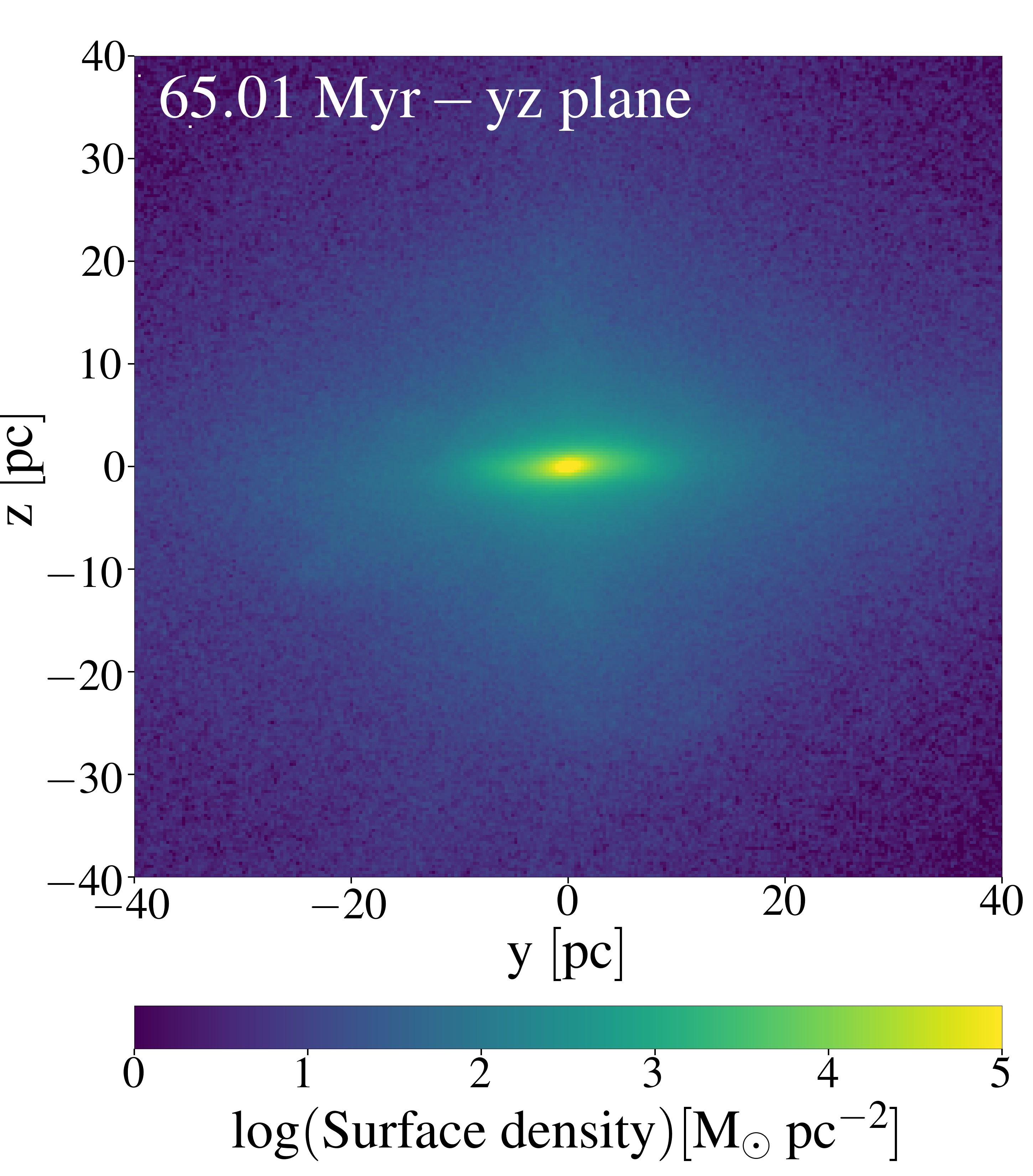}
        \includegraphics[width=0.324\textwidth]{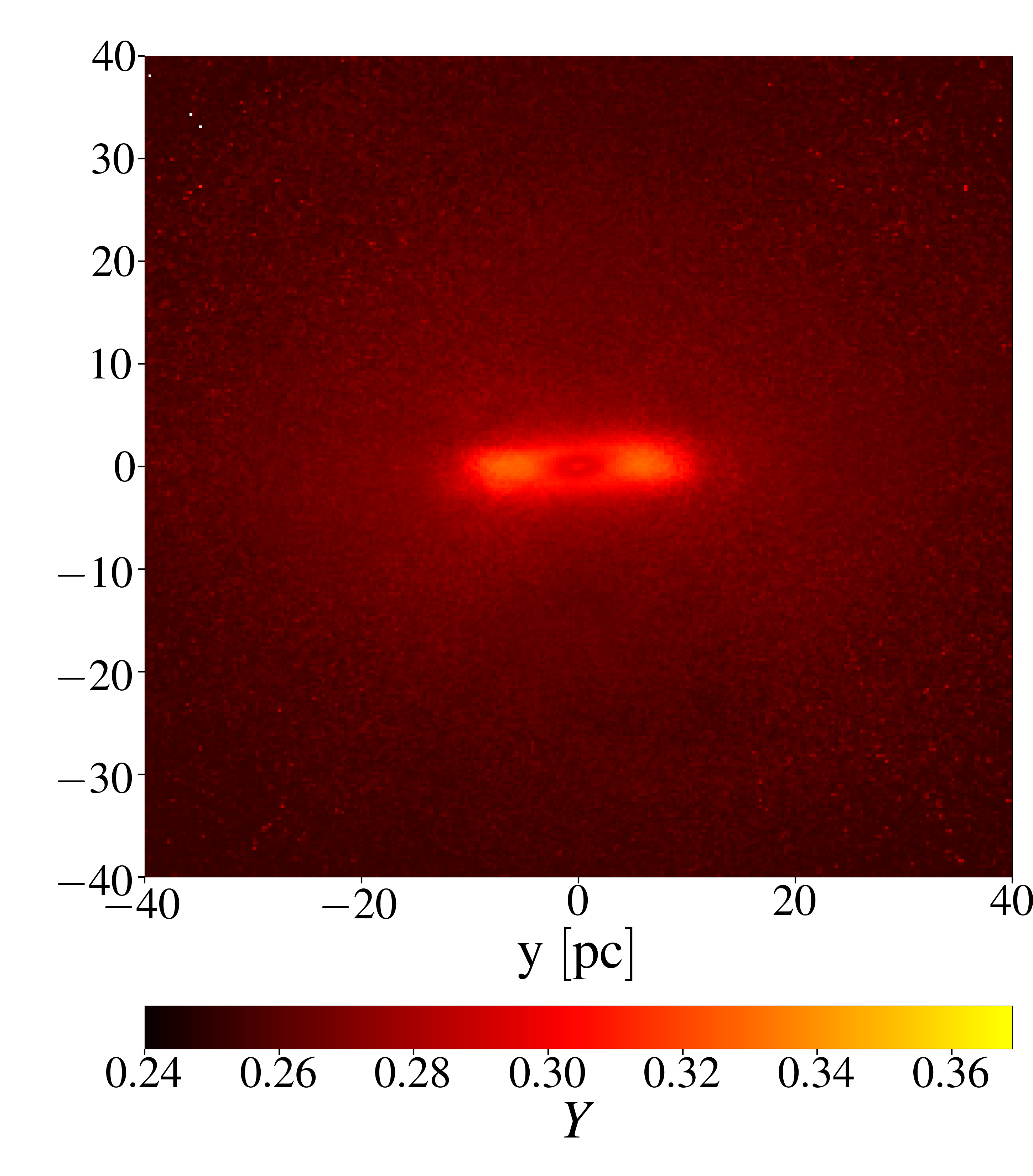}
        \includegraphics[width=0.324\textwidth]{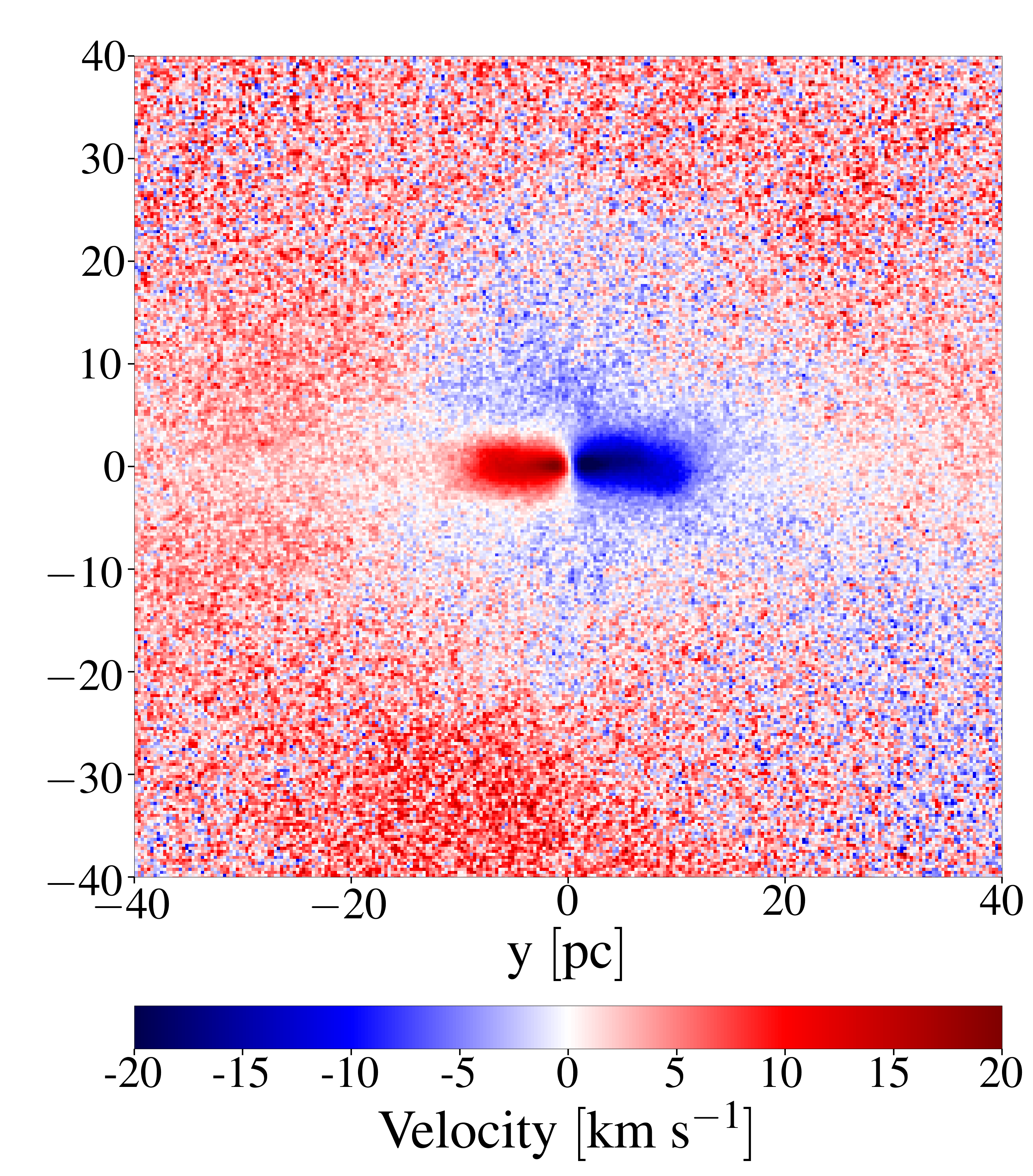}

\caption{Two-dimensional maps of the stellar component at two evolutionary times for the {\tt LDanaz} simulation on the x-y plane and y-z plane at two evolutionary times (reported on the left panels). The first column shows the surface density, the second represents the helium mass fraction $Y$, while the third the line-of-sight velocity of the stars.}
  \label{fig:maps_part_LDanaz}
\end{figure*}

\begin{figure}
        \centering
        \includegraphics[width=0.454\textwidth]{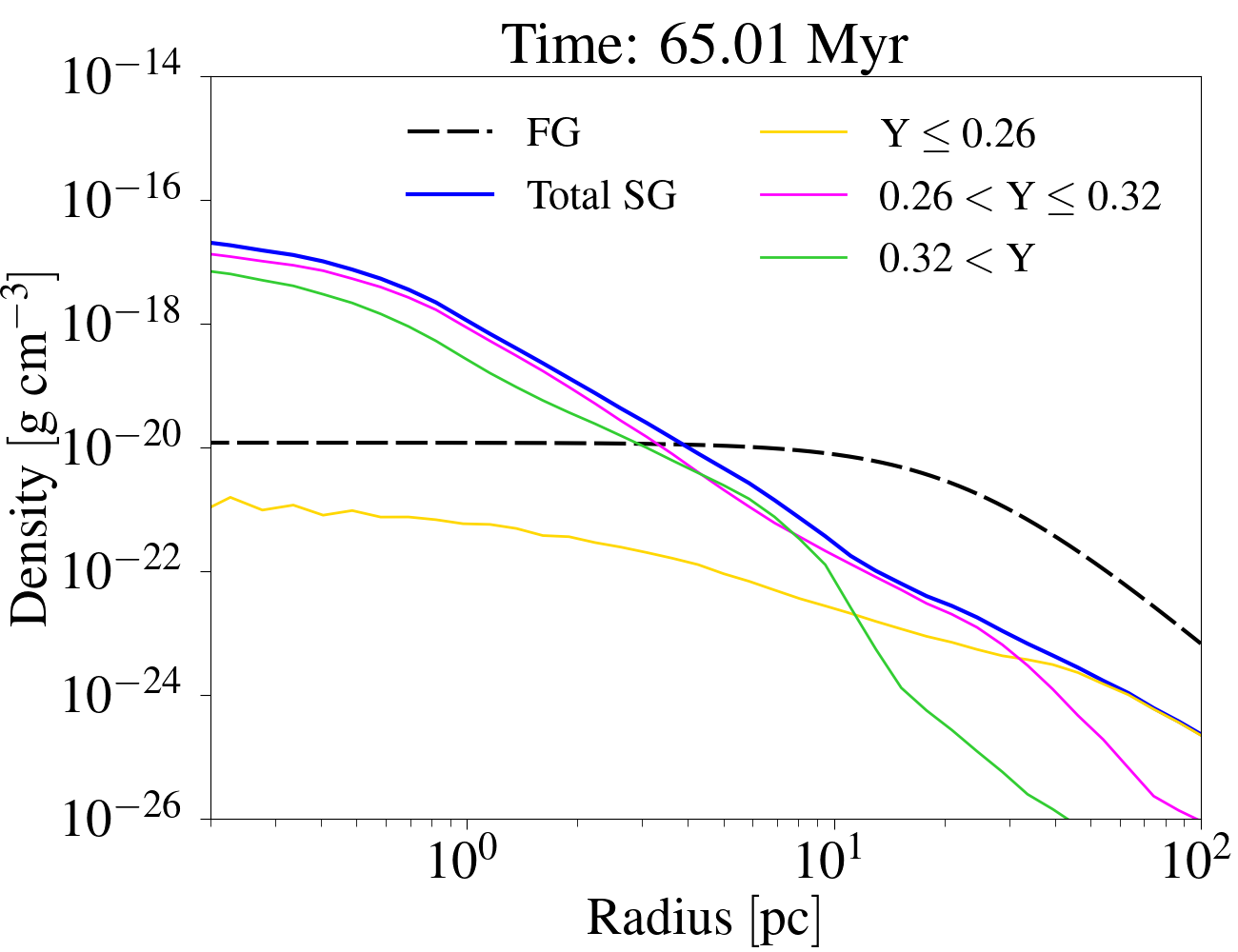}    
        \caption{Total SG density profile at the end of the simulation and SG density profiles for three intervals of the helium mass fraction $Y$, for the low-density model {\tt LDanaz}. The density profile of FG stars is also shown (see the legend for the details).}
  \label{fig:densprof_LDanaz}
\end{figure}

\begin{figure*}
        \centering

        \includegraphics[width=0.454\textwidth,trim={0 0 0 8.0cm},clip]{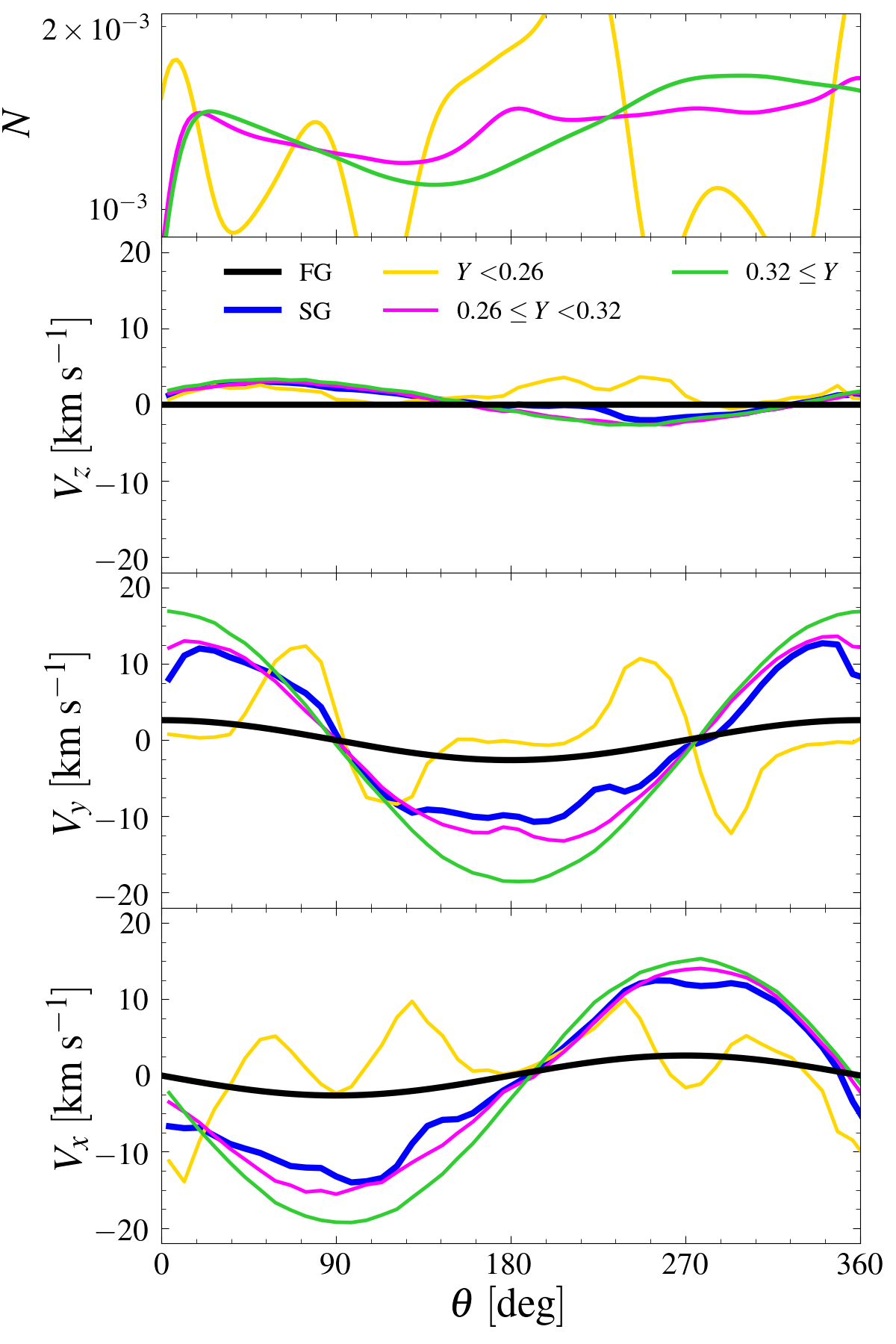}    
         %\hspace{0.01cm}
        \includegraphics[width=0.454\textwidth,trim={0 0 0 8.0cm},clip]{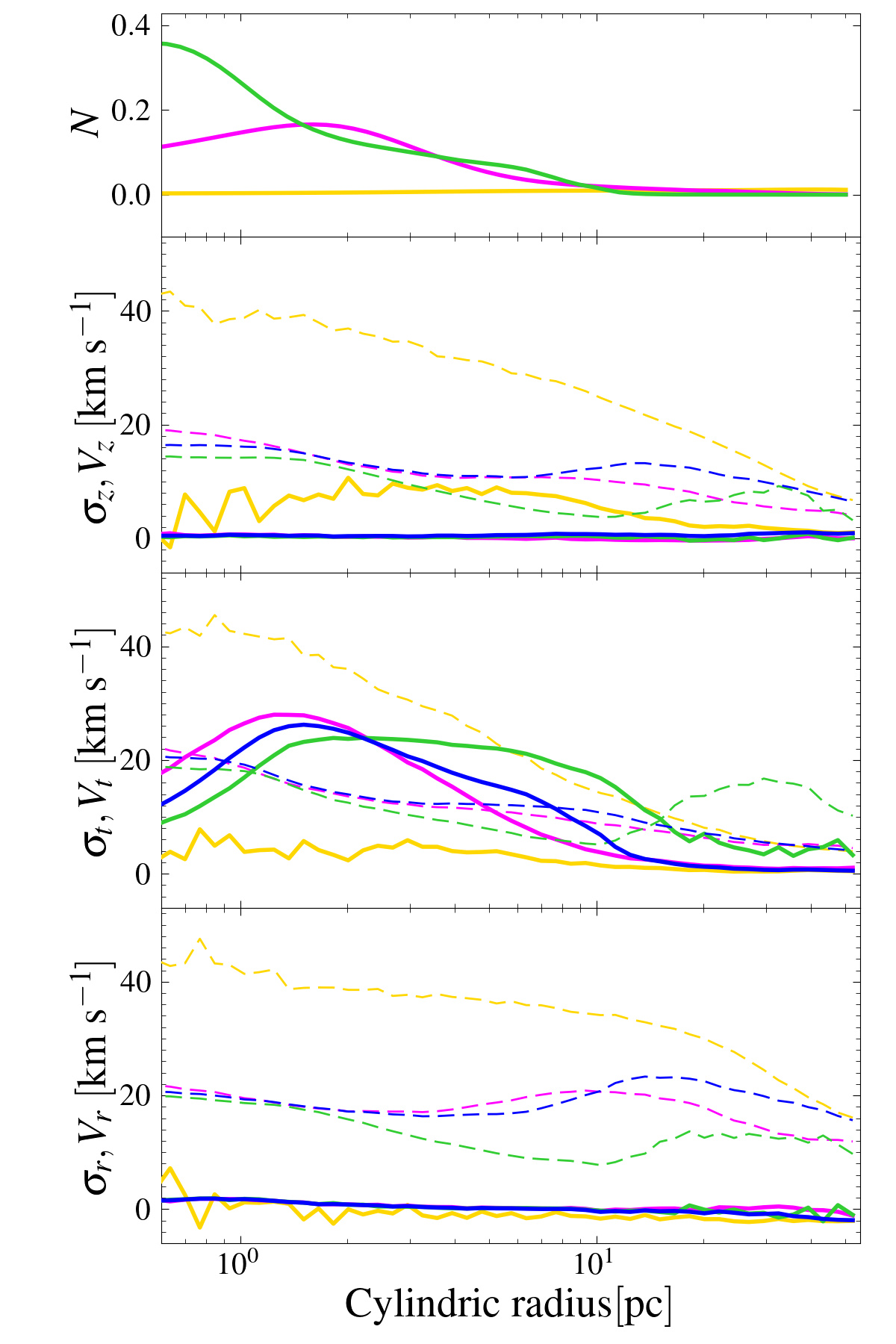}     
        \caption{Stellar rotation profiles for the model {\tt  LDanaz} at 65Myr. On the left: Rotation amplitude of the Cartesian velocity components for SG stars as a function of $\theta$, defined as the angle between the direction of each star projected on the $xy$ plane and the $x$ axis, for three bins of the helium mass fraction $Y$ (see the legend for more details). The FG rotation amplitudes are also shown by the black line. On the right: Radial profiles of the cylindrical velocity components (solid lines) and velocity dispersions $\sigma$ (dashed lines) for SG stars for three bins of the helium mass fraction $Y$.}
  \label{fig:sigmav_LDanaz}
\end{figure*}

%%%%%%%%%%%%%%%%%%%%%%%%% HIGH DENSITY %%%%%%%%%%%%%%%%%%%%%%%%%%%%

\begin{figure*}
        \centering

        \includegraphics[width=0.493\textwidth]{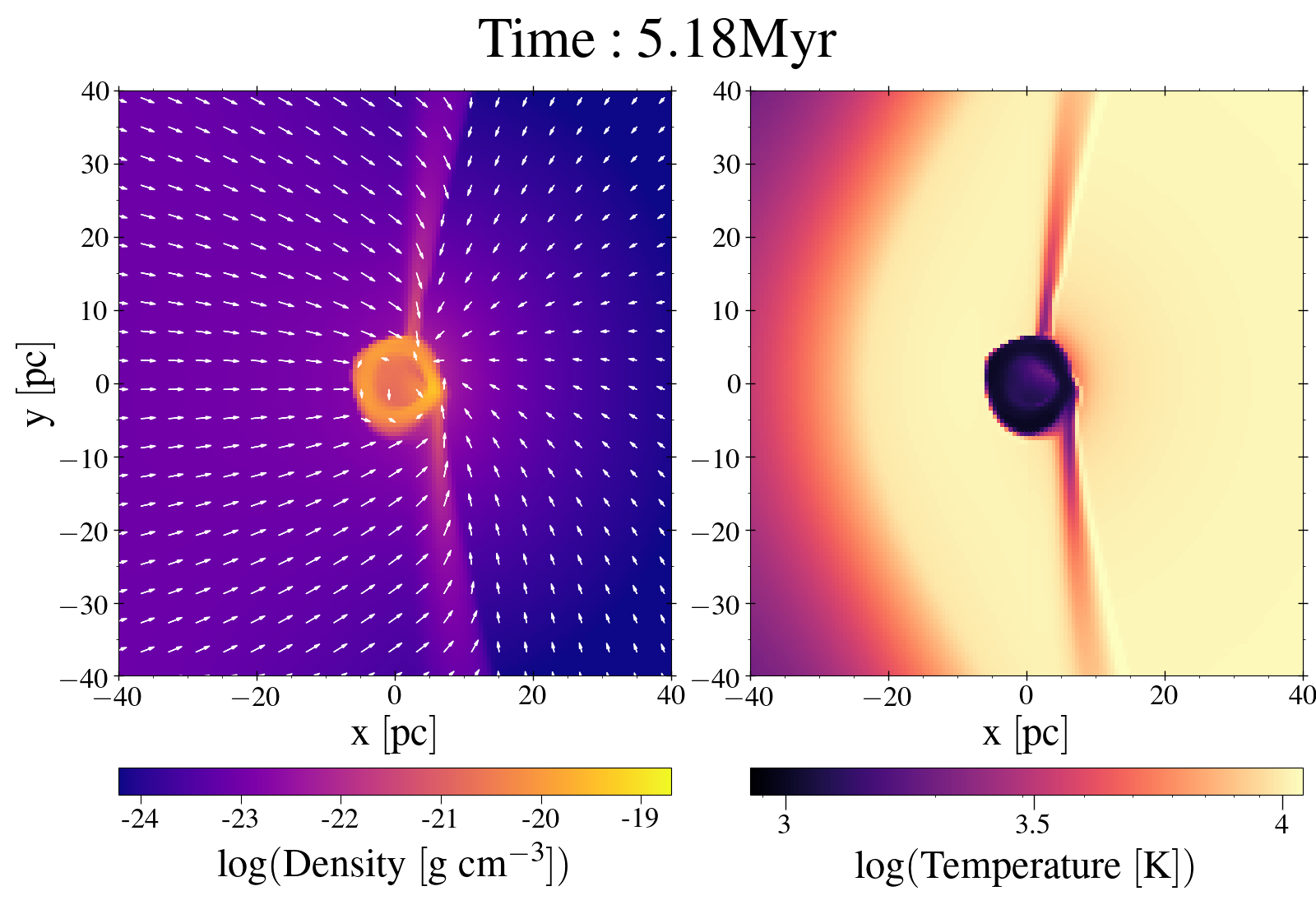}
        \hspace{0.1cm}
        \includegraphics[width=0.493\textwidth]{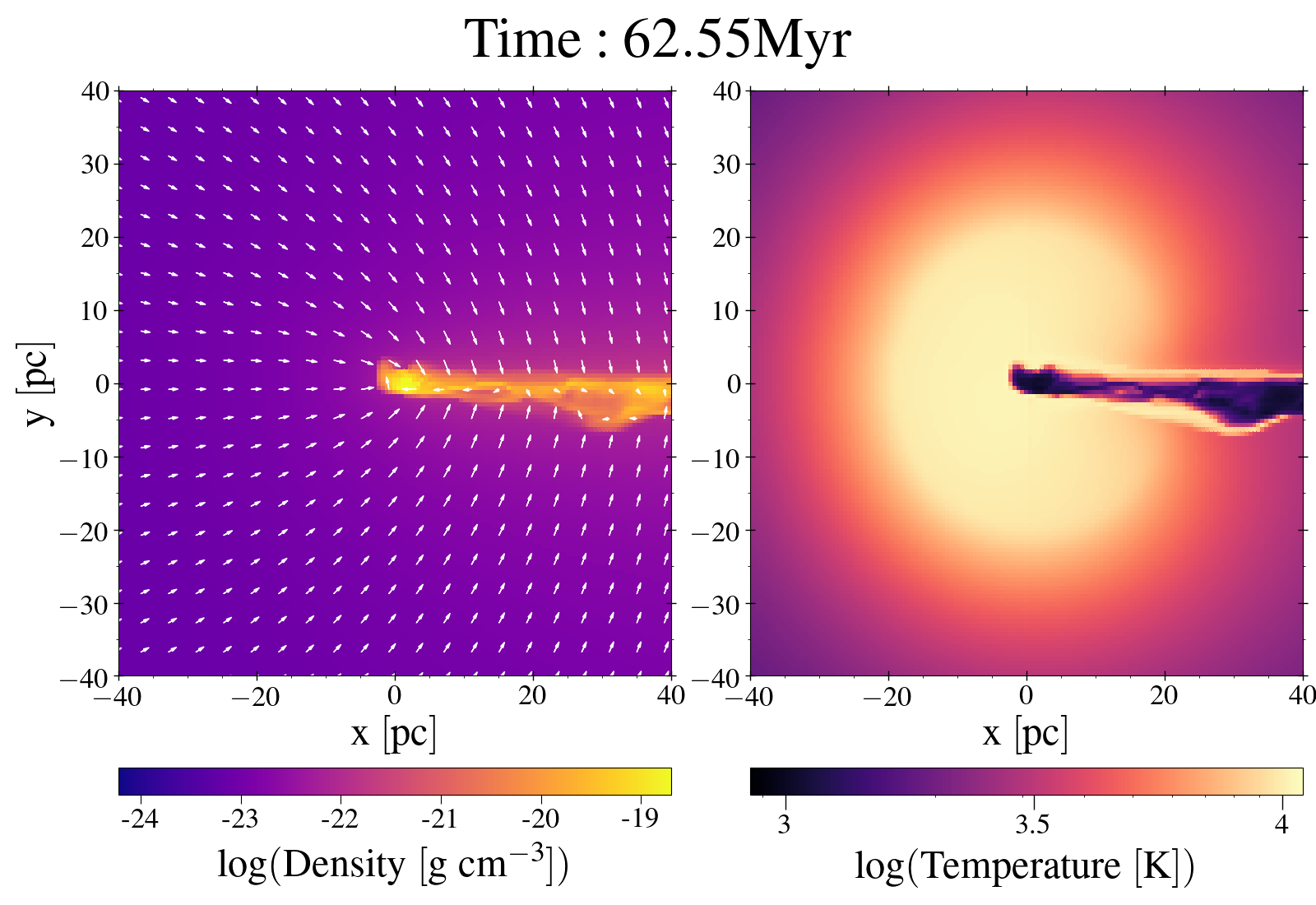}

\caption{Two-dimensional slices of the gas density (left-hand panels) and temperature (right-hand panels) on the x-y plane for the {\tt HDanaz} simulation. The corresponding evolutionary time is reported on the top of each set of panels. The white arrows in the density maps represent the velocity field of the gas.}
  \label{fig:maps_HDanaz}
\end{figure*}

%%%%%%%%%%%%%%%%%%%%%%%%%%%%%%%%%%%%%%%%%%%%%%%%%%%%%

%%%%%%%%%%%%%%%%%%%%%%%%%%%%%%%%%%%%%%%%%%%%%%%%%%%%%%%%%%%%%%%%%%%%%%%%%%%%%%%%%%%%%%%%%%%%%%%%
\subsection{Rotation perpendicular to the infall}
In this section we present the results obtained assuming that the rotation axis of the FG system is perpendicular to the direction of the external gas infall and therefore lies along the $z$ axis.

%%%%%%%%%%%%%%%%%%%%%%%%%%%%%%%%%%%%%%%%%%%%%%%%%%%%%%%%%%%%%%%%%%%%%%%%%%%%%%%%%%%%%%%%%%%%%%%%

\subsubsection{Low-density model}

\subsubsection*{Gas evolution}

In Fig. \ref{fig:maps_LDanaz}, we show the 2D density, temperature and helium mass fraction slices for the gas component at 26 and 65 Myr, for the model with a low pristine gas density. To build the slices, we select all cells on the $xy$ plane passing through the centre of the computational box.

We have displayed, on top of the helium mass fraction map, the gas velocity field as white arrows. For a comparison with \citetalias{calura2019}, we show here the maps at similar evolutionary times. 

In the first snapshot, at 26 Myr, the infall has already started. The shock front has just crossed the system creating two \enquote{wings} which are both visible in all the three maps. Contrary to \citetalias{calura2019}, the wings are not perfectly symmetric because of the rotation of FG stars and therefore of their ejecta on the xy plane. The presence of the infalling gas is even more clearly visible in the maps of the helium mass fraction where the helium-rich gas ejected by FG stars, marked in red, is located downstream to the cluster and in the central regions while the blue helium-poor pristine gas of the infall is dominant upstream to the system. In the central part, the system is composed of several cold (T$\sim 10^3$K) and dense (${\rm\rho=10^{-19} g \ cm^{-3}}$) clumps. These structures are the result of the fragmentation of a torus which has been recently described in detail by \citet{inoue2021} through a linear perturbation analysis. In codes which exploit the AMR technique, the computational box is discretized in multiple grids, which in turn leads to a discretization of the hydrodynamic equations. Such modellization can lead to the so called \enquote{artificial fragmentation} which takes place when small perturbations grow to form fragments. \citet{truelove1997} studied the conditions that give rise to this phenomenon, by means of a 3D hydrodynamic AMR code, finding that, to avoid artificial fragmentation, the Jeans length ${\rm \lambda_J}$ \citep{jeans1902} should be at least four times greater than the resolution element. For the gas properties in our simulation, we end up with a ${\rm \lambda_J}$ is equal to $2.6$ pc and therefore satisfies the Truelove criterion for avoiding artificial fragmentation. 
The clumps found in our simulation are thus not numerical artifacts but rather represent the result of the actual dynamics of the gas out of which SG stars form. A detailed investigation of the formation and evolution of these clumps is beyond the scope of the present work and will be further investigated in a future study.

At the end of the simulation, $\sim 65$  Myr, a tail of cold and dense gas is pointing towards the cluster centre, the so called \enquote{accretion column}. This helium-poor gas is generated out of the infall event through the Bondi-Hoyle-Litterton accretion mechanism \citep{bondi1944,moeckel2009}.

As shown in \citet{lacchin2021}, the structure and dynamics of the accretion column may be significantly affected by the feedback of SNe Ia.
%\citet{lacchin2021}  have shown that this phenomenon could be significantly affected by the presence of feedback sources such as prompt Type Ia SNe and that, in the low density models, such energy sources can completely prevent its formation. 

\subsubsection*{Evolution of the stellar component}

In Fig \ref{fig:maps_part_LDanaz}, we plot the 2D surface density, helium mass fraction and line-of-sight velocity maps on the $xy$ and $yz$ planes for the stellar component at the same evolutionary times shown in Fig. \ref{fig:maps_LDanaz}.

In the 26 Myr snapshots, stellar particles are preferentially lying on a disk in the $xy$ plane; this is the expected distribution of stars forming out of the gas released from a FG system rotating about the $z$-axis. The surface density map in the $xy$ plane clearly reveals the presence of five clumps in the very central part of the system. Their average mass is ${\rm 4\times 10^4 M_{\odot}}$ and more than $90\%$ of all the particle mass is confined in these clumps. %Clumps are visible also when star formation is switched off and therefore the only dynamical component is the gas one.
The five clumps are formed and orbit the centre of the system, and eventually merge after a few Myr. %These clumps are formed also in the model with solid body rotation even though they are less dense (to check quantitatively) and more streamed. %In addition, we have tested whether the clumps were formed even when star formation is switched off founding that the clumps of gas are still present, meaning that they are formed firstly in the gas component giving later birth to the stellar ones.
%\textcolor{red}{say something on the simulation without star formation. And maybe also about the formation of the clumps??}

All stars at this stage are strongly helium-enhanced, given that the infall has just started and has not had enough time to significantly dilute the AGB ejecta. These stars are concentrated in the cluster's innermost regions (their half-mass radius is $r_{h}= 3.7$ pc ) and distributed in a toroidal structure; they are characterized by a significantly more rapid rotation than that of the FG population. Specifically, the SG peak rotational velocity is equal to about $V_{\rm peak}={\rm 16\ km\ s^{-1}} $ at a distance from the cluster's centre of $R_{\rm pk}=6.7$ pc.

At $65$ Myr the clumps have already merged and collected in the centre. The disk is still clearly visible in all maps and contains $80\%$ of the total mass of the SG. It extends for $\sim 8$ pc and is composed of helium-enhanced stars that are still significantly rotating. It is interesting to note the presence of a small region in the innermost part of the disk characterized by a helium abundance smaller than in the rest of the disk.
This feature is a consequence of the early dynamics of the SG formation: as discussed above, the most He-rich AGB ejecta are first collected in a clumpy toroidal structure, which later forms a stellar disc; the non-rotating pristine gas, on the other hand, flows directly in the centre and creates this small region with slightly lower helium abundance. This is also illustrated by the density profiles presented in Fig. \ref{fig:densprof_LDanaz} showing a slightly lower central density of the most He-rich population.
In order to link these fine structural and chemo-dynamical features within the SG system with the present-day properties of old GCs, it is necessary to carry out further explorations of the long-term dynamical evolution with initial conditions similar to those emerging from our simulations. 

%{\bf The reason for this is that the initial, most He-rich AGB ejecta is first collected in a torus, which later forms a stellar disc. As a consequence, in the centre, the density of stars formed out of the most He-rich AGB ejecta is lower than the one of stars born by less helium-enhanced gas, as shown also in Fig. \ref{fig:densprof_LDanaz}, at variance with the results obtained for its analogue, non-rotating system \citetalias{calura2019}.  
%The most He-rich stars preserve their positions and kinematics and they remain located in correspondence of the initial location of the torus. Such finding is not in contraddiction with observations of present-day GCs \citep{cordero2014,simioni2016}, because we are here missing all the long term evolution of the system which is expected to lead to a decrease of the angular momentum as derived both through $N$-body simulations \citep{tiongco2017} and observations \citep{kamann2018,bianchini2018}}. 

In Fig. \ref{fig:sigmav_LDanaz} we show, for the {\tt LDanaz} model, the angular variation of the Cartesian components of the velocity of the SG stars (left panels) and the radial profiles of the mean velocity and velocity dispersions in cylindrical coordinates (right panels) measured at the end of the simulation (t=65 Myr).
This figure further illustrates the kinematical properties of the SG population and the various SG subgroups. %For comparison, we show also the FG trend with a rotational amplitude of ${\rm 2.5\ km\ s^{-1}}$, meaning that it rotates slower than the SG, whose amplitude is ${\rm \sim 12\ km\ s^{-1}}$.{ \bf FG stars has been found to rotate slower than SG ones even in several present-day clusters as it has been reported by several observational studies \citep{lee2017,lee2018,dalessandro2021,szigeti2021}, but with significantly lower amplitude than the ones we find here. The long-term evolution, in fact, has been shown to lead to a weakening of the rotational intensity. The higher rotational velocity of the SG with respect of the FG follows from the angular momentum conservation, given that SG stars, that form at least in part from the rotating gas lost by FG stars, are much more concentrated than FG ones. 
SG stars rotate more rapidly, with an amplitude of ${\rm \sim 12\ km\ s^{-1}}$, than the FG population, whose amplitude is equal to ${\rm 2.5\ km\ s^{-1}}$. Such difference in the rotational intensity between the two populations has been observed in several present-day clusters \citep{lee2017,lee2018,dalessandro2021,szigeti2021}, although the amplitudes are significantly smaller than the once we obtain, since our simulations do not include the long-term evolution of the system.
In addition, more helium-enriched stars are found to rotate faster than weakly enriched ones, qualitatively in agreement with recent observational works on M13 \citep{cordero2017} and M80 \citep{kamann2020} (with the exception, already discussed above, of the rotation in the very inner regions, where the intermediate SG group rotates slightly faster). The small angular variation in the mean value of $V_z$ is a consequence of the fact that the SG disc is slightly tilted (see also Fig. \ref{fig:maps_part_LDanaz}).
Moreover, we found that FG and SG stars are rotating in phase, a feature often found in present-day GCs \citep{milone2018b,cordoni2020b}. %It is worth noting that, contrary to the FG whose $V_z$ is equal to zero, the SG rotates with an amplitude of ${\rm 3\ km\ s^{-1}}$, meaning that the SG population and therefore the disk is weakly tilted. In addition, the rotational axis of the system is no more along the $z$ axis, but it is precessing about it by a few degrees.

%Concerning the radial distribution of the cylindric velocity components in the right hand panels of Fig \ref{fig:sigmav_LDanaz}, we find that the $V_t$ component shows differences depending on the helium composition. Helium poor stars are rotating faster within 2pc, while at larger radii their velocity drops significantly. The very high velocity of stars with $Y<0.27$ are found also in the non-rotating model %($V_t$ is mostly negative in {\tt LD} due to the clockwise rotation, while in the {\tt LDanaz} model the rotation is counter-clockwise)
%and therefore the rotation of these stars does not have to be interpreted as an effect of the imprinted rotation of the FG stars, at variance with the most extremely helium-enhanced stars. %These stars show a much slower rotation (around half in the velocity amplitude) in the non-rotating model {\tt LD}. 

Regarding the velocity profiles in Fig. \ref{fig:sigmav_LDanaz}, it is not surprising that the peak of the SG rotational velocity $V_t$ is located approximately at the SG half-mass radius equal to 1.5 pc, as suggested by observational studies of very young stellar clusters \citep{henaultbrunet2012}, but also of older globular clusters \citep{bianchini2018}.
%It is worth noting that helium-poor stars are rotating slower than all other subpopulations once we look at the velocities versus angle (left panel of Fig. \ref{fig:sigmav_LDanaz}), while they show the strongest rotation once looking at the velocity profile (the $V_t$ in the right panel). The peak of tangential velocity of helium-poor stars is located in the very centre of the system, where their density is the lowest, while where they become the dominant component, their velocity is approximately zero (see Fig. \ref{fig:densprof_LDanaz}). 

As for the radial profiles of the velocity dispersion shown in the right panels of Fig. \ref{fig:sigmav_LDanaz}, it is interesting to note the presence of a few bumps indicating that the SG system is still in the process of settling into a dynamical quasi-equilibrium.
Along the tangential direction, $|V_{rot}/\sigma|$ is greater than unity in the disk region, with a peak of 3.5 at the edge of the stellar disk, confirming that the system is rotationally supported. This ratio is much larger than the observed one, but, as for the rotational amplitudes of the two populations, the long-term evolution has been shown to lead to a significant decrease of this quantity \citep{tiongco2017}.
Finally, we point out that here we have explored a single case for the initial FG rotational profile; an extension of this study to explore different initial FG rotational properties will be carried out in the future. The early study on the formation of SG in rotating clusters by \citet{bekki2010} suggests that the strength of the SG rotation emerging at the end of the formation phase is correlated with that of the FG system.

%\textcolor{red}{con i nuovi bin, la dispersione di velocità della pop più He poor è molto più alta. Quindi più dispersione che ci sta essendo che proviene dall'infall. Scrivere una frase? da notare che nel caso LDanax lungo la dir tangenziale non è alta}

%%%%%%%%%%%%%%%%%%%%%%%%%%%%%%%%%%%%%%%%%%%%%%%%%%%%%%%%%%%%%
\begin{figure*}
        \centering

        \includegraphics[width=0.962\textwidth]{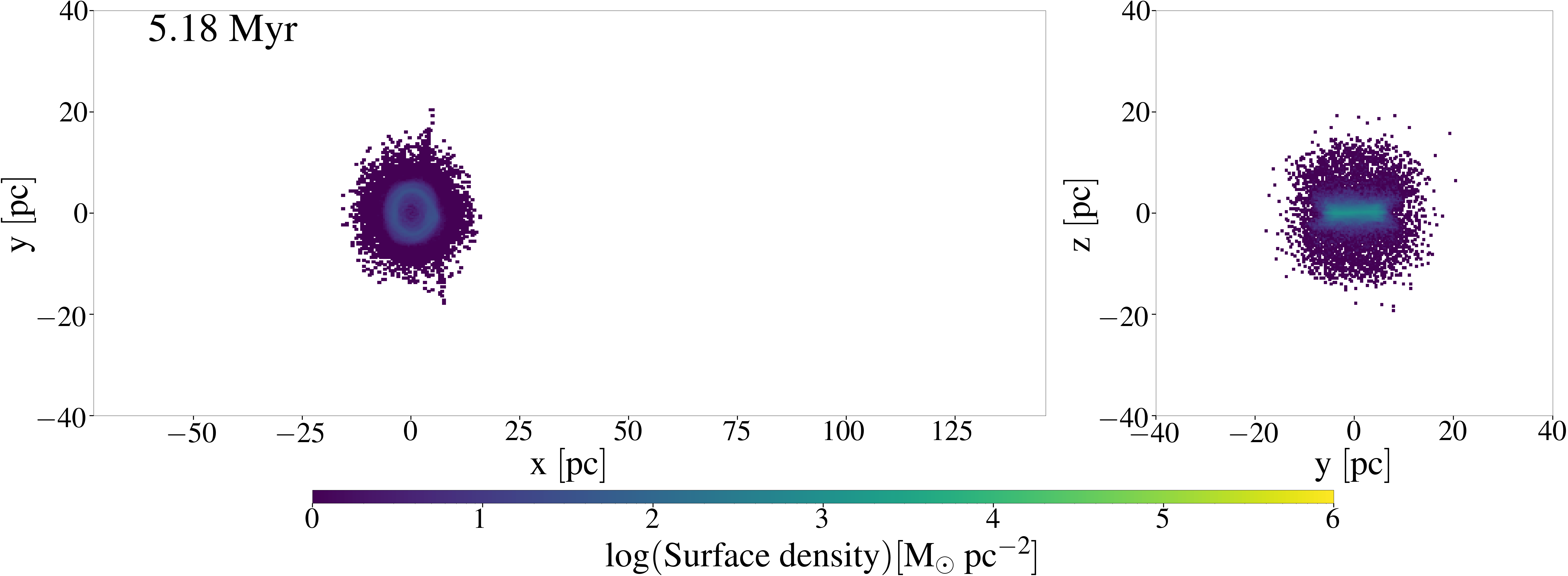}
        \\
        \vspace{-1.46cm}
        \includegraphics[width=0.962\textwidth]{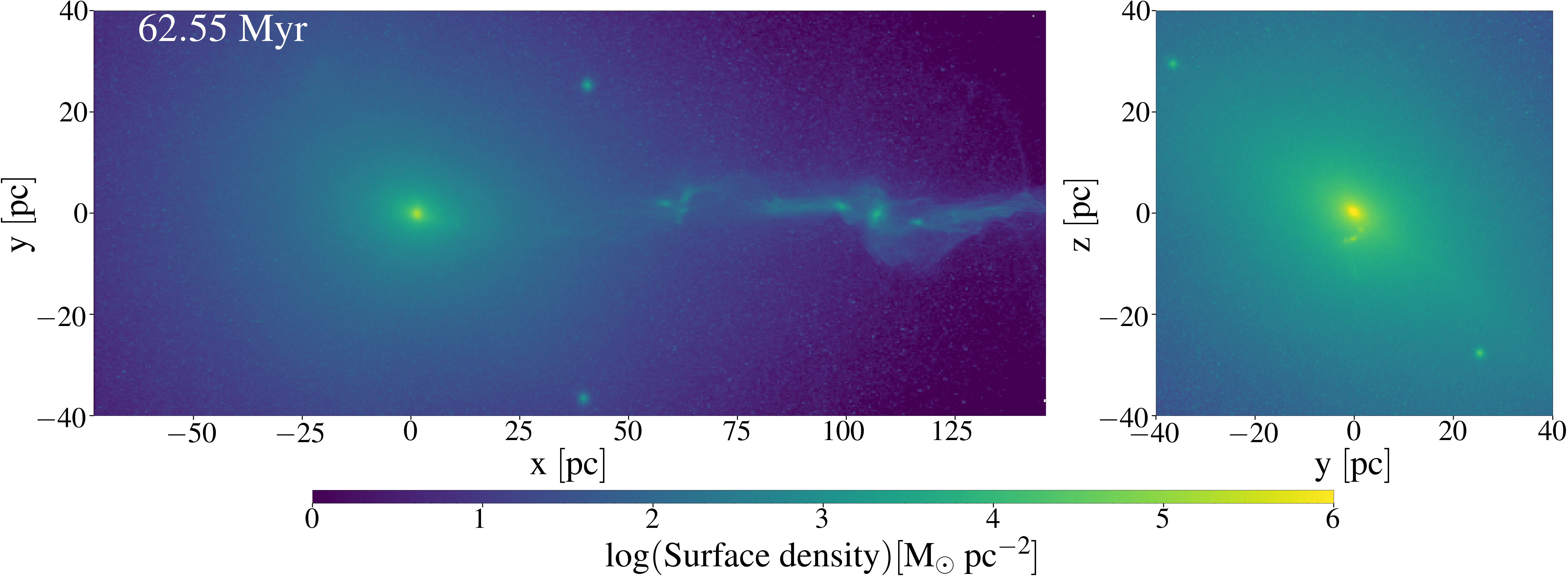}

\caption{Two-dimensional surface density maps of the stellar component at two evolutionary times for the {\tt HDanaz} simulation on the x-y plane, on the left, and y-z plane, on the right.  }
  \label{fig:maps_part_HDanaz}
\end{figure*}

%%%%%%%%%%%%%%%%%%%%%%%%%%%%%%%%%%%%%%%%%%%%%%%%%%%%%%%%%%%%%%%%%%%%%%%%%%%%%%%%%%%%%%%%%%%%%%%%
\subsubsection{High-density model}
\subsubsection*{Gas evolution}

Fig. \ref{fig:maps_HDanaz} represents the 2D density and temperature slices at 5 and 62 Myr for the {\tt HDanaz} model. At 5Myr the system is encountering the infalling gas which creates a shock front that is, as in Fig. \ref{fig:maps_LDanaz}, not perfectly symmetric because of the rotation of the already present AGB ejecta. The presence of a bow shock is also visible in the temperature map, at a distance comparable to the Plummer radius of the cluster (see also \citealt{Naiman2011}).
In the central 10 pc the gas is rotating counter-clockwise due to the imposed rotation of the FG stellar component. In particular, the gas is mainly confined in a torus due to the balancing effects of gravity and rotation. 

At 62 Myr the gas is not rotating anymore and, similarly to the results obtained by \citetalias{calura2019}, the accretion column is carrying pristine gas towards the cluster centre.

We do not report here the maps of the gas helium mass fraction since, at 5 Myr, the map is similar to the one of the {\tt LDanaz} model at 26 Myr and, at the  end of the simulation, no feature is visible since the entire computational box is dominated by gas with moderate helium enhancement.

\subsubsection*{Evolution of the stellar component}

\begin{figure*}
        \centering

        \includegraphics[width=0.454\textwidth,trim={0 0 0 8.0cm},clip]{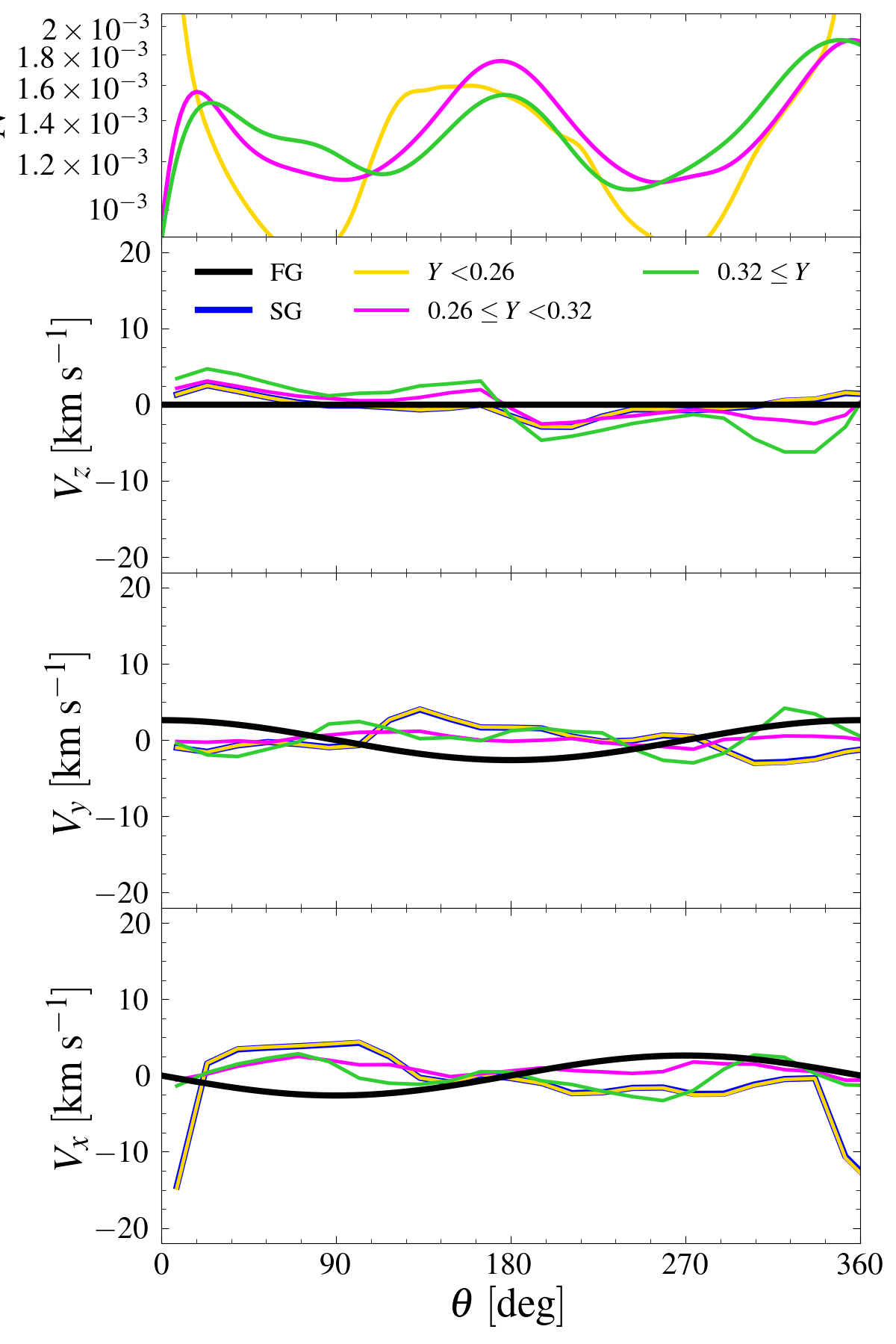}    
         %\hspace{0.01cm}
        \includegraphics[width=0.454\textwidth,trim={0 0 0 8.0cm},clip]{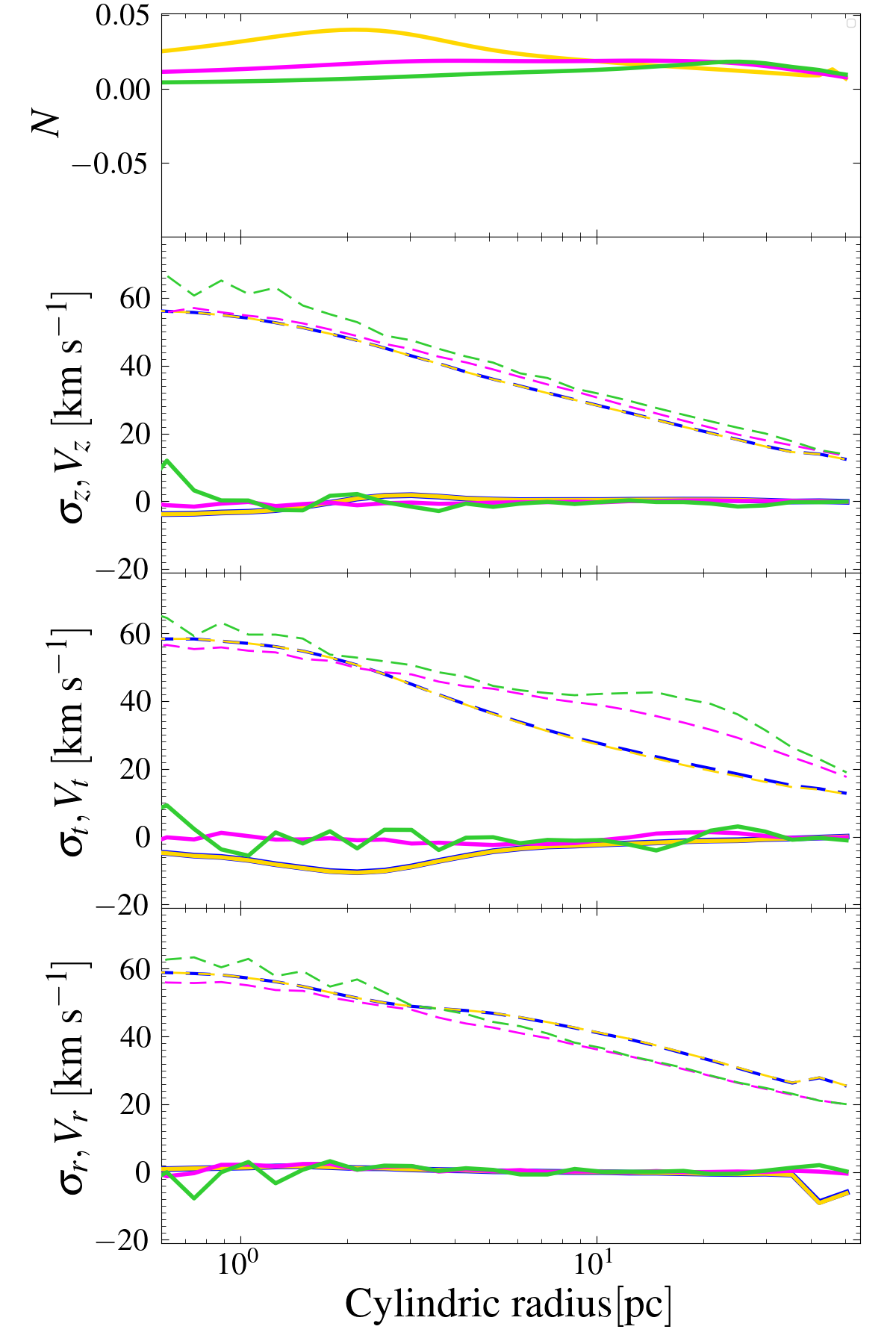}     
        \caption{ Stellar rotation profiles of the model {\tt HDanaz} at 65 Myr. The reported quantities are as in Fig. \ref{fig:sigmav_LDanaz}. }
  \label{fig:sigmav_HDanaz}
\end{figure*}

Contrary to the {\tt LDanaz} model, here the earlier incoming of the infall prevents the clump formation while the disk is still formed as shown in Fig. \ref{fig:maps_part_HDanaz} in the snapshot at 5 Myr. Not surprisingly, SG stars are mainly located in a torus resembling the gas distribution. 
% \textcolor{red}{perchè c'è un blob di circa 25 pc attorno al centro? perchè le stelle cominciano a formarsi da lì a sx? stessa cose del blob nella temperatura del gas?}.

At 62 Myr the stellar component is much more extended than at the previous snapshot and a significant number of stars form along the accretion column. %, at variance with the {\tt LDanaz} model where the accretion column is not detectable from the stellar surface density. 
Small clumps form along the accretion column, move towards the cluster centre and merge with the central component. Instead, a minority of them is kicked away.

In contrast to the {\tt LDanaz} model, in the {\tt HDanaz} one the disk does not survive once the infall starts, both because of the higher density of the accreted gas, and, in turn, its stronger ram pressure, but also of because of its earlier arrival when the disk is still forming. The system is dominated by moderately-helium enriched stars which are formed out of the AGB ejecta and a large amount of the infalling pristine gas.
Nevertheless, a small fraction of highly enhanced helium stars preserves the
rotation acquired at birth.

Fig. \ref{fig:sigmav_HDanaz} shows the velocity profiles as a function of the $\theta$ angle (left, defined as in Fig. \ref{fig:sigmav_LDanaz}) and radius (right) for the {\tt HDanaz} model. SG stars rotate with an amplitude similar to that of the FG with peaks in the $x$ component. This feature is due to the stars born along the accretion column and consequently possessing a significant velocity towards -$x$. Along $z$, the velocity is not zero due to the gas motions produced by the infall and the subsequent formation of the accretion column.

At variance with the low-density model, here the dispersion velocities of all the various SG subgroups are similar. The maximum value of  $|V_{rot}/\sigma|$ is 0.2, much smaller than in the low density models confirming that the system is dispersion-supported. %Nevertheless, this model could explain the presence of non-rotating GCs as found by \citet{cordoni2020}.

%Io direi che in questo caso non vedi significative differenze tra le disp delle varie componenti, cosa che indica che il gas e' mescolato, al contrario del caso a bassa densita'. Vedo anche che le sigma >> vrot, quindi il sistema non e' chiaramente rotationally supported.

%%%%%%%%%%%%%%%%%%%%%%%%%%%%%%%%%%%%%%%%%%%%%%%%%%%%%%%%%%%%%%%%%%%%%%%%%%%%%%%%%%%%%%%%%%%%%%%%%%%
%%%%%%%%%%%%%%%%%%%%%%%%%%%%%%%%%%%%%%%%%%%%%%%%%%%%%%%%%%%%%%%%%%%%%%%%%%%%%%%%%%%%%%%%%%%%%
\subsection{Rotation parallel to the infall}
We present here the results for the model where the FG is assumed to rotate about the $x$-axis and therefore with a rotational axis parallel to the direction of the infall. We describe here only the low density model ({\tt LDanax}), given the similar results obtained for the {\tt HDanaz} and {\tt HDanax} models. In both cases a disc is formed both in the gas and stellar components, but, once the infall starts, it is disrupted and no significant rotation is found.

%\textcolor{red}{how can a cluster start rotating in this direction?}
%%%%%%%%%%%%%%%%%%%%%%%%%%%%%%%%%%%%%%%%%%%%%%%%%%%%%%%%%%%%%%%%%%%%%%%%%%%%%%%%%%%%%%%%%%%%%%%%
\subsubsection{Low-density model}
\subsubsection*{Gas evolution}

\begin{figure*}
        \centering

        \includegraphics[width=0.953\textwidth]{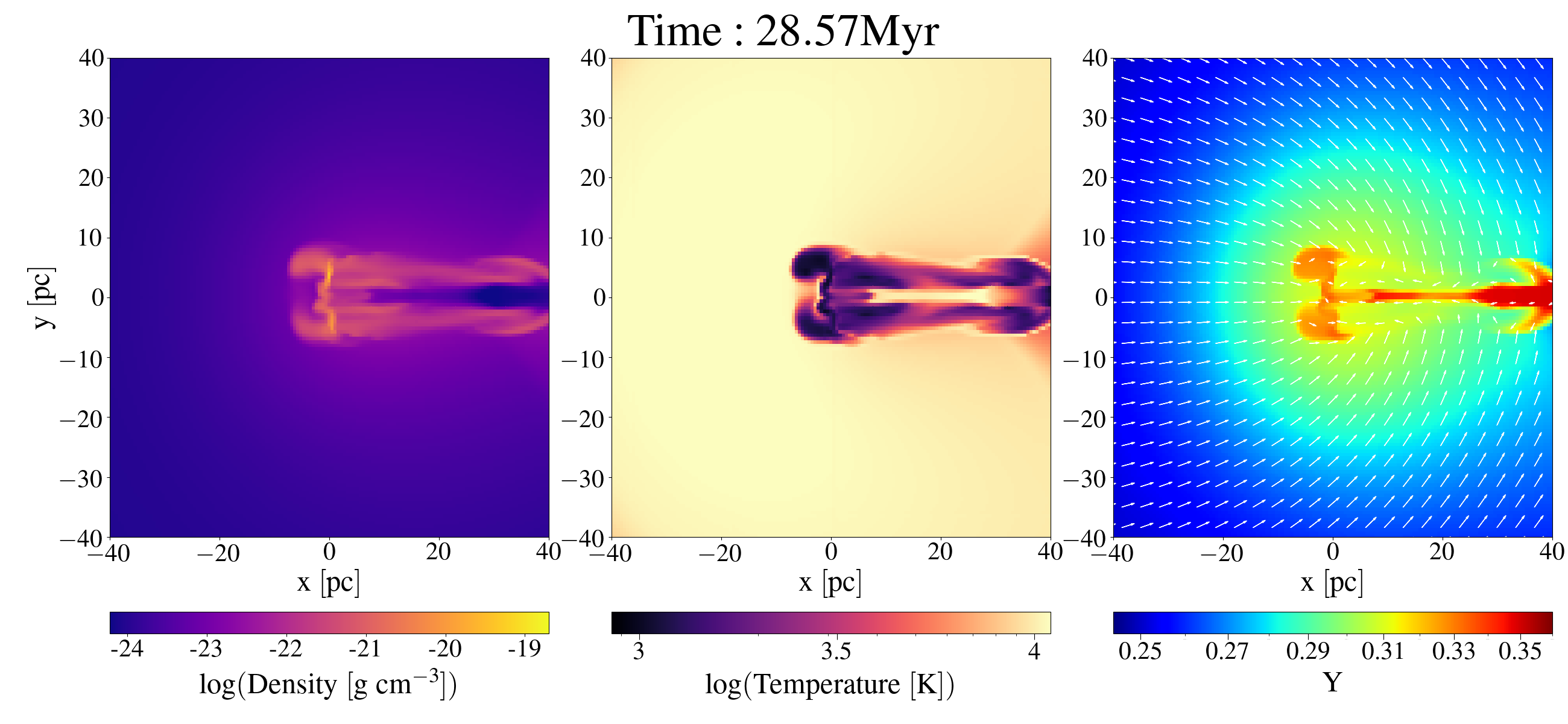}
        \\

        \includegraphics[width=0.953\textwidth]{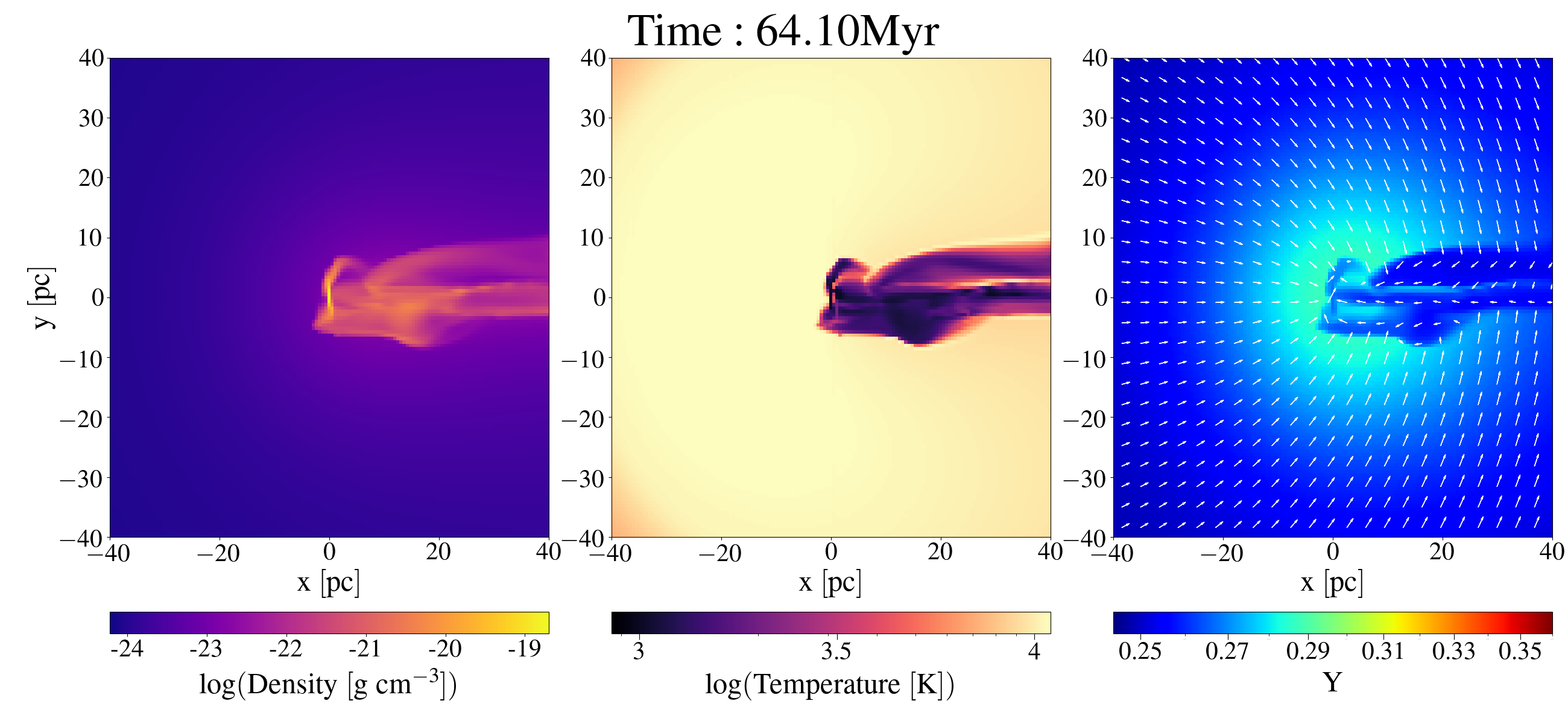}

\caption{Two-dimensional maps of the gas density (left-hand panels), of the temperature (central panels) and helium mass fraction (right-hand panels) on the x-y plane for the {\tt LDanax} simulation. The time is reported on the top of each set of panels. The white arrows in the helium mass fraction maps show the gas velocity field.}
  \label{fig:maps_LDanax}
\end{figure*}

In Fig. \ref{fig:maps_LDanax} we show the 2D density, temperature and helium mass fraction slices at almost the same times as in Fig. \ref{fig:maps_LDanaz}.
At 28 Myr, the shock front due to the infalling event has already overcome the cluster and the accretion column has formed. However, the shape of the accretion tail is very different from the one in Fig. \ref{fig:maps_LDanaz} since here the rotation leads to the formation of a disk on the $yz$ plane. 

The infalling gas is colliding with the disk face-on and is forced to circulate around the disk giving rise to two accretion columns divided by a very narrow tail, still not polluted by the pristine gas. This feature is even more clear in the helium mass fraction map, where a tight tail of extremely helium enhanced gas is present downstream of the system. The two helium-enriched blobs near the centre are instead due to the bounce of the helium-rich gas first pushed by the infalling gas downwards. to explain  

At 64 Myr, the tail of extremely helium-enhanced gas has disappeared, while the accretion column has acquired its standard shape. In the central regions, the gas is still enriched in helium, whereas in the outskirts and along the accretion column the AGB ejecta have been strongly diluted by the infalling gas and have an helium abundance very similar to that of the pristine gas.

%Therefore, an extended area is hit by the infalling gas, which is forced to circulate around the disk giving rise to two accretion columns divided by a very narrow tail, still not polluted by the pristine gas. Such distinction is even more clear looking at the helium mass fraction map, where a tight tail of extremely helium enhanced gas is present downstream of the system.{\bf The two helium-enriched blobs near the centre are instead due to the bounce of the helium-rich gas first pushed by the infalling gas downwards. to explain }

%At 64 Myr, the tail of extremely helium-enhanced gas is disappeared while the accretion column has acquired its standard shape. In the central 10 pc the gas is still enriched in helium, whereas in the outskirts and along the accretion column the gas has essentially pristine helium composition.
%{The bubble of gas of intermediate helium composition surrounding the system is due to the formation of the bow shock, which has a radius comparable to the Plummer radius. }

\subsubsection*{Evolution of the stellar component}

\begin{figure*}
\centering

        \includegraphics[width=0.324\textwidth]{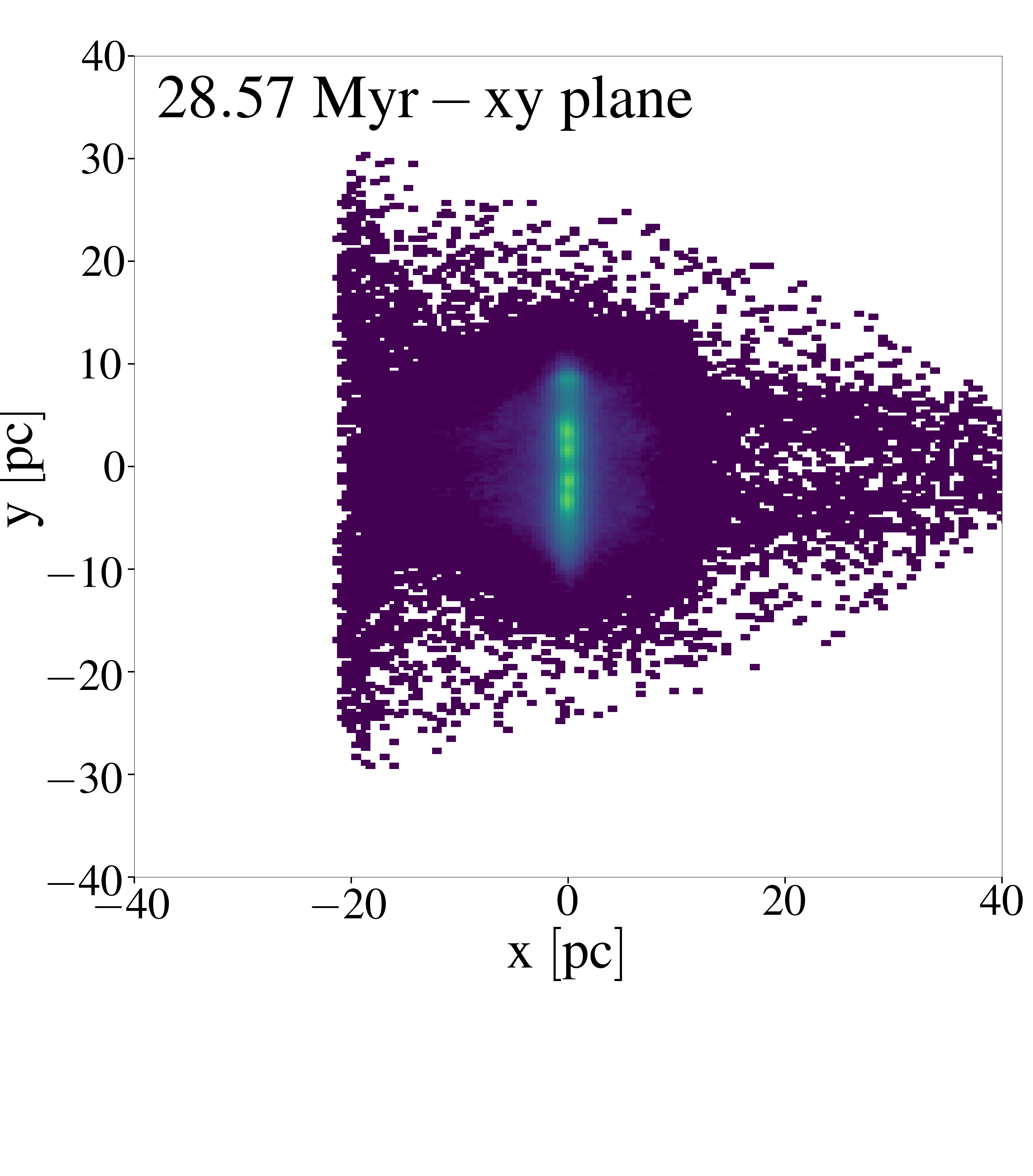}        
        \includegraphics[width=0.324\textwidth]{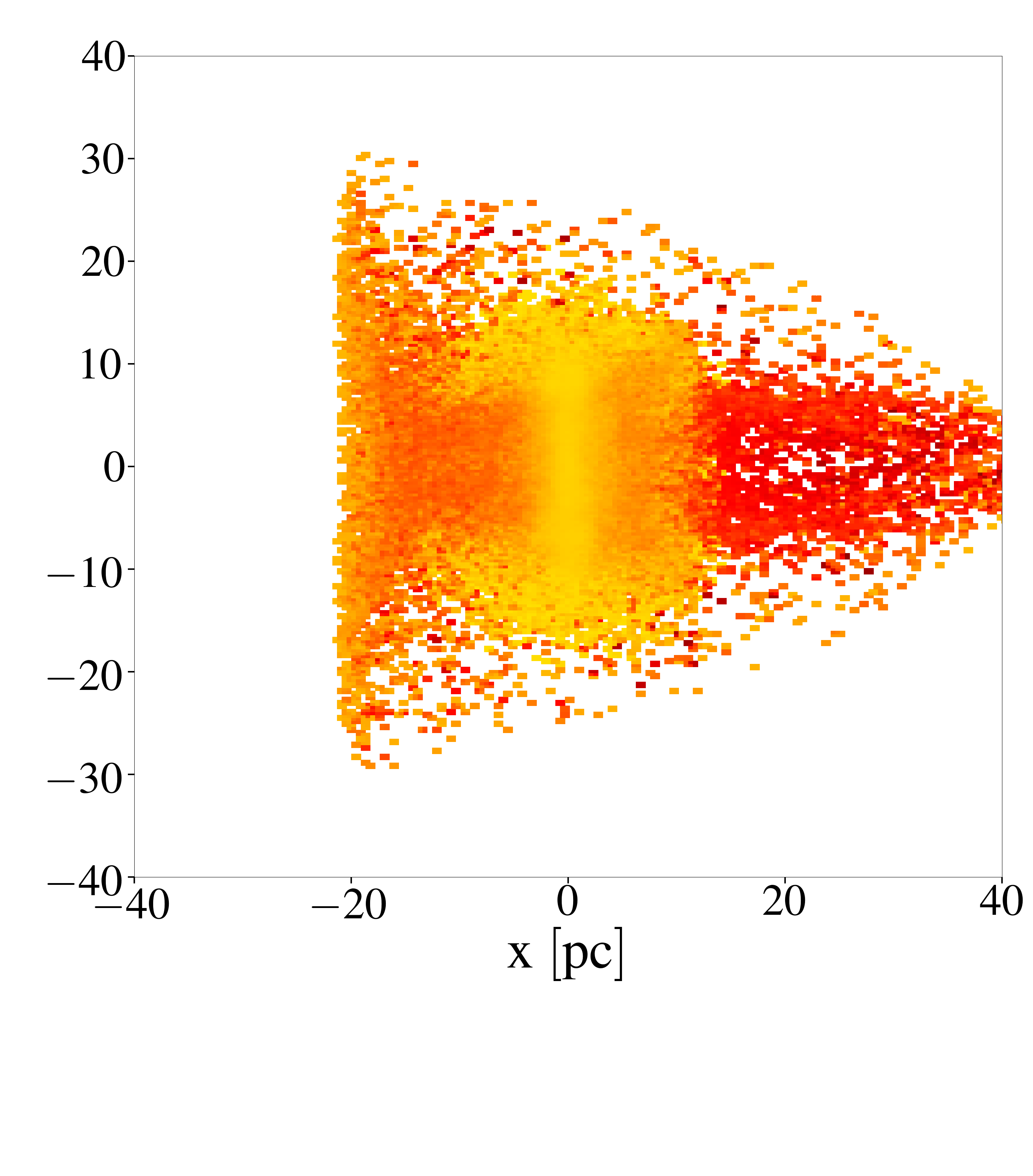}
         \includegraphics[width=0.324\textwidth]{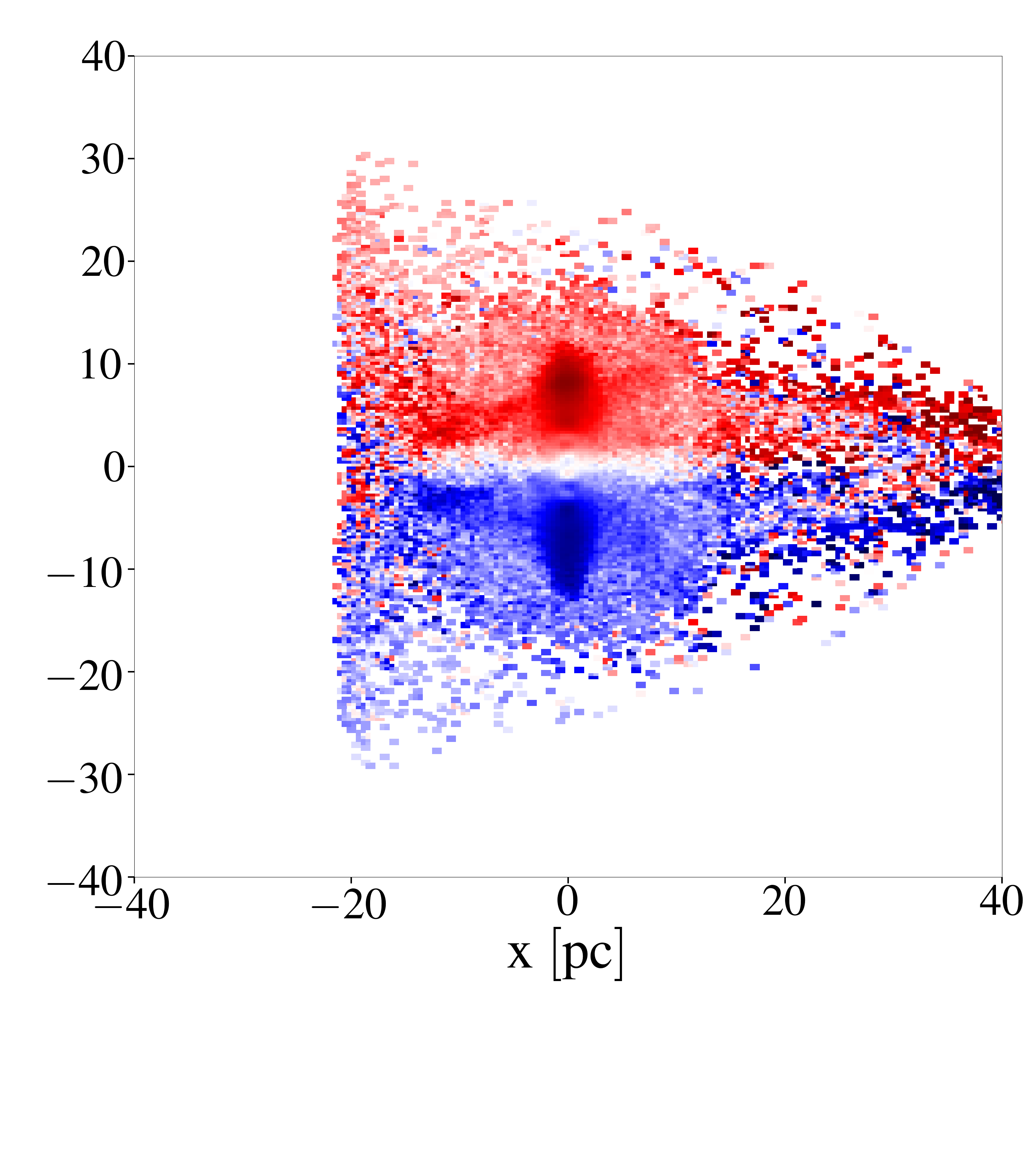}
      \\
       \vspace{-1.1cm}
      
        \includegraphics[width=0.324\textwidth]{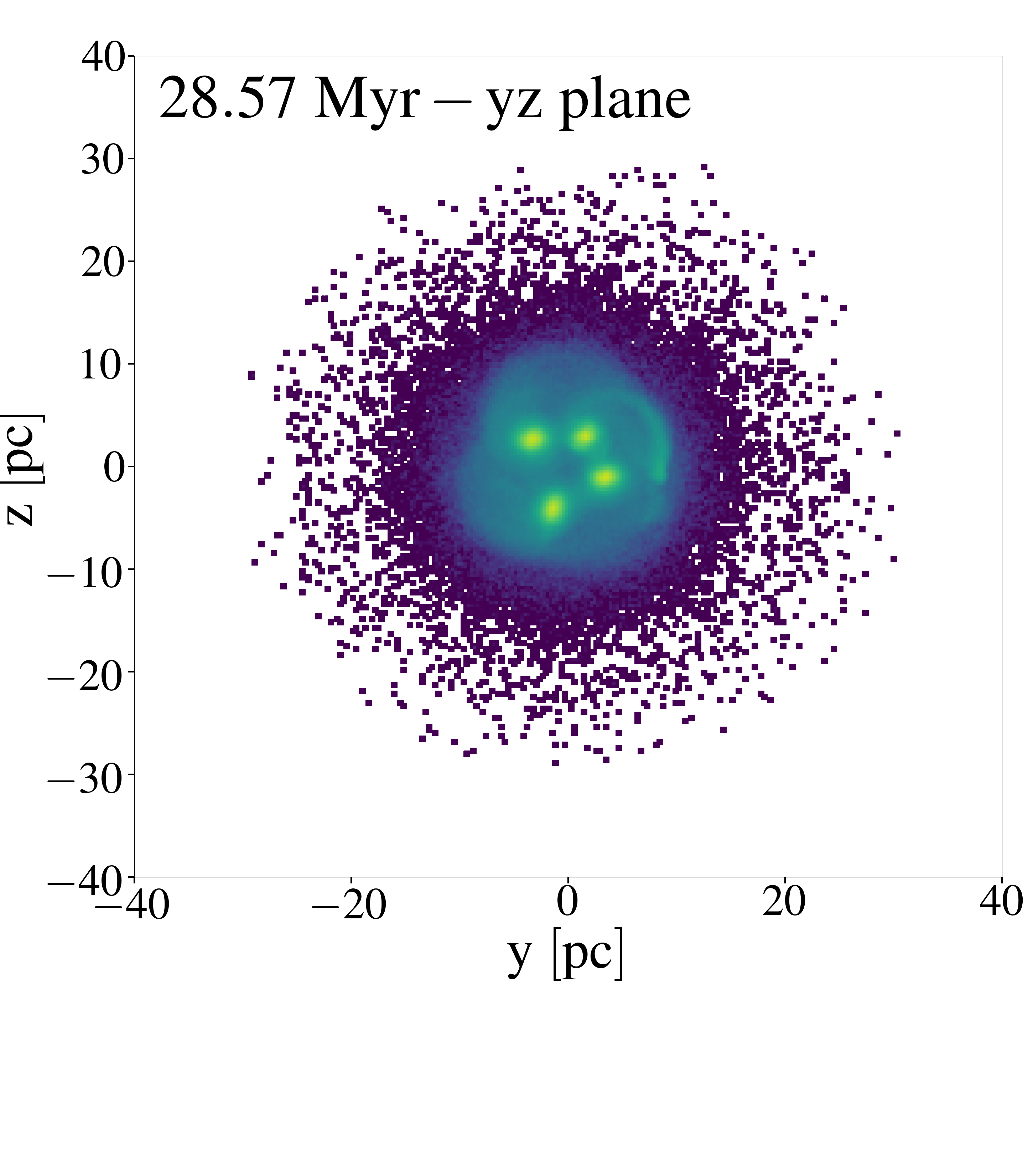}
        \includegraphics[width=0.324\textwidth]{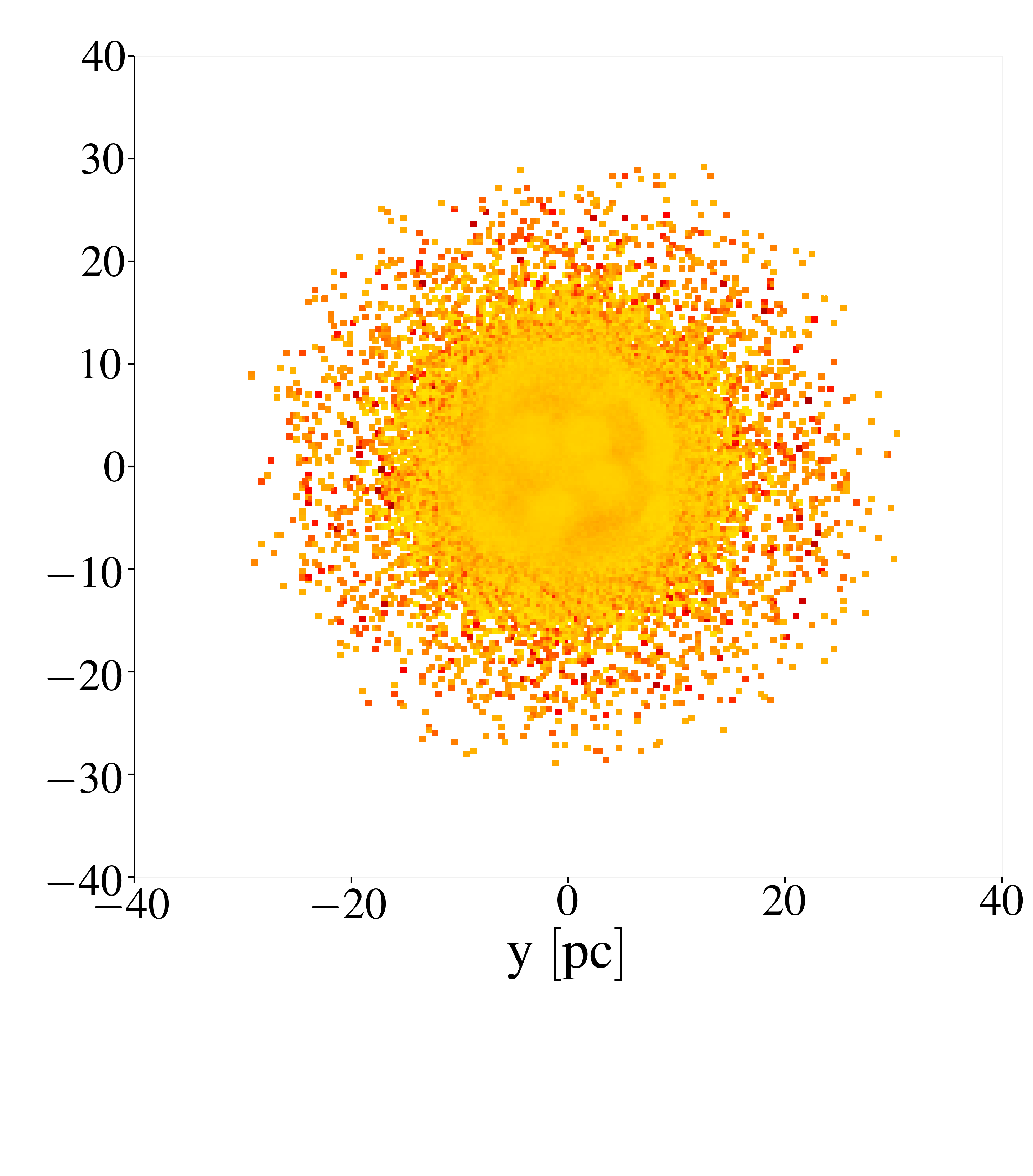}
           \includegraphics[width=0.324\textwidth]{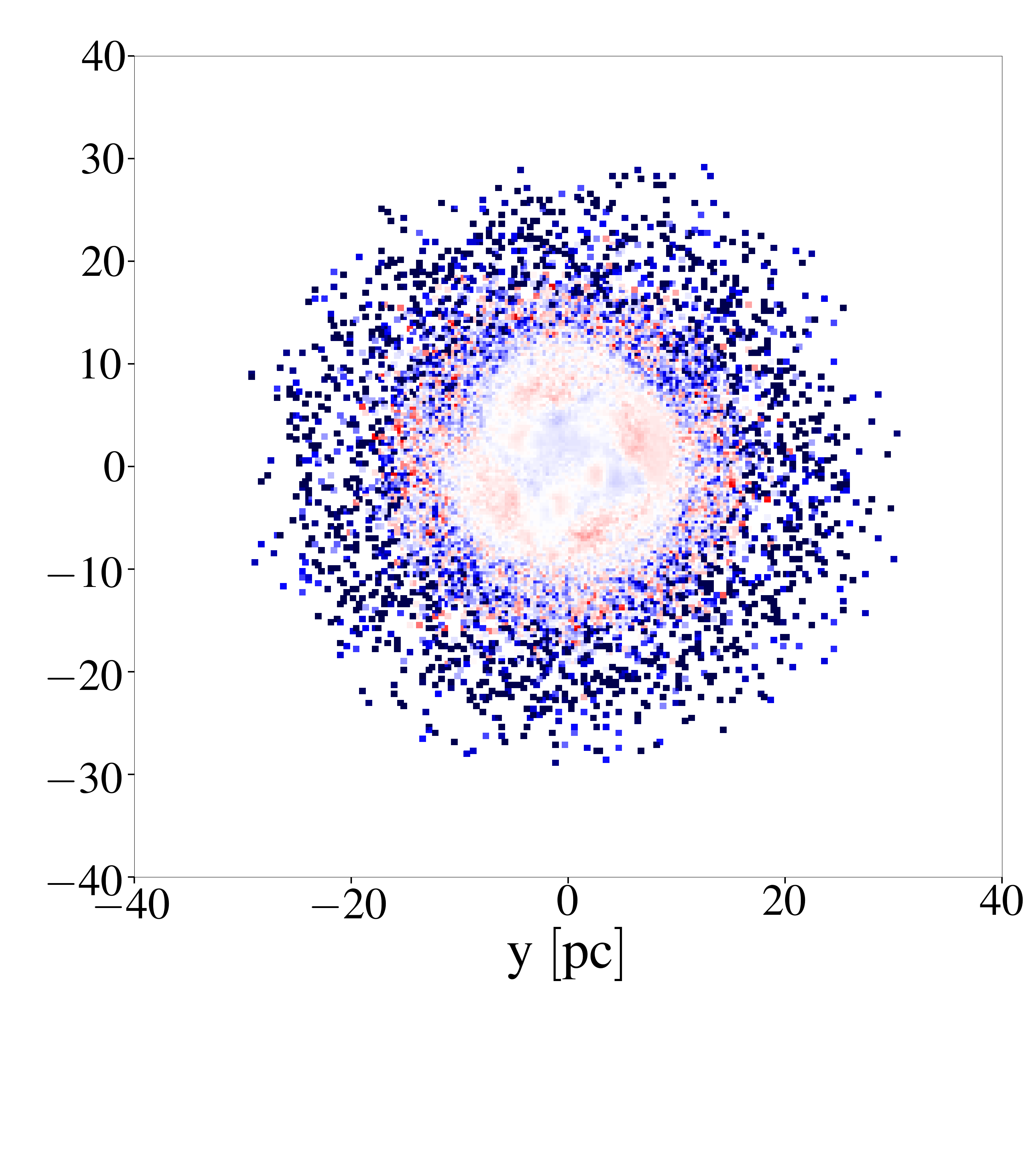}
        \\
        \vspace{-1.1cm}
          \includegraphics[width=0.324\textwidth]{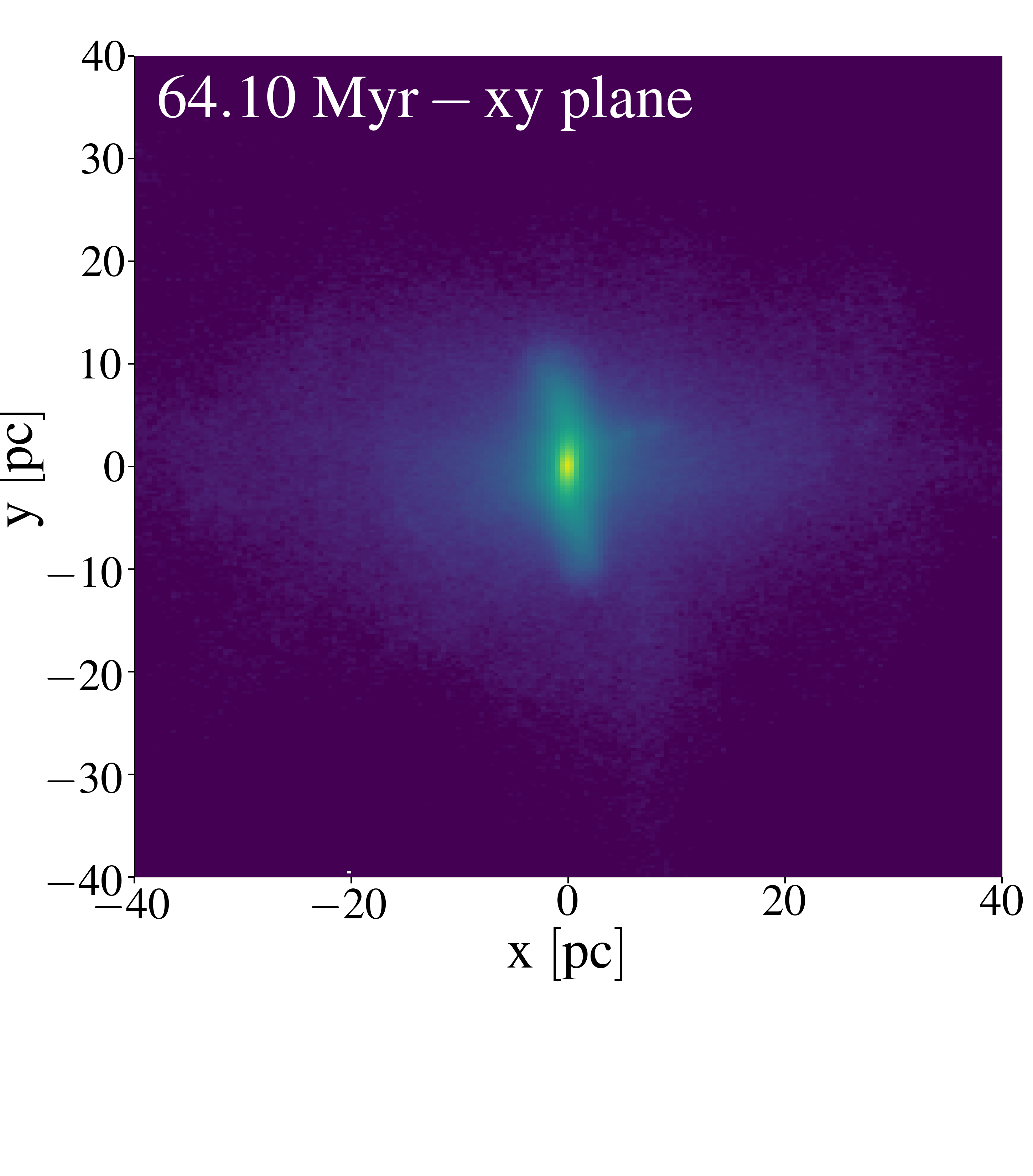}
        \includegraphics[width=0.324\textwidth]{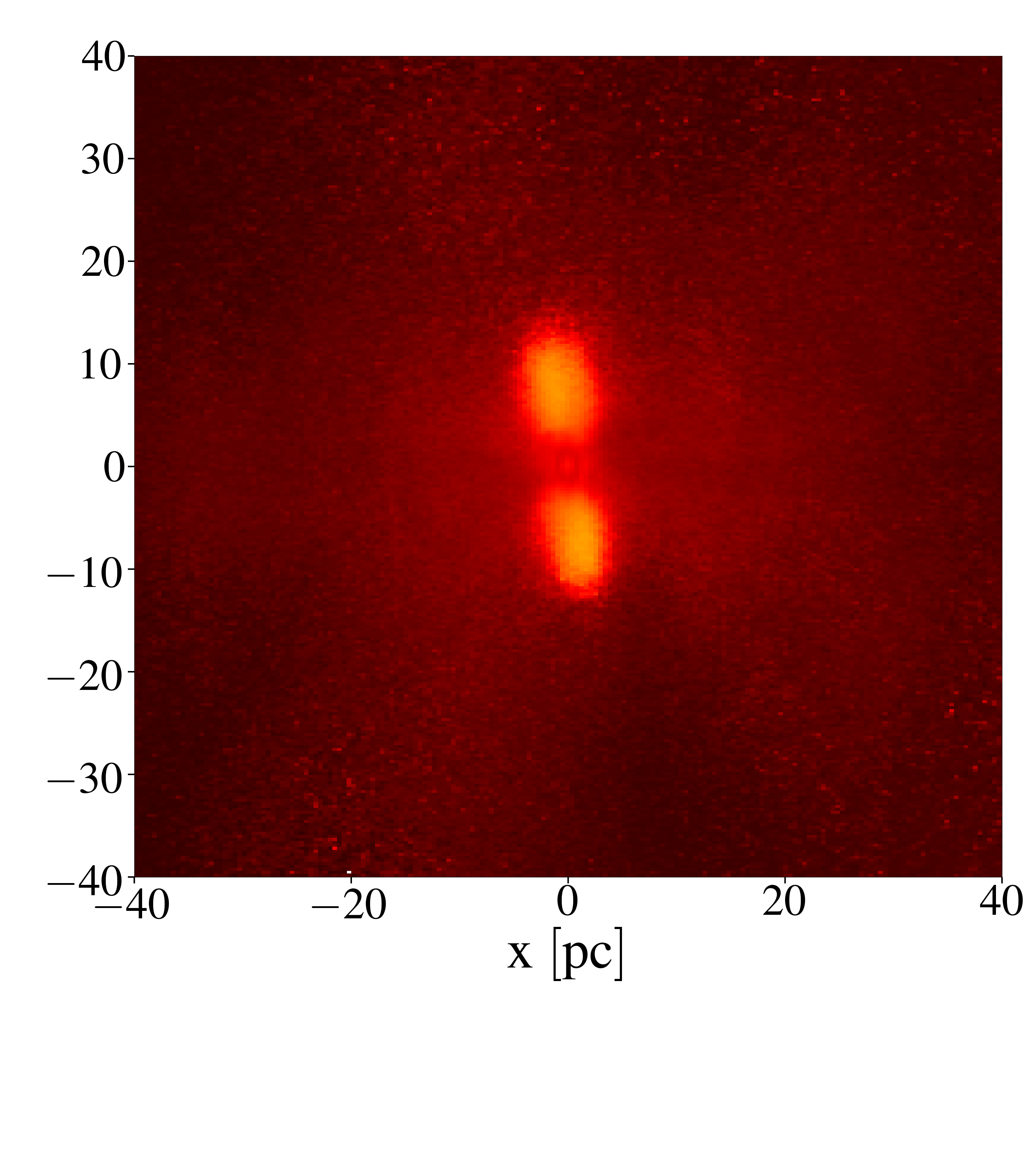}
        \includegraphics[width=0.324\textwidth]{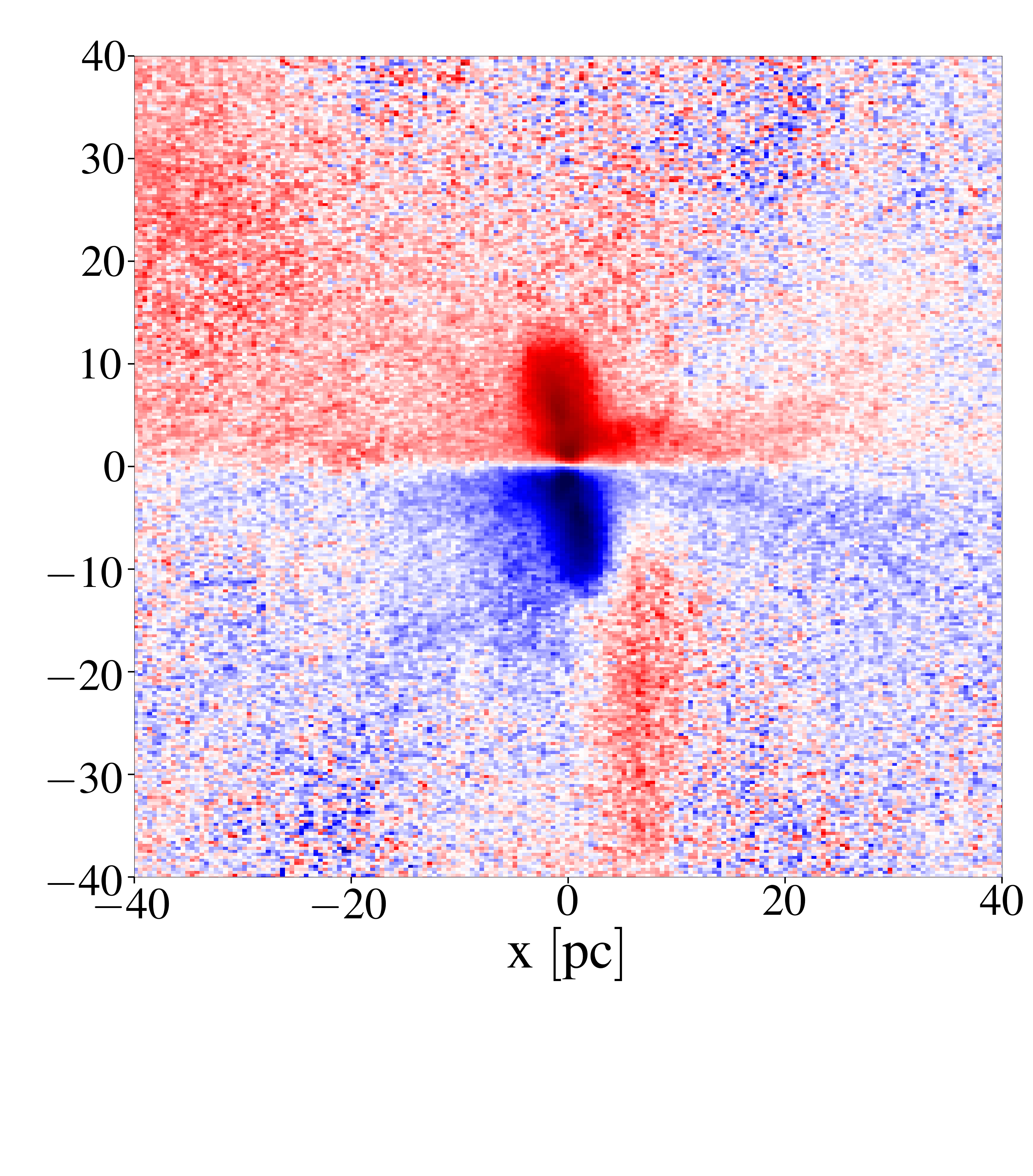}
        \\
        \vspace{-1.1cm}
        \includegraphics[width=0.324\textwidth]{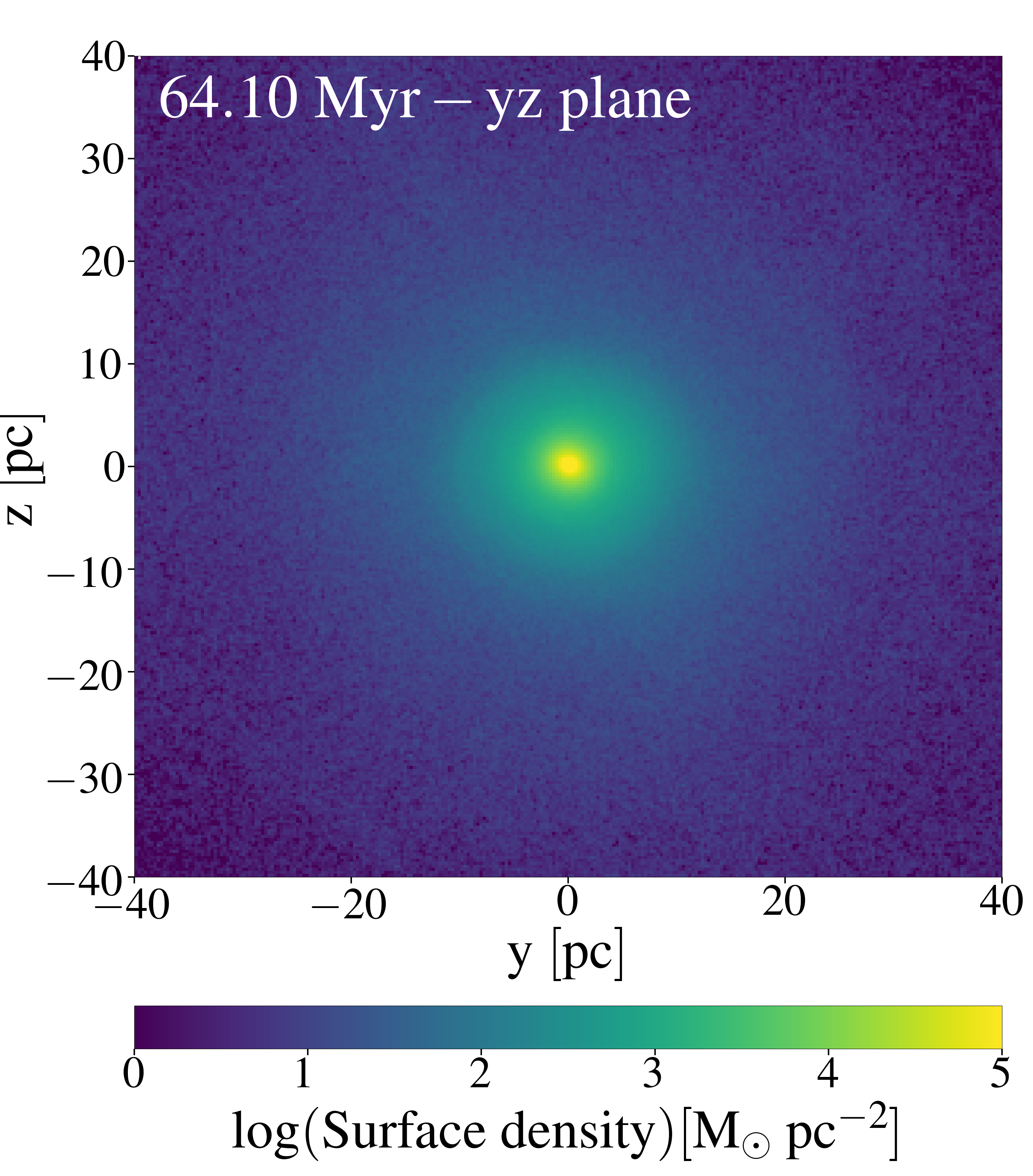}
        \includegraphics[width=0.324\textwidth]{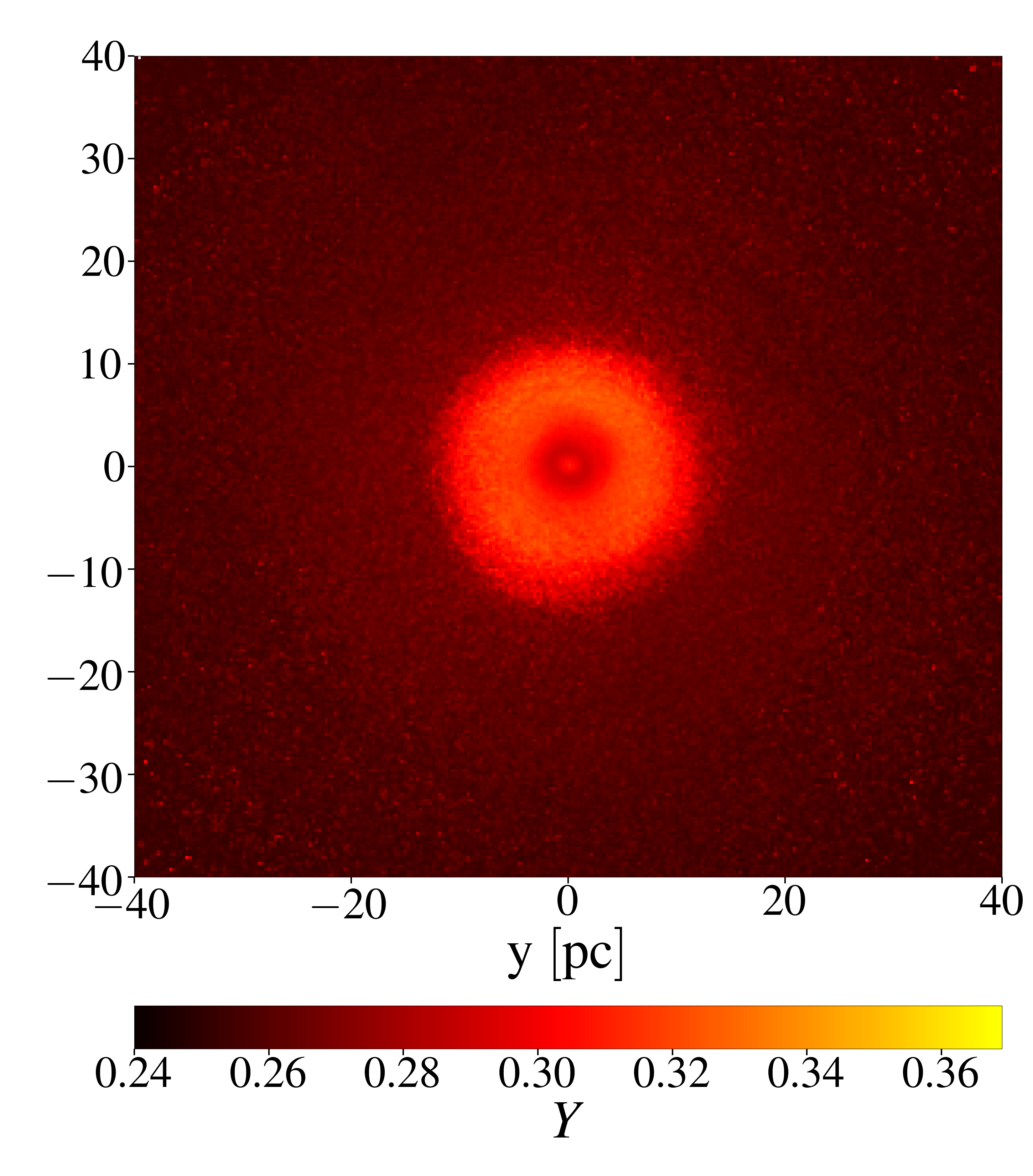}
        \includegraphics[width=0.324\textwidth]{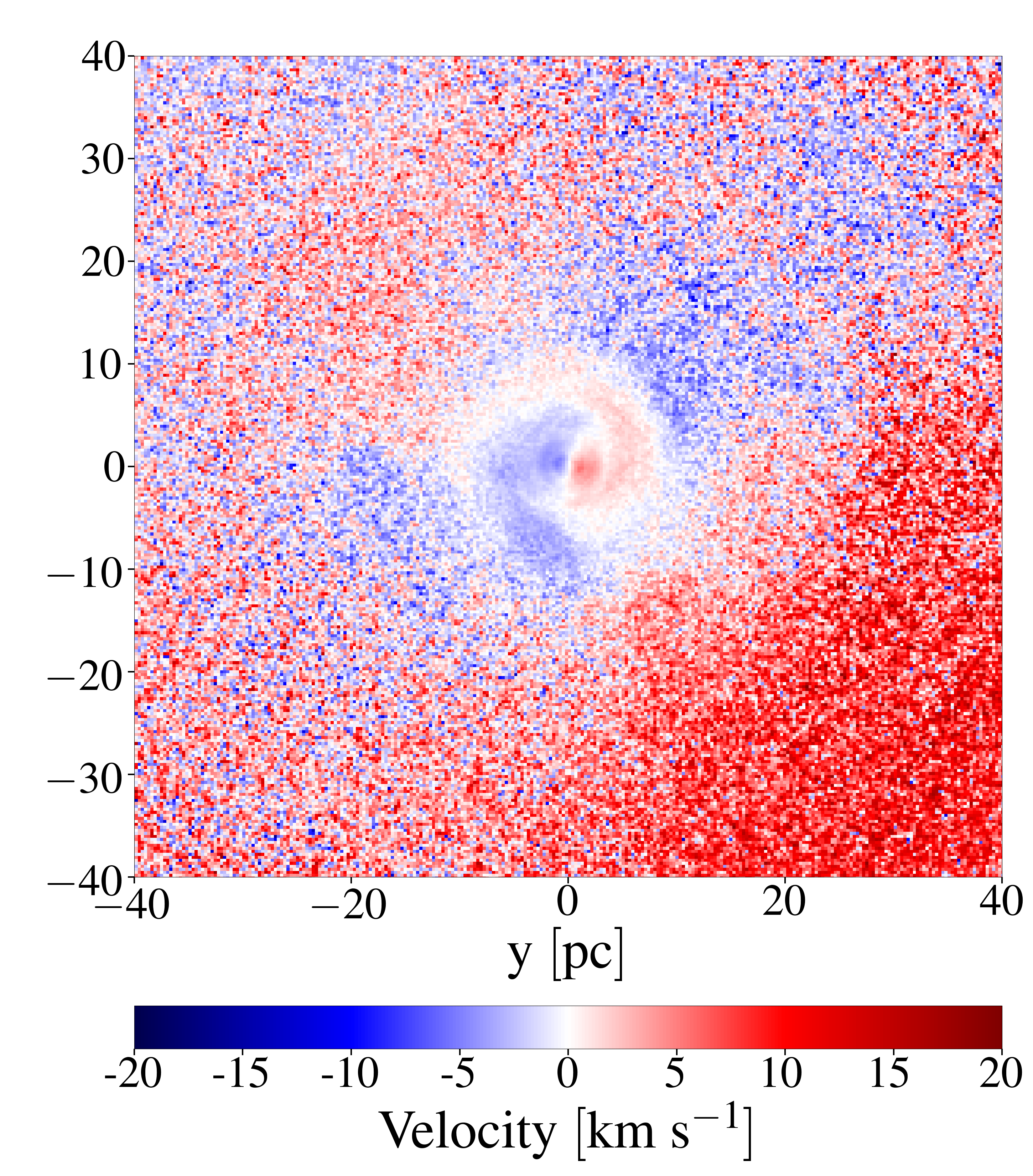}

\caption{Two-dimensional maps of the stellar component for the {\tt  LDanax} model. The reported quantities are as in Fig. \ref{fig:maps_part_LDanaz}.}
  \label{fig:maps_part_LDanax}
\end{figure*}

\begin{figure*}
        \centering

        \includegraphics[width=0.454\textwidth,trim={0 0 0 8.0cm},clip] {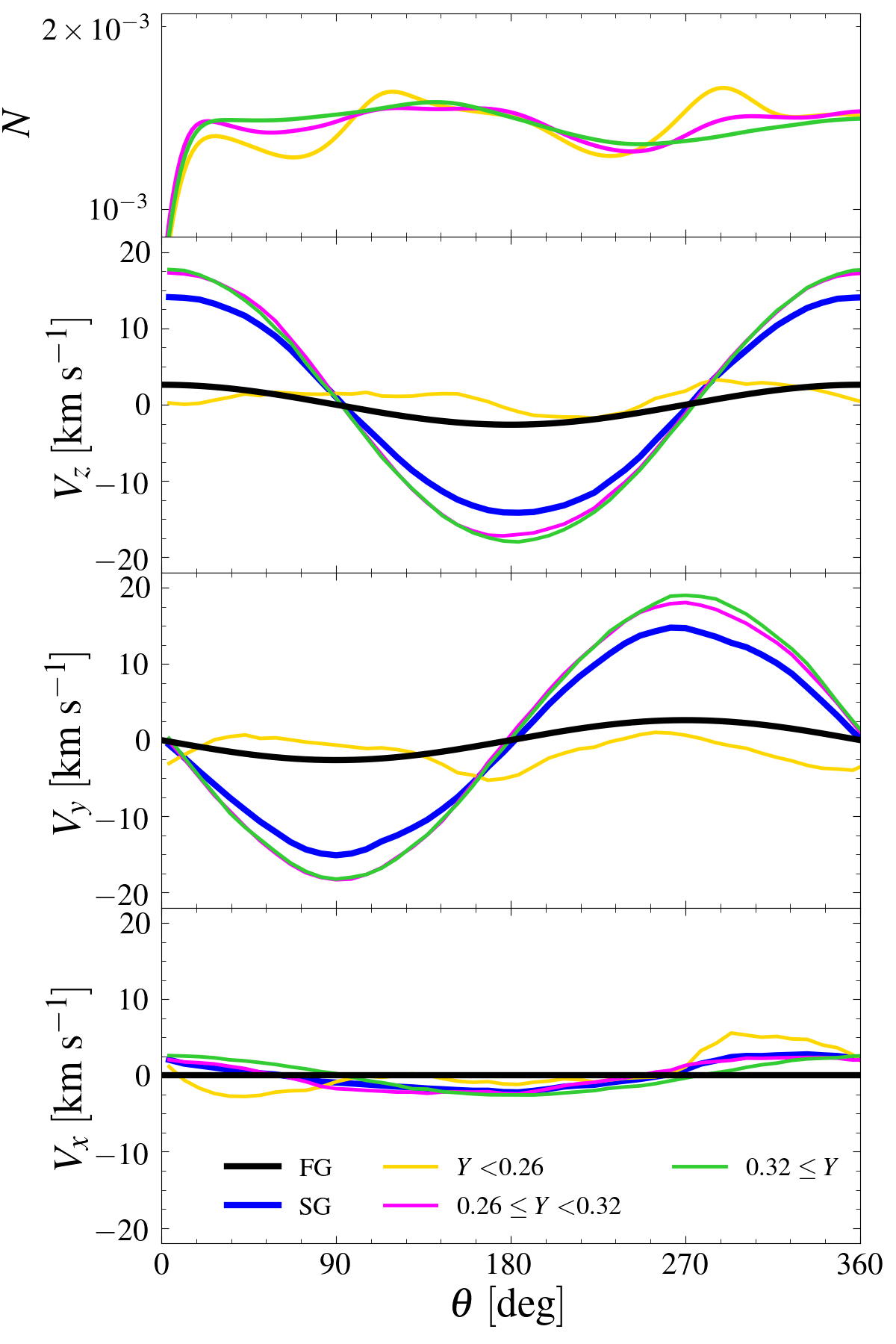}    
         %\hspace{0.01cm}
        \includegraphics[width=0.454\textwidth,trim={0 0 0 8.0cm},clip]{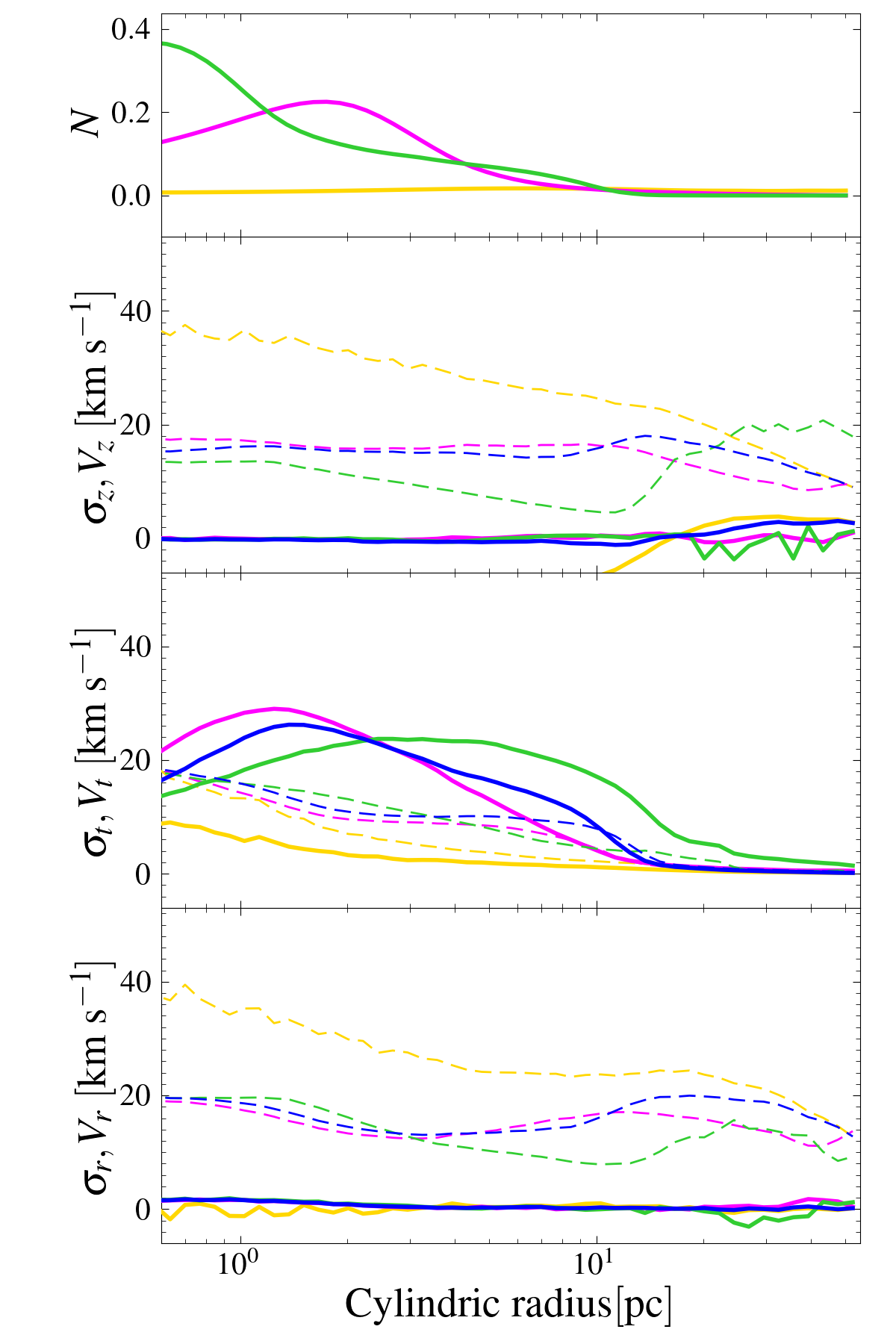}     
        \caption{Stellar rotation profiles of the model {\tt LDanax} at 65 Myr. The reported quantities are as in Fig. \ref{fig:sigmav_LDanaz}. Note that the $z$ component is the one parallel to the rotational axis that here is along the $x$ axis. }
  \label{fig:sigmav_LDanax}
\end{figure*}

Fig. \ref{fig:maps_part_LDanax} shows the 2D maps for the stellar component at the same evolutionary times of Fig. \ref{fig:maps_LDanax}. 

At 28 Myr, stars with low helium enhancement start to form both downstream of the system along the two cold and dense tails clearly seen in Fig. \ref{fig:maps_LDanax} and upstream of it at around 20pc from the cluster centre. At this distance, a shock is formed due to the interaction of the infalling gas with the potential well of the cluster, which induces the formation of new stars. Four clumps of extremely helium enhanced stars are formed in the very centre and located in the disk similarly to what has been obtained for the {\tt LDanaz} model.%{\bf In the $yz$ plane of the velocity map it is visible the difference between the rotating stars, which possess very low velocity along the line of sight, and stars mainly formed out of the infalling gas, which instead have a high line of sight velocity}

 At 64 Myr, the stellar disc is still present and particularly evident in all the maps in Fig. \ref{fig:maps_part_LDanax}. Similarly to the {\tt LDanaz} model, we find that the very inner regions host stars slightly less enriched in helium. The disk is slightly thicker than in the {\tt LDanaz} model; this is due to the dynamical effect of the gas infalling along the accretion column in the direction perpendicular to the plane of the disk. In addition, the disk appears slightly tilted by about 10 degrees relative to the x-y plane.
The left panel of Fig. \ref{fig:sigmav_LDanax} shows that the SG component has a higher rotational amplitude than the FG one, with a peak value similar to the one found for the {\tt LDanaz} model, and, in addition, SG stars with high helium composition are rotating faster than those with moderate helium enrichment. %In this model, the SG has an overall rotational amplitude (i.e. the peak of the mean values computed for bins of $\theta$) of ${\rm 14\ km\ s^{-1}}$, which is slightly higher than in the {\tt LDanaz} model. 
The different orientation of the rotation axis with respect to the direction of the infall implies that the external gas motions do not affect the rotation plane of the disk, leading to a smoother rotation profile as shown in the left panel of Fig. \ref{fig:sigmav_LDanax}.
 %Before the arrival of the infalling gas, in fact, the rotation amplitude is the same in both the two models. Moreover, the $V_x$ component of the SG is not zero, which confirms that the disk is tilted and, in particular, it preceeds about the $x$ axis similarly to the {\tt LDanaz} model.
 %Even in this model, the very central part of the cluster is dominated by stars with modest helium enhancement.%, which are rotating at almost ${30 \rm km \ s^{-1}}$. 
Similarly to the {\tt Ldanaz}, $|V_{rot}/\sigma|$ is increasing inside the disk region with a peak of 2, therefore the system is rotationally supported even at large radii.%, while in the low-density case with a rotation about the $z$-axis the dispersion $\sigma_t$ significantly increases in the outskirts, due to the gas coming from the infalling event and the accretion column.} 

\section{DISCUSSION}
\label{sec:discussion}

\begin{figure*}
        \centering

        \includegraphics[width=0.434\textwidth]{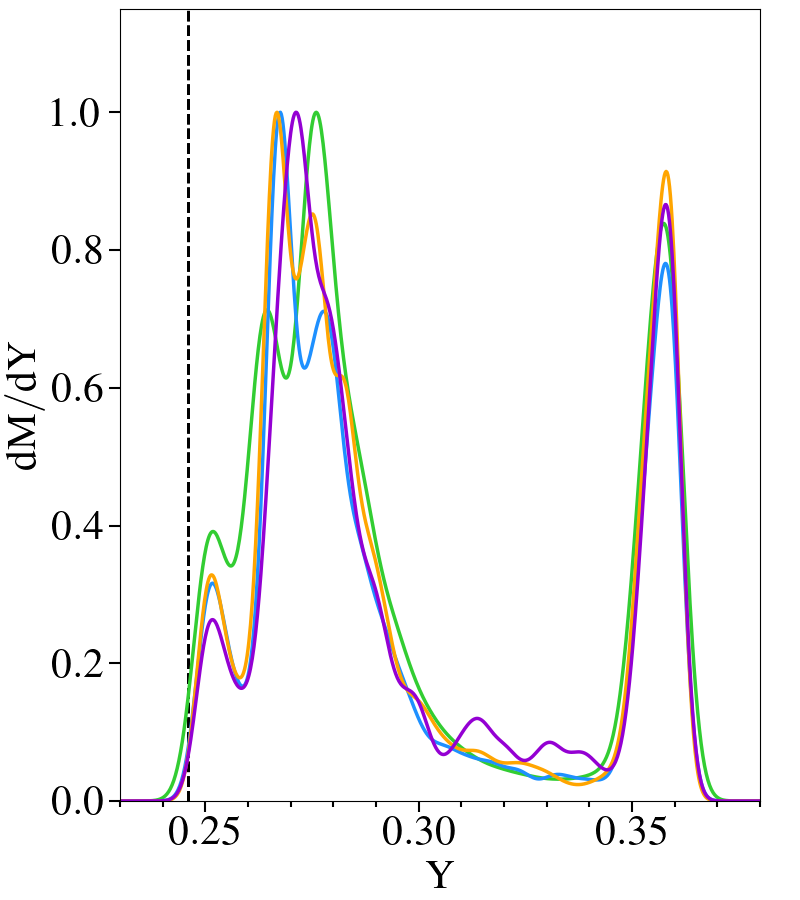}        
        \includegraphics[width=0.434\textwidth]{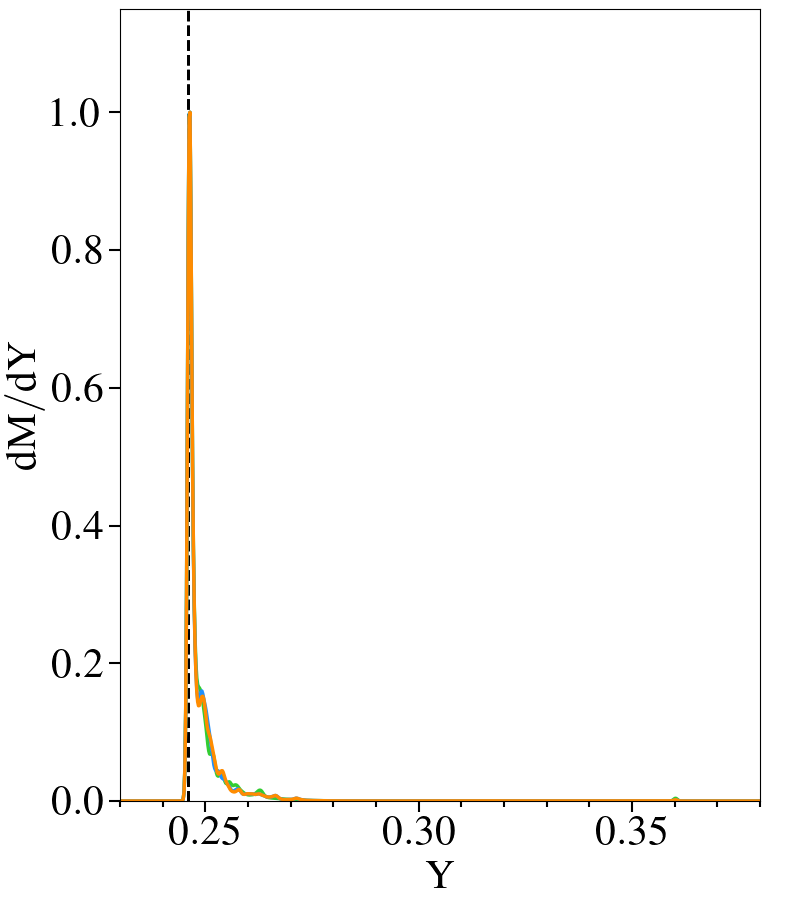}

\caption{The mass distributions of Y in SG stars at the end of the simulations ($\sim 65 $ Myr) for different models. On the left: low-density models: {\tt LDanaz} (blue), {\tt LDanax} (red), {\tt LDsbz} (dots) and {\tt LD} (green line). On the right: high-density models: {\tt HDanaz} (blue), {\tt HDanax} (red) and {\tt HD} (green line). Each distribution is normalised to its maximum value. The pristine gas helium mass fraction Y is shown by the black dashed line. }
  \label{fig:mdfs}
\end{figure*}

Our simulations have explored the role of cluster rotation on the formation of SG stars in a massive proto-GC. We have shown how different inclinations and pristine gas densities affect the kinematical and structural evolution not only of the FG and SG, but for the first time, of different subpopulations. We here discuss our results, connecting them with the relevant literature both from the observational and theoretical side.

\subsection{On the chemical composition of SG stars}
While cluster rotation has a significant effect on the morphological and kinematical properties of SG stars, it does not have a strong impact on the overall chemical composition of the system and its final stellar mass. In Fig. \ref{fig:mdfs}, the helium distribution functions of both the low and high density models are compared. As shown in this figure, for both the low-density and high-density models, the helium distribution is not significantly affected by the orientation of the rotation axis relative to the direction of motion of the cluster in the external gas environment. The two panels are also in general agreement with the helium distribution found in the non-rotating models by \citetalias{calura2019}.

 %A similar result is found once comparing the SG density profiles for various ranges of $Y$. The only difference is the presence of the disk, which leads to a lower density of extremely helium enhanced stars in the very centre, and their predominance between 3 and 10 pc, where the disk lies. {\bf The half-mass radius of the SG component is $\sim 1.5$ pc for all models with rotation,} independently on the rotational axis and density profile. It is however 1.5 time larger in comparison with the case without rotation.

\subsection{Comparison with literature}

Recently, \citet{mckenzie2021} studied the formation of MPs in proto-GCs within the context of their parent galaxy, through 3D hydrodynamic simulations. In all their models, the cluster is assumed to rotate as a solid body. At variance with our simulations, the infalling gas is modelled as clumpy. For their fiducial model, which assumes a FG of mass $M_{\rm FG}=10^6 {\rm M_{\odot}}$, an order of magnitude lower than in our case, they found a flattened distribution of the SG, similarly to \citet{bekki2010,bekki2011} and the present work. Disks, however, are not a common feature in GCs, which are generally described as spheroidal systems, although some degree of ellipticity has been reported by \citet{frenk1982} who found an age-ellipticity relation in Galactic and LMC GCs. It is however verified by $N$-body simulations that SG stars born in a disk mix with FG ones through angular momentum exchange, ending up with a flattened system after 12 Gyr of evolution \citep[see also \citealt{tiongco2021}]{mastrobuonobattisti2016}.
%Their strength depends both on the mass of the SG disk and the relaxation time of the system.  
At variance with our results, instead, \citet{mckenzie2021} found a less concentrated SG compared with the FG, a configuration found only in few GCs. On the other hand, in both our studies, FG and SG have been derived to rotate in phase, with the SG rotating significantly faster than the FG. Even though we are exploring here the evolution of a much more massive system, our resulting SG rotational amplitude is in agreement with the mass-amplitude relation found by \citet{mckenzie2021}. %More massive clusters are generally assumed to rotate faster, explaining the discrepancy in our results. Since our FG is modelled as a static Plummer profile, we cannot compare the velocity dispersions of the two populations. 

As in \citet{mckenzie2021}, we have tested different inclinations of the GC rotational axis with respect to the one of the host galaxy (which is here identified by the direction of the infalling gas). However, in the low-density models, where the disk survives, the inclination of the disk is not significantly affected by the infalling gas and, therefore, it does not align with the host galaxy as it instead happens in their simulations. The models presented here and those of \citet{mckenzie2021} differ in various aspects (FG initial mass, external infalling gas, time interval explored). Further investigation is necessary to clarify the origin of the differences between the final SG system orientation found in the two works. %Clusters with higher mass would develop a stronger rotation and also give birth to a higher fraction of SG stars, slowing down the alignment.%\textcolor{red}{ (citare qualcosa)}

 \subsection{Dynamical and kinematical features of present-day globular clusters }
%\subsection{Comparison with observations}
 Our simulations follow the formation and very early evolutionary phases of a cluster composed by multiple stellar populations. For a close comparison with the properties of present-day old and intermediate-age GCs, it would be necessary to take into account the effects of the subsequent long-term dynamical evolution. A number of studies have shown that the long-term evolution gradually modifies the properties imprinted by the formation processes. Although, in some dynamically old clusters, the differences between the dynamical properties of SG and FG stars may be completely erased during the clusters' long-term evolution, in dynamically younger ones some memory of these differences may still be preserved (see e.g. \citealt{mastrobuonobattisti2013,mastrobuonobattisti2016,vesperini2013,henaultbrunet2015,vesperini2021,tiongco2019,sollima2021,hypki2022}).
Indeed, many of the observational studies cited in the Sec. \ref{sec:Intro} have found that SG stars are more centrally concentrated than the FG population, in general agreement with what found in this paper and in previous theoretical studies (e.g. \citealt{dercole2008,bekki2010,calura2019}).

As for the kinematical properties on which this paper is focused on, a number of observational studies (see the discussion in the Sec. \ref{sec:Intro}) revealed that in some clusters the SG system rotates more rapidly than the FG, in agreement with what found in our work. The observational study of the kinematics of multiple populations is still in its early stages and much work remains to be done to build a comprehensive observational picture of the kinematical differences between FG and SG stars and, for clusters with multiple SG groups, between the various SG groups. Concerning the possible difference between various SG subgroups, our study predicts that the more extreme SG population is initially rotating more rapidly than the intermediate (moderately enriched) SG groups; this result is consistent with the trend found by \citet{cordero2017} and \citet{kamann2020} in  M13  and M80, respectively. As for the morphology of different stellar populations, a few early studies have found that in some clusters the SG subsystem is more flattened than the FG one (see \citealt{lee2017,cordoni2020,cordoni2020b}; see also \citealt{lee2018} for a cluster showing instead an opposite trend) providing possible examples of clusters retaining some memory of the initial differences emerging from our simulations.  
Additional observational studies and numerical simulations exploring the long-term evolution of clusters with initial conditions informed from our study will be necessary to further explore these aspects of the structure and kinematics of multiple stellar populations.

Finally, we conclude this section with a brief remark on the mass of SG stars forming in our simulations. The total mass of SG stars formed by the end of our simulation is about 0.1 the mass of the FG system, in the low-density models, and about equal to the mass of FG stars, in the high-density models. As shown in a number of previous studies (see e.g. \citealt{dercole2008,vesperini2021,sollima2021}), the fraction of the total cluster mass in SG stars may significantly increase as a result of the preferential loss of FG stars during the cluster's early and long-term evolution but, as already pointed out in \citet{calura2019}, the extent of mass loss required to reach the values of the SG mass fraction observed in present-day globular clusters (see e.g. \citealt{milone2017}) is much less extreme than sometimes reported in the literature. More extended surveys of simulations exploring the formation of multiple populations are necessary to test the role of various physical ingredients (e.g. FG structural properties and initial stellar mass function, stellar feedback, external environment) and shed further light on the initial mass of the SG population and its dependence on the initial FG properties (see e.g. \citealt{mckenzie2021,yaghoobi2022} for two recent studies addressing these issues).

\section{CONCLUSIONS}
\label{sec:conclusions}

An increasing number of observational studies of the kinematics of stars clusters are revealing that internal rotation is a common kinematic feature of these stellar systems. It is therefore important to include rotation in theoretical studies of star clusters and explore its role on their formation and dynamical evolution.

In this paper, by means of 3D hydrodynamic simulations, we have studied the role of rotation on the formation and early dynamics of multiple stellar populations in GCs.

Our models follow the formation of SG stars out of the ejecta released by AGB stars in a rotating FG system and the external pristine gas accreted by the cluster. We have  explored the resulting structural and kinematical properties of the SG subsystem, and studied the differences between the dynamical properties of FG stars and those of the various SG subgroups.

In our simulations, we have modelled a massive proto-GC with a FG mass of $10^7 {\rm M_{\odot}}$, moving through a uniform gas distribution for which we have assumed two different density values: a low-density one characterized by $\rho_{\rm pg}=10^{-24} {\rm g \ cm^{-3}}$ and a high-density one with $\rho_{\rm pg}=10^{-23} {\rm g \ cm^{-3}}$. In order to explore the interplay between the internal rotational dynamics of the AGB ejecta and that of the external infalling gas, we have considered two different configurations: one in which the FG rotational axis is perpendicular to the motion of the cluster through the external medium and one in which it is parallel to it. 

The main results of the paper are the following:

\begin{itemize}
    \item Our simulations have revealed the complex hydrodynamics/stellar dynamics of the SG formation phase in the presence of a rotational FG system. As derived in previous investigations, we find that the SG forms concentrated in the innermost regions of the FG cluster. Both the SG morphology and kinematics are significantly affected by rotation and the interaction between rotating AGB ejecta and non-rotation infalling external gas. The AGB ejecta initially collect in a disk of gas in the inner regions of the FG cluster and form a rotating disk of helium-enhanced SG stars. The disk survives in the model of a cluster moving in a low-density external medium where the infalling, non-rotating gas has a minor effect on the overall evolution of the system. An inner SG disk also forms in the high-density model but, as a consequence of the earlier arrival of the infalling gas and its higher density, the disk is disrupted before the end of the simulation. Although the long-term dynamical processes may gradually alter the SG disk and drive it towards an increasingly spherical spatial distribution, massive clusters forming a helium-enhanced population in a low-density external environment are those where some memory of an initial flattened SG subsystem might be found. 
    
    \item The SG populations forming in a rotating FG cluster embedded in low-density pristine gas are characterized by a rotational amplitude larger than that of the FG population. The differences between the SG and the FG rotational kinematics we find in our simulations are generally consistent with the findings of previous theoretical \citep{bekki2010,bekki2011,mckenzie2021} and observational investigations \citep{cordero2017,cordoni2020,dalessandro2021,szigeti2021}.

    \item The more He-rich SG subgroups forming earlier, mainly out of AGB ejecta, rotate around the cluster's centre more rapidly than SG stars formed later, out of a mix of rotating AGB ejecta and non-rotating infalling pristine gas. However, for a comparison with the present-day properties of old GCs, it is necessary to take into account the effects of the cluster long-term dynamical evolution  \citep{mastrobuonobattisti2013,mastrobuonobattisti2016,henaultbrunet2015,tiongco2019,mastrobuonobattisti2021,sollima2022,vesperini2021}. The findings of our simulations are generally consistent with those of the first observational studies that have explored the kinematics of different SG subgroups and found that the more helium-enhanced SG stars rotate faster than the helium-poor ones \citep{cordero2017, kamann2020}.  When SG stars are formed from a rotating FG component embedded in high-density pristine gas, no significant differences are found both between the SG subgroups and the FG and SG systems.

    \item  Very minor differences have been found between models assuming a FG solid-body rotational profile and those assuming the analytic profile of Eq. \ref{eq:anarot} or changing the orientation of the rotational axis with respect to the direction of the infalling external gas.
    \item The more complex hydrodynamics of SG formation in a rotating FG cluster does not significantly affect the final distribution of the helium abundances of the SG populations.
\end{itemize}

In future studies, we will further expand the investigation presented here to fully explore the dependence of our results on the initial properties of the FG clusters (e.g. initial mass, structure, strength of initial rotation) and those of the external environment (e.g. clumpy ISM).

%%%%%%%%%%%%%%%%%%%%%%%%%%%%%%%%%%%%%%%%%%%%%%%%%%%%%%%%%%%%%%%%%

\section*{Acknowledgements}

We are grateful to the anonymous referee for the useful suggestions.
AMB acknowledges funding from the European Union’s Horizon 2020 research and innovation programme under the Marie Sk\l{}odowska-Curie grant agreement No 895174.
FC acknowledges support from grant PRIN MIUR 2017- 20173ML3WW 001, from the INAF main-stream (1.05.01.86.31) and from PRIN INAF 1.05.01.85.01. EV acknowledges support from NSF grant AST-2009193. 
We acknowledge PRACE for awarding us access to Discoverer at Sofia Tech Park, Bulgaria.
We acknowledge the computing centre of Cineca and INAF, under the coordination of the "Accordo Quadro MoU per lo svolgimento di attività congiunta di ricerca Nuove frontiere in Astrofisica: HPC e Data Exploration di nuova generazione", for the availability of computing resources and support. We acknowledge the use of computational resources from the parallel computing cluster of the Open Physics Hub (https://site.unibo.it/openphysicshub/en) at the Physics and Astronomy Department in Bologna. This research was supported in part by Lilly Endowment, Inc., through its support for the Indiana University Pervasive Technology Institute, and in part by the Indiana METACyt Initiative. The Indiana METACyt Initiative at IU is also supported in part by Lilly Endowment, Inc.

%%%%%%%%%%%%%%%%%%%%%%%%%%%%%%%%%%%%%%%%%%%%%%%%%%
\section*{Data Availability}

The data underlying this article will be shared on reasonable request to the corresponding author.



\bibliographystyle{mnras}
\bibliography{biblio} % if your bibtex file is called example.bib

%\begin{thebibliography}{99}
%\end{thebibliography}

%%%%%%%%%%%%%%%%%%%%%%%%%%%%%%%%%%%%%%%%%%%%%%%%%%

%%%%%%%%%%%%%%%%% APPENDICES %%%%%%%%%%%%%%%%%%%%%

\appendix
\section{Model with solid body rotation}

{\label{sec:appendix}

\begin{figure*}
        \centering

        \includegraphics[width=0.324\textwidth]{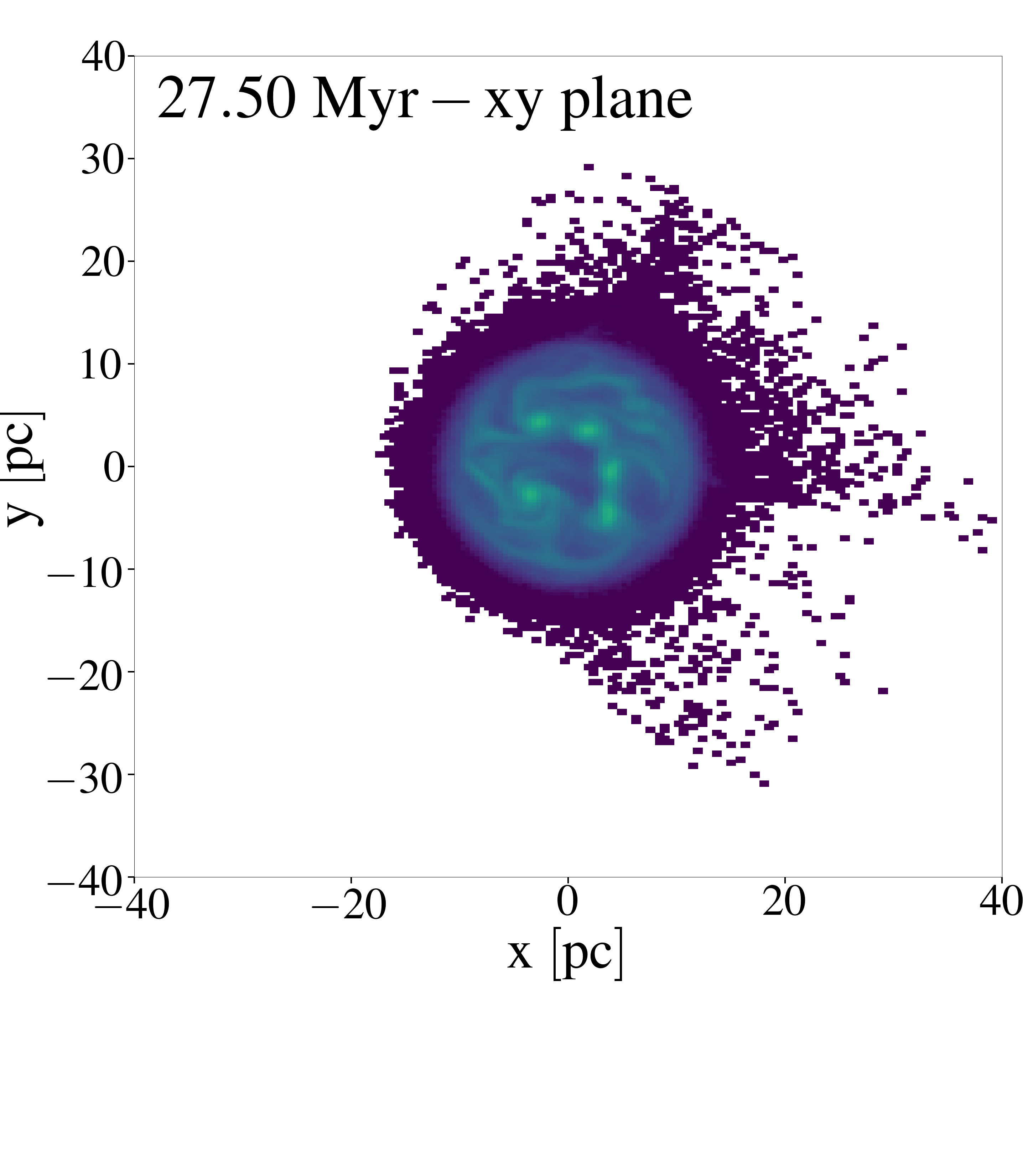}        
        \includegraphics[width=0.324\textwidth]{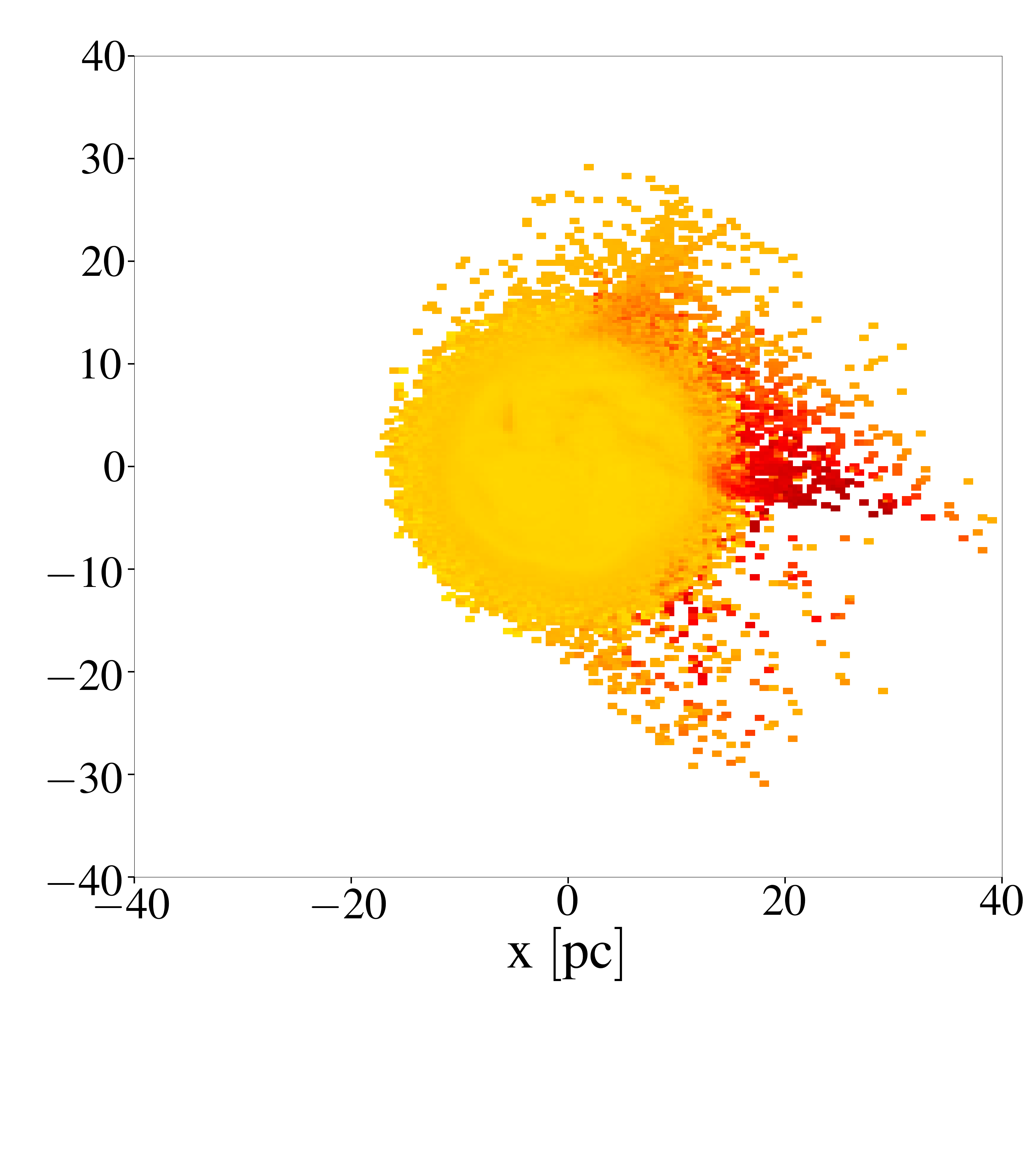}
         \includegraphics[width=0.324\textwidth]{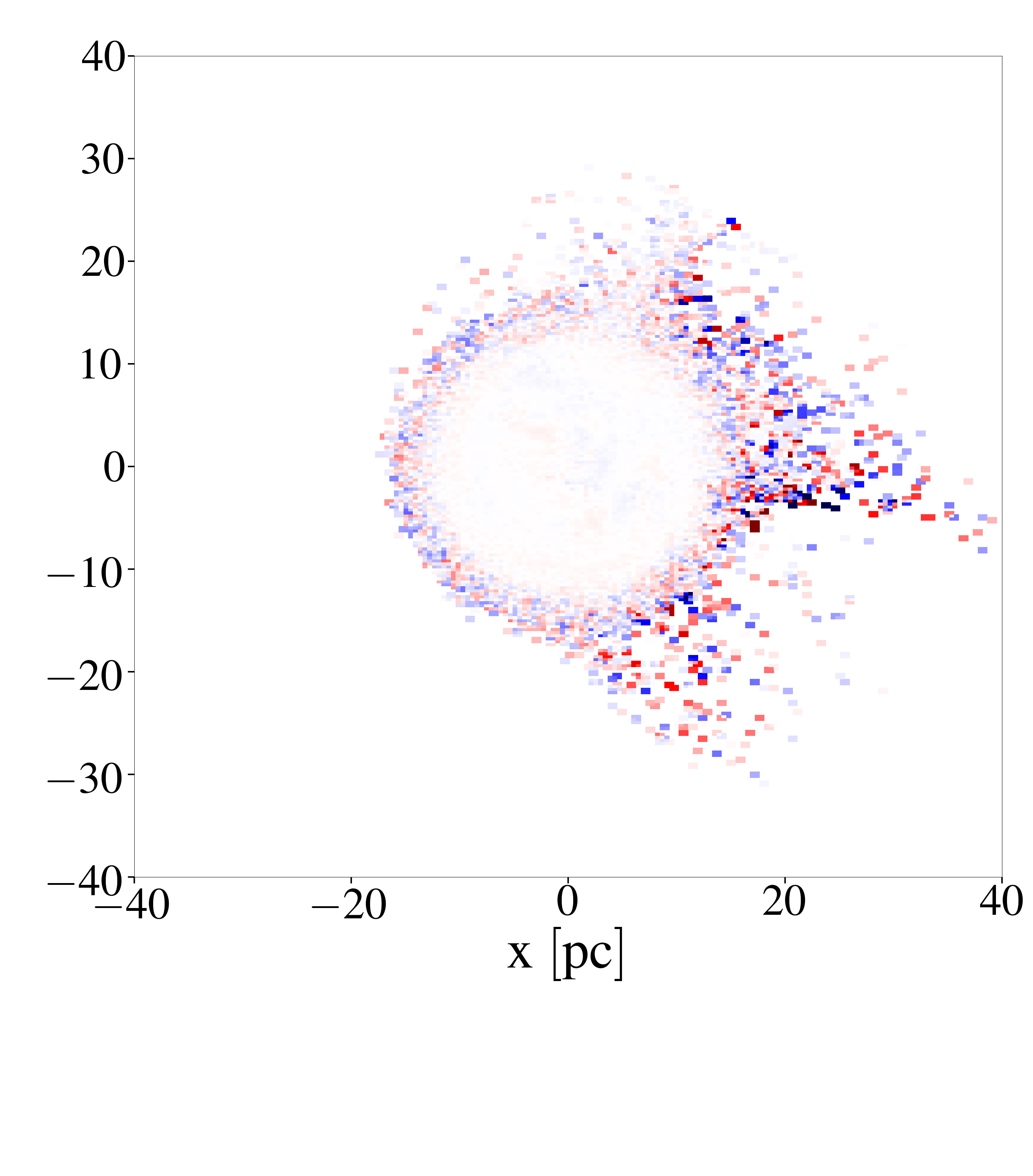}
      \\
       \vspace{-1.1cm}
      
        \includegraphics[width=0.324\textwidth]{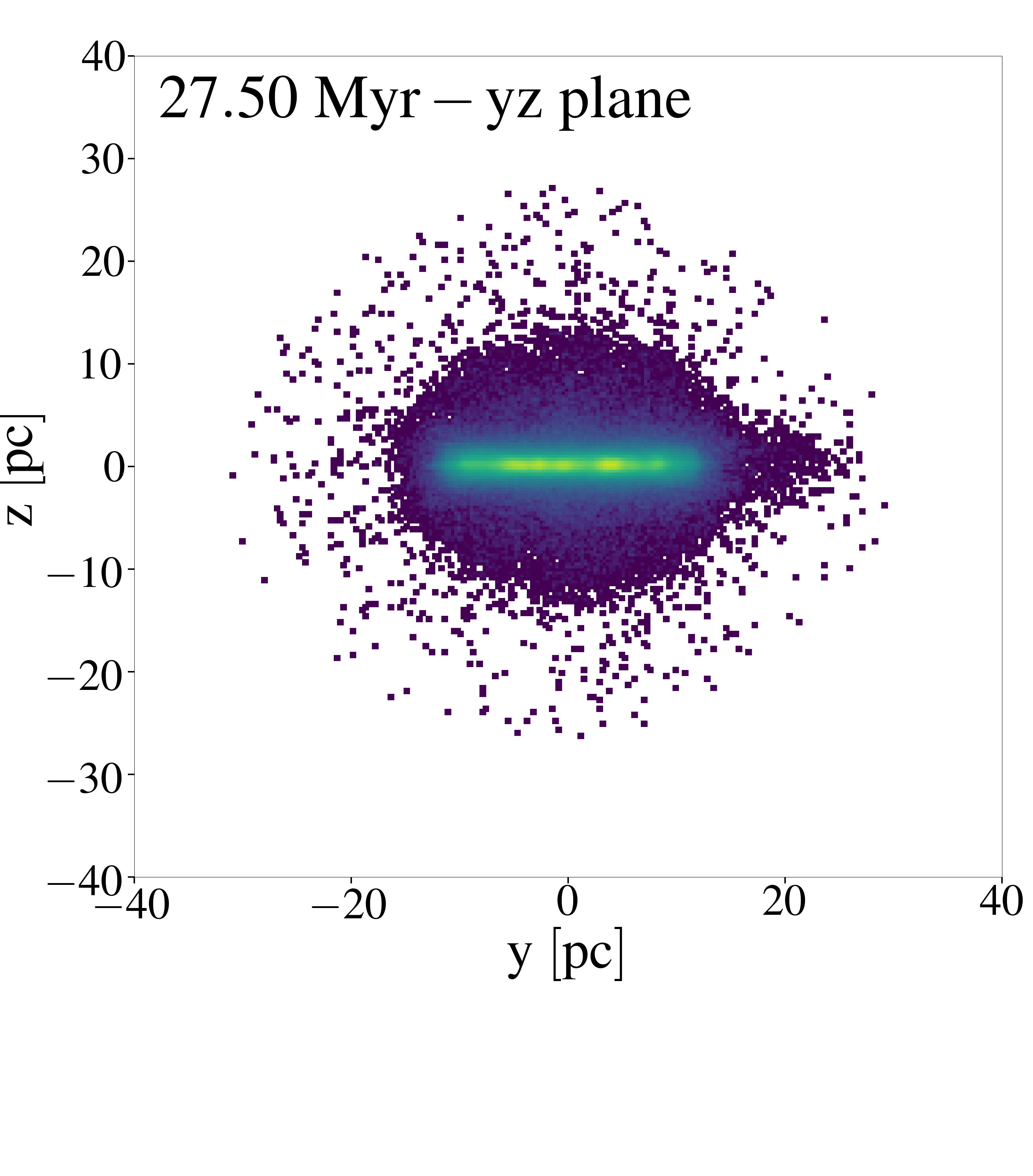}
        \includegraphics[width=0.324\textwidth]{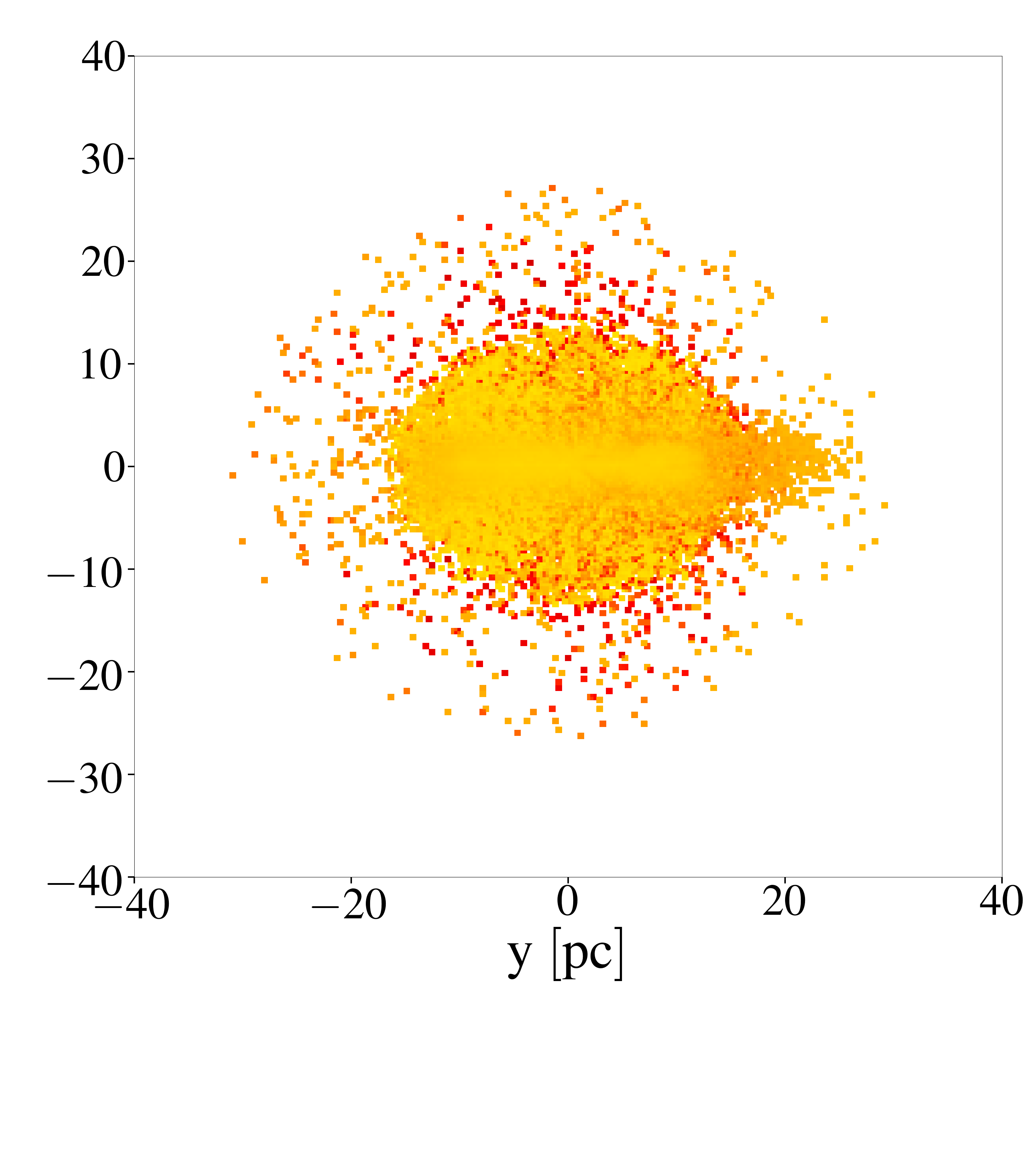}
           \includegraphics[width=0.324\textwidth]{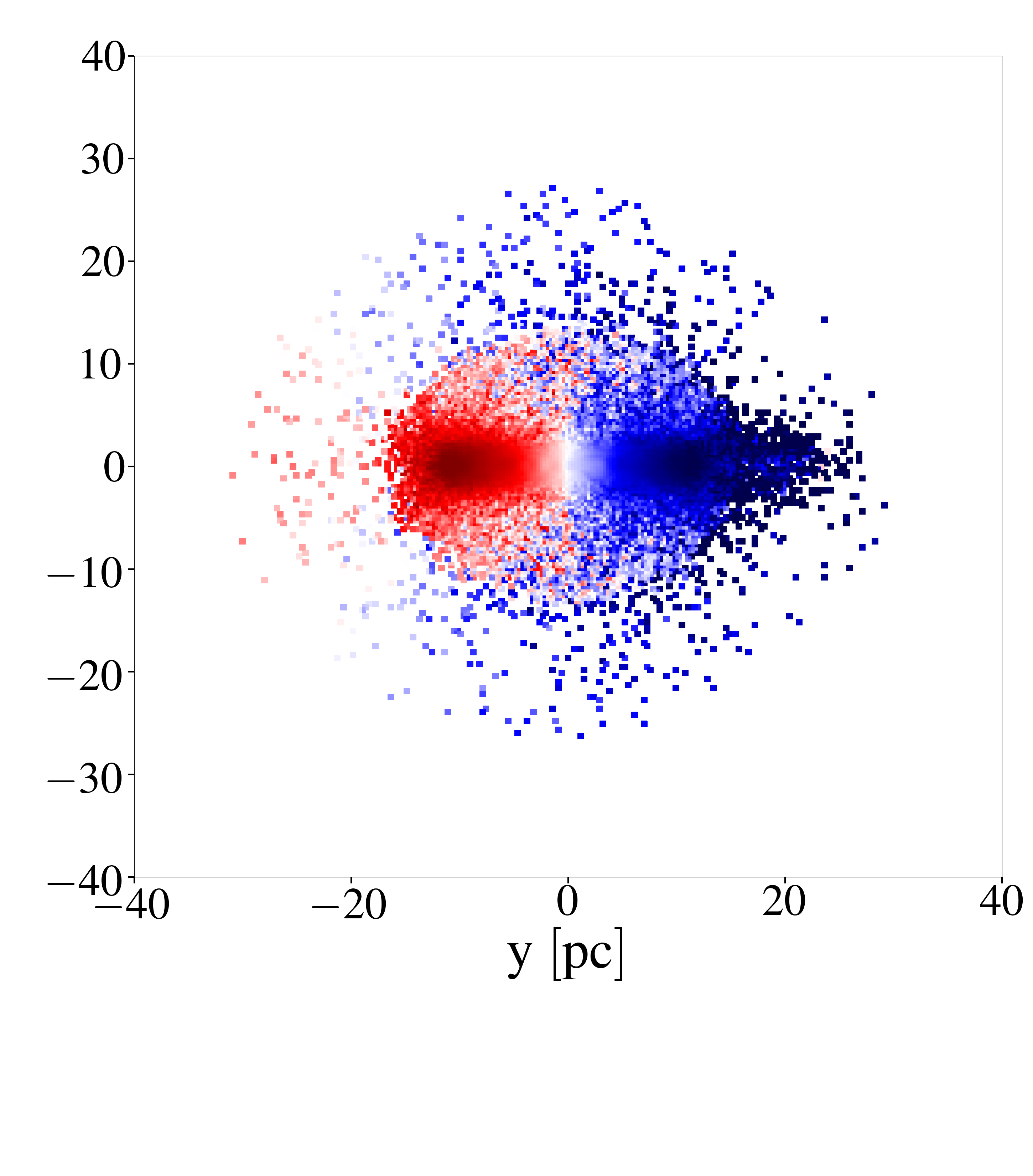}
        \\
        \vspace{-1.1cm}
          \includegraphics[width=0.324\textwidth]{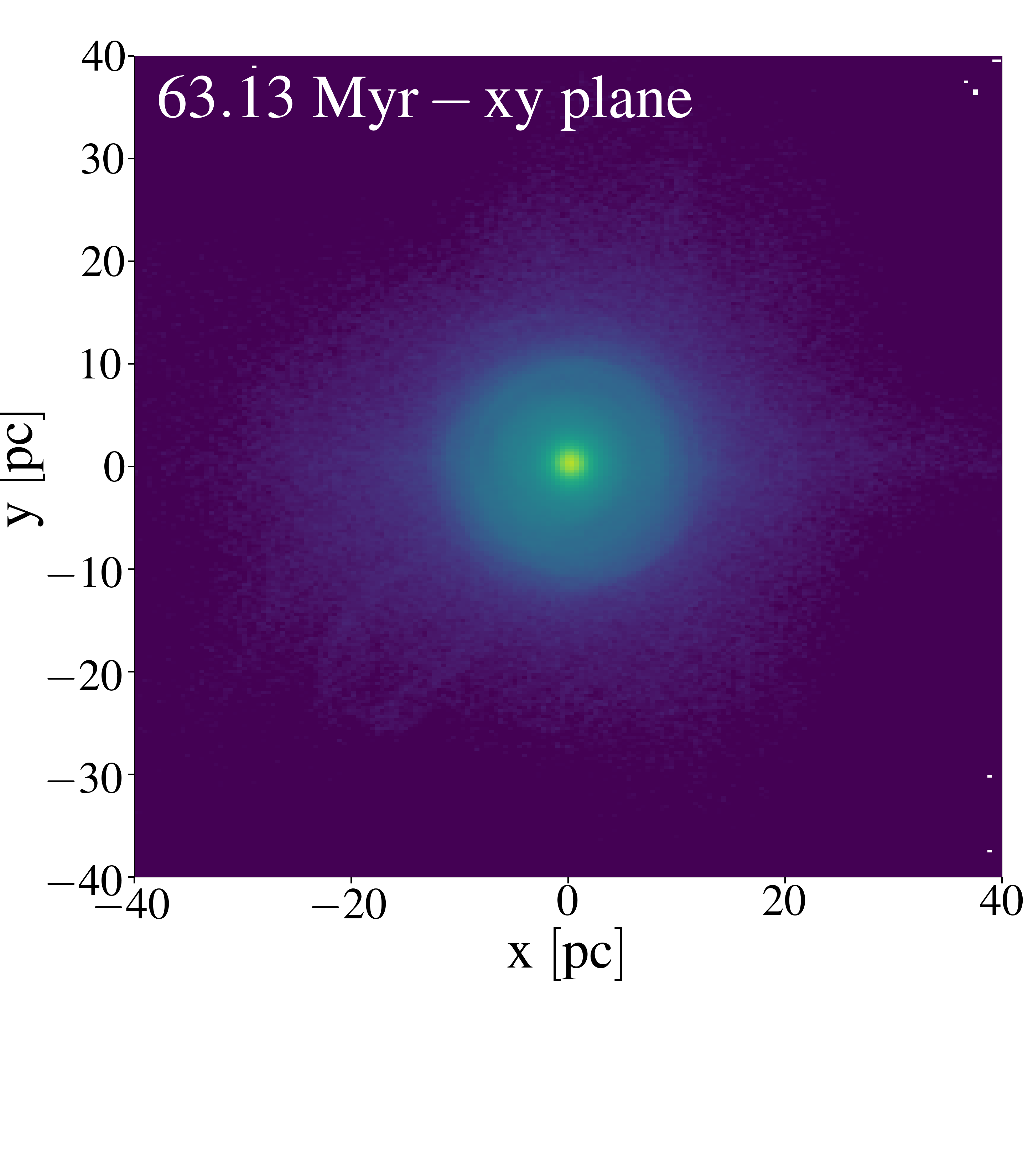}
        \includegraphics[width=0.324\textwidth]{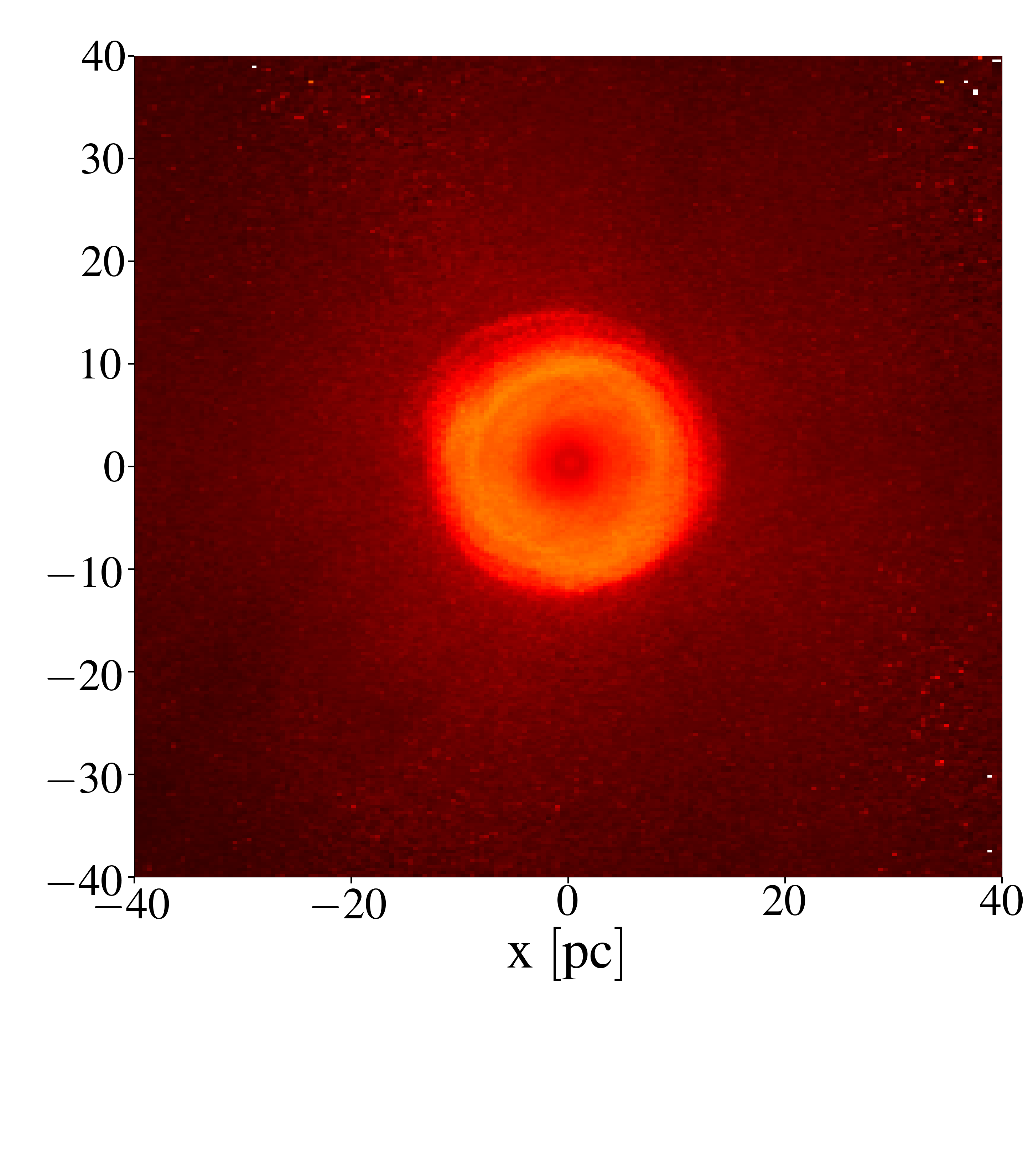}
        \includegraphics[width=0.324\textwidth]{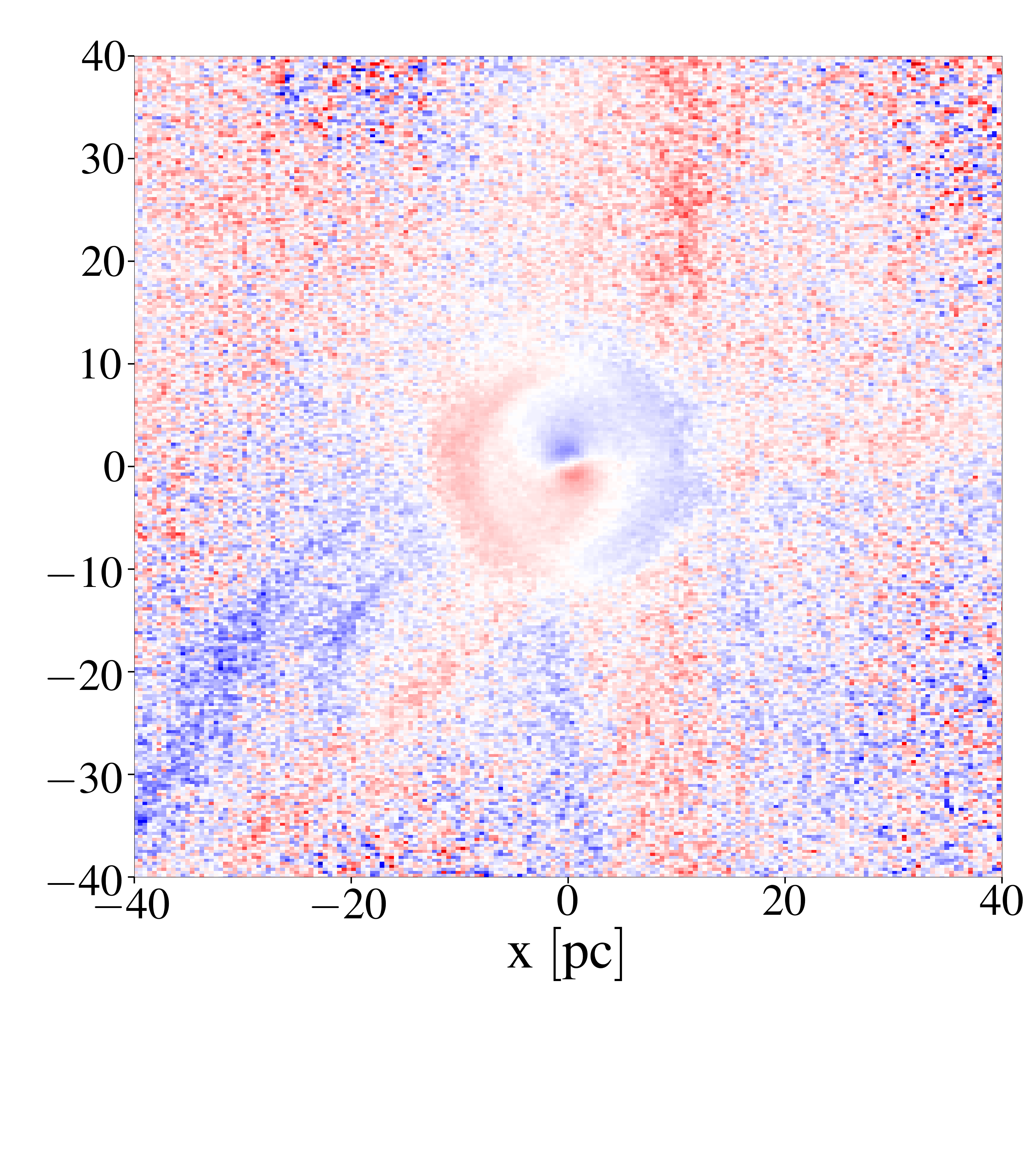}
        \\
        \vspace{-1.1cm}
        \includegraphics[width=0.324\textwidth]{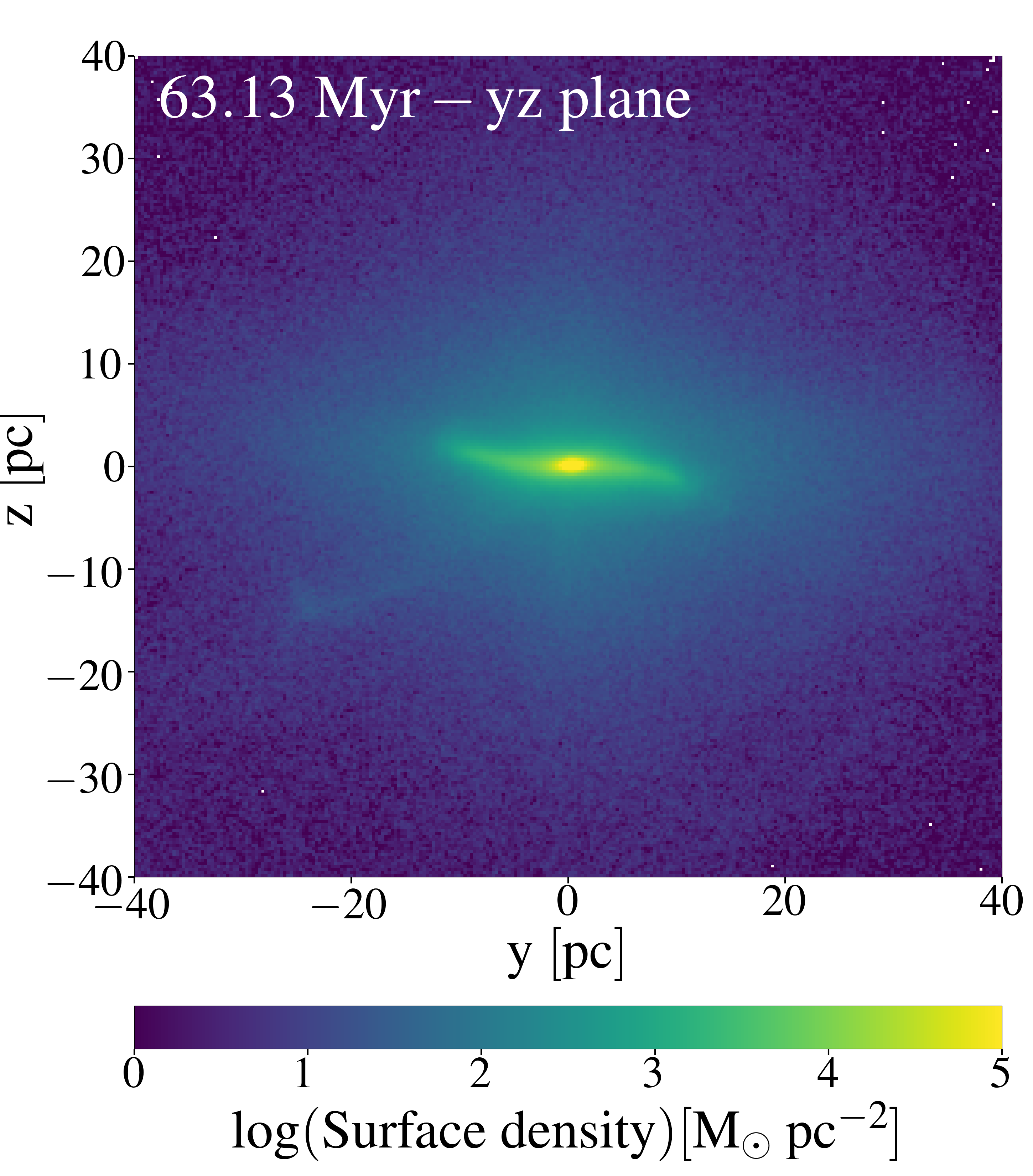}
        \includegraphics[width=0.324\textwidth]{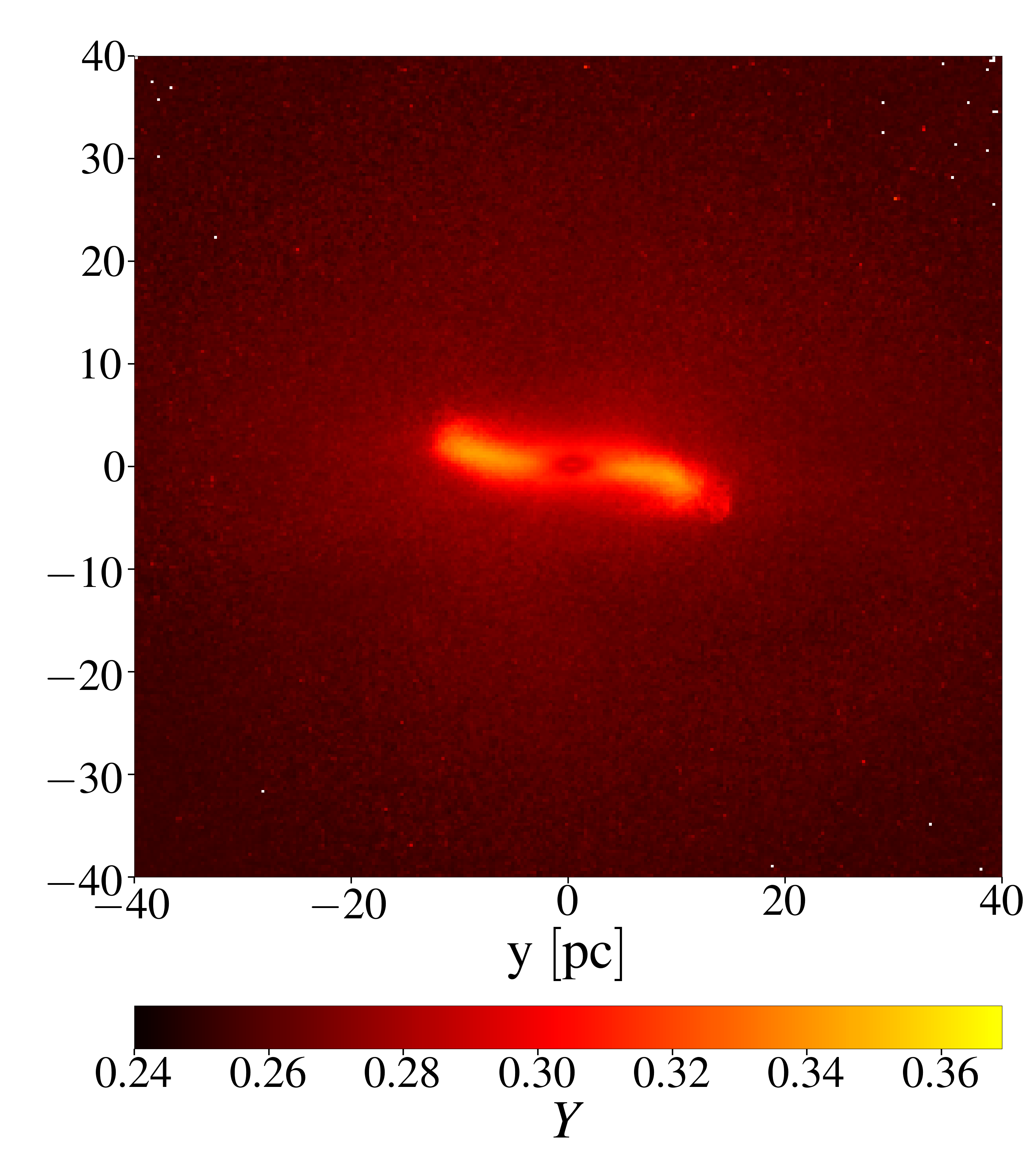}
        \includegraphics[width=0.324\textwidth]{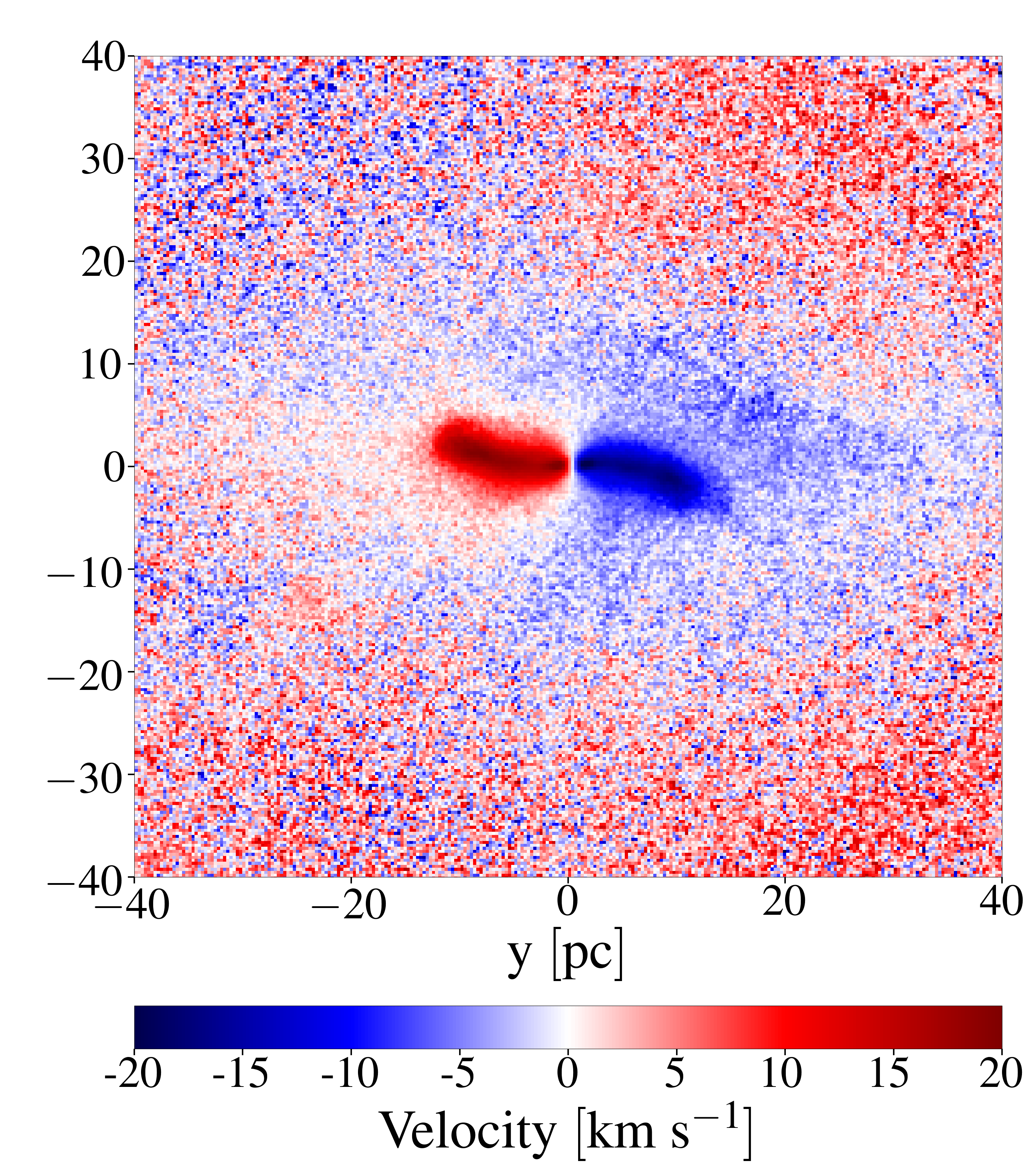}

\caption{Two-dimensional maps of the stellar component for the {\tt LDsbz} model. The reported quantities are as in Fig. \ref{fig:maps_part_LDanaz}.}
  \label{fig:maps_part_LDsbz}
\end{figure*}

\begin{figure*}
        \centering

        \includegraphics[width=0.454\textwidth,trim={0 0 0 8.0cm},clip]{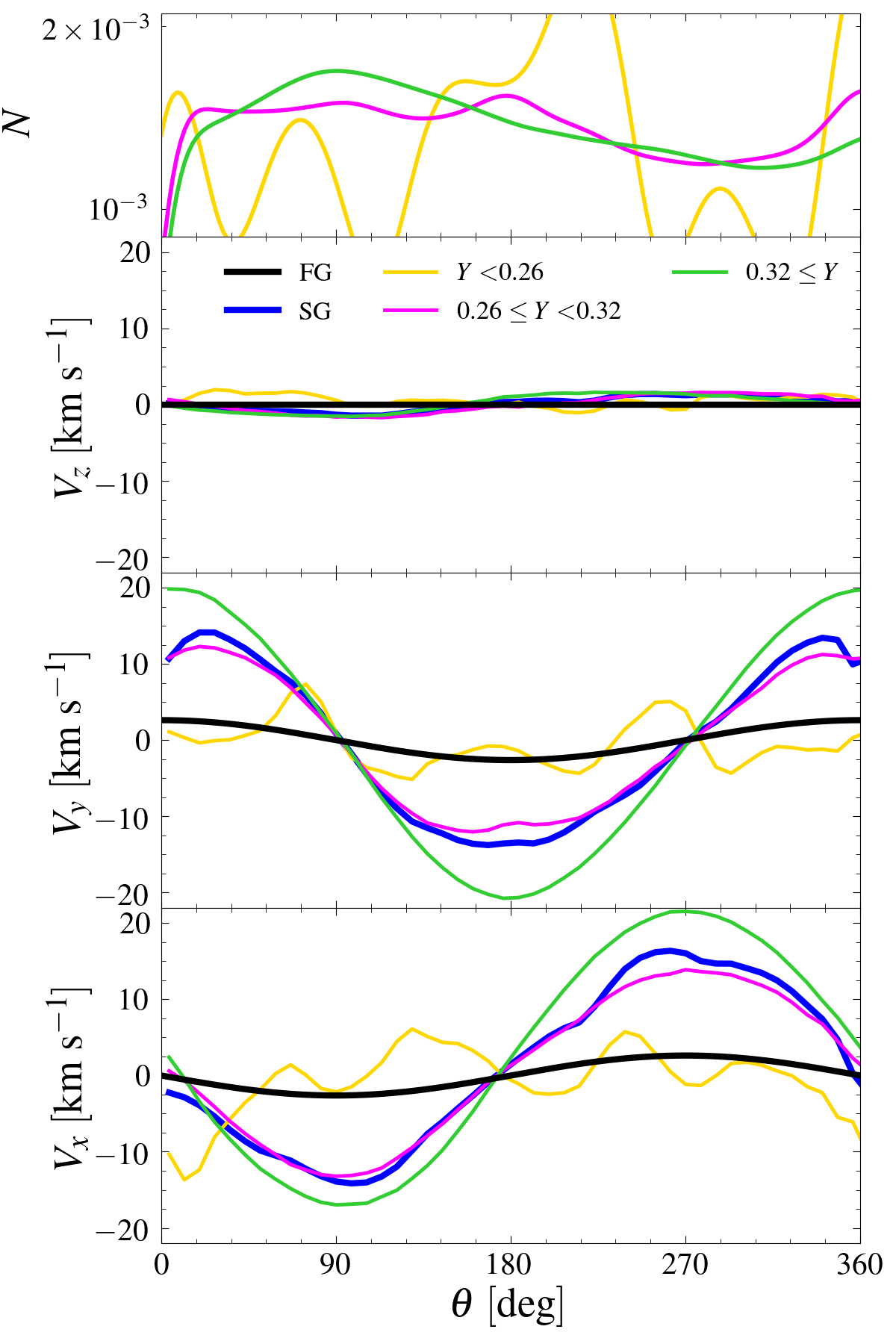}    
         %\hspace{0.01cm}
        \includegraphics[width=0.454\textwidth,trim={0 0 0 8.0cm},clip]{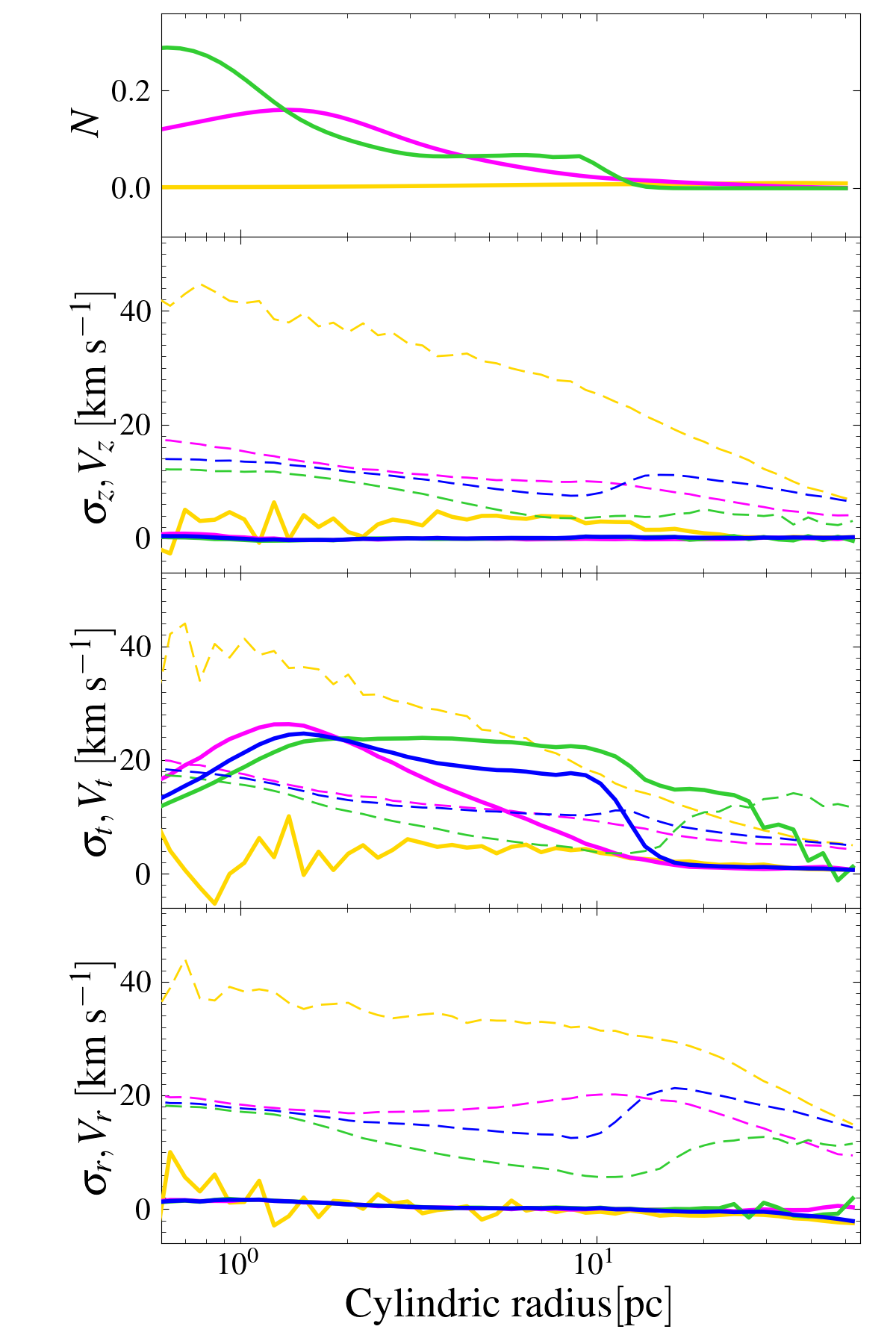}     
        \caption{ Stellar rotation profiles of the model {\tt LDsbz} at 65 Myr. The reported quantities are as in Fig. \ref{fig:sigmav_LDanaz}.}
  \label{fig:sigmav_LDsbz}
\end{figure*}

Until now, all presented results have been obtained assuming the analytic velocity profile of Eq. \ref{eq:anarot} for the FG component. For the low-density model, which is the one where the rotation is much stronger, we have performed, for a comparison, an additional run where we have adopted a solid body rotation profile ({\tt LDsbz}, see Table \ref{tab:simu} for the model description). Such velocity profile was earlier adopted in the 3D hydrodynamic simulations of  \citet{bekki2010}, \citet{bekki2011} and \citet{mckenzie2021}. 

Even in this model, several clumps are formed at around 10 Myr, and clearly visible also at 27 Myr in Fig. \ref{fig:maps_part_LDsbz}, however their stellar surface density is 3 times lower than in the simulation assuming the analytic velocity profile ({\tt LDanaz} model). {In addition, stars in the disk possess a stronger line of sight velocity in the $yz$ plane, as seen comparing it to Fig. \ref{fig:maps_part_LDanaz}, a consequence of the higher velocity imparted to FG ejecta in the outskirt in the solid body model.}
At the end of the simulation, the SG velocity is overall larger than the one obtained for the {\tt LDanaz} model due to the slightly larger velocity imposed to the AGB ejecta with the solid body profile. Helium-enhanced stars are, in fact, rotating faster in the outskirts, a consequence of the more extended disk ($\sim 12 $ pc), obtained for the model with a solid body profile. In particular, Figure \ref{fig:sigmav_LDsbz} shows the rotational amplitudes along the three Cartesian coordinates with an average value of $\sim 15{\rm\ km\ s^{-1}}$ along both the $x$ and $y$ axes. The $z$ component is instead lower than in the $\tt LDanaz$ model, even though the disk is tilted of about $15\degree$.% but does not show any sign of precession.

It is worth noticing that, even though the FG is set to rotate with a solid body profile, the SG has a pattern much more similar to Eq. \ref{eq:anarot}, with a peak at around the half-mass radius, which is compatible with the one obtained for the {\tt LDanaz} model.
}

%%%%%%%%%%%%%%%%%%%%%%%%%%%%%%%%%%%%%%%%%%%%%%%%%%

% Don't change these lines
\bsp	% typesetting comment
\label{lastpage}
\end{document}